\documentclass[a4paper,fleqn]{cas-sc}
\renewcommand{\printorcid}{}
\usepackage[authoryear,longnamesfirst]{natbib}
\usepackage{amsmath}
\usepackage{booktabs} 
\usepackage{subcaption} 
\usepackage{graphicx}
\usepackage{pgfplots}
\pgfplotsset{compat=1.6}
\usepackage{tabstackengine}
\usepackage[all]{nowidow}
\usepackage[utf8]{inputenc}
\usepackage{tikz}
\usetikzlibrary{spy}
\usetikzlibrary{er,positioning,bayesnet}
\usepackage{multicol}
\usepackage{algpseudocode,algorithm}
\usepackage{hyperref}
\usepackage[inline]{enumitem} 
\usepackage{mwe}
\usepackage{multirow}
\usepackage{ragged2e}
\usepackage{diagbox}
\usepackage{gensymb}
\usepackage{soul}
\renewcommand{\baselinestretch}{1.5}
\newcommand{\change}{}

\def\tsc#1{\csdef{#1}{\textsc{\lowercase{#1}}\xspace}}
\tsc{WGM}
\tsc{QE}
\tsc{EP}
\tsc{PMS}
\tsc{BEC}
\tsc{DE}

\begin{document}
\let\WriteBookmarks\relax
\def\floatpagepagefraction{1}
\def\textpagefraction{.001}
\shortauthors{Sahar Chehrazad et~al.}

\title [mode = title]{A fast and scalable \change{bottom-left-fill} algorithm to solve nesting problems
using a semi-discrete representation}           
\author[1]{Sahar Chehrazad}
\ead{Sahar.chehrazad@kuleuven.be}
\address[1]{KU Leuven, Department of Computer Science, campus Leuven, Belgium}

\author[1]{Dirk Roose}
\ead{Dirk.Roose@kuleuven.be}
\ead[URL]{https://people.cs.kuleuven.be/~dirk.roose/Site/Home.html}
\author[2]{Tony Wauters}
\ead{Tony.Wauters@kuleuven.be}
\address[2]{KU Leuven, Department of Computer Science, Technologiecampus Gent, Belgium}



\begin{abstract}
We present a fast algorithm to solve nesting problems based on a semi-discrete representation of both the 2D non-convex pieces and the strip.
The pieces and the strip are represented by a set of equidistant vertical line segments. The discretization algorithm uses a sweep-line method and applies minimal extensions to the line segments of a piece to ensure that non-overlapping placement of the segments, representing two pieces, cannot cause overlap of the original pieces.
We implemented a bottom-left-fill greedy placement procedure, using an optimised ordering of the segments overlap tests. 
The C++ implementation of our algorithm uses appropriate data structures that allow fast execution. It executes the bottom-left-fill algorithm for typical ESICUP data sets in a few milliseconds, even when rotation of the pieces is considered, and thus provides a suitable `building block' for integration in metaheuristics. Moreover, we show that the algorithm scales well when the number of pieces increases.

\end{abstract}


\maketitle

\section{Introduction}\label{section1}

Nesting problems, a branch of cutting and packing problems, are important for many industries, e.g.\,textile, sheet metal, leather, glass and paper industries, see \cite{1} and also in additive manufacturing. 
\change{In this paper the goal is to place 2D, possibly non-convex, pieces without overlap on a strip, a rectangular sheet with a fixed width and a variable length, while minimizing the used length of the strip. This is known as the irregular strip packing problem, which is a branch of `open dimension' problems, see \cite{wascher}.}

For a classification of nesting problems and for an overview of the many solution strategies we refer to review papers such as \cite{1, 27, 2}.
The solution strategies vary from a combination of specific heuristics to using a general-purpose MIP solver, based on a mathematical model of the problem.
Here we only highlight two aspects of the problem and its solution, namely a) continuous vs.\ discrete representation of the pieces and the strip and b) the quality of the solution vs.\ the run-time and the scalability of the algorithms.

Many methods use a continuous representation of the pieces and the strip. We refer to the use of the no-fit polygon, see \cite{20, 21}, or the use of `direct trigonometry' as in \cite{14}. Several methods combine a continuous representation of the pieces with a discretized strip, as for example in \cite{14} and in the dotted board model, see for example \cite{dotted-board}. A two-dimensional discretization of both the pieces and the strip is used in e.g.\,\cite{26, raster}. Of course, any discretization or approximation of the geometry reduces the potential quality of the solution, but the effect on the actually achieved solution quality can be negative or positive, since nearly all methods use heuristics because the nesting problem is NP-hard. The discretization effect will be smaller if the pieces and/or strip are discretized in only one direction, instead of in both directions, as mentioned in \cite{14}. However, such a semi-discretization has not received much attention so far.

Due to the iterative nature of most heuristic methods, the solution quality and the run-time are conflicting aspects. It is important that the basic operation in the iteration, e.g.\,a bottom-left placement of pieces in a given order or testing the constraints in a MIP method, is executed in a short time, since this allows, within a given computing budget, to perform more iterations, allowing potentially a better solution. The run-time of an algorithm not only depends on the number of operations performed, but also on the data structures and the data access pattern. Indeed, due to the availability of several pipelined functional units on a single core to perform floating point operations, with increasing length of the pipeline, and the availability of SIMD vector instructions, the gap between the time to perform a (floating point) operation and the time to access data from (main) memory has increased over time and is substantial in current processors. 
This so called `DRAM gap' is alleviated by the presence of several levels of cache memory, with different access times, due to the different latency and bandwidth to transfer data to the next level in the hierarchy.
To achieve high performance, the computer code must exploit the functionality of this memory hierarchy. When the data used in consecutive steps of the algorithm is stored in consecutive memory locations, the data can often be fetched from the cache, i.e.\,a high cache hit ratio is achieved. For more information we refer to \cite{Hager2011}.
Further, the computational complexity as a function of the number of pieces determines the scalability of the method, but this aspect is often not discussed in detail.  

The aim of this paper is to explore the potential of a semi-discretization of both the pieces and the strip to achieve high performance, and thus a short run-time, for a basic placement method that can be used as a basic operation or `building block' in a heuristic method to efficiently solve the nesting problem. We also aim to assess the scalability of the approach.
We expect that semi-discretization can lead to a good compromise between accuracy and computational complexity. Semi-discretization of both pieces and strip transforms the problem of placing pieces without overlap into the problem of placing line segments without overlap.
We will see that the core of the algorithm consists of simple comparisons of floating point numbers that are stored in regular data structures with a rather regular data access pattern, which allows to achieve high performance on current processors.

Only a few papers have explored the use of semi-discretization of pieces and/or strip.

In \cite{6}, a semi-discrete representation discretizes the pieces and the strip by a number of vertical rectangles of the same width. The lower and upper bound of each rectangle is equal to the lower and upper bound of the part of the piece lying in that rectangle. Each vertical rectangle of the board represents empty space on the board. A feasible position for each piece is determined by checking whether the empty rectangles of the strip can contain all the rectangles of the piece. Even though the proposed representation guarantees that overlap can never happen, it results in wasted area.

\cite{5} proposed another semi-discrete representation for the pieces and the strip. Here the discretization does not consist of vertical rectangles but consists of a set of equidistant vertical line segments. The semi-discretization of a piece is computed by traversing the circumference of the piece and intersecting the edges of the piece at \textit{x}-coordinates that are multiples of the distance between the vertical lines. The intersection points are stored in a table and the vertical line segments are derived from this table. 
They propose an extension procedure to avoid overlap of pieces during placement. However, the extension rules \change{do not} cover all cases, for example, when a whole edge of the piece lies between two of the vertical equidistant lines. Also, the implementation of the placement heuristic (a bottom-left-fill algorithm) is not sufficiently detailed to be able to re-implement the algorithm.

In \cite{12}, the semi-discrete representation discretizes the pieces and the board along the \textit{y}-axis while keeping the \textit{x}-axis continuous. A mixed integer programming model (MIP) is used to solve the nesting problem. The NFP is used to guarantee that the pieces do not overlap while placing them in the board. The results indicate that for data sets with more than ten pieces the execution time is prohibitively high.

Since the strip is also semi-discretized and represented by vertical line segments, the number of potential locations to place a piece is infinite, due to its continuous \textit{y}-axis property, as the `variable shift' approach in \cite{14}. To test whether a piece can be placed in a certain position, we must test whether all line segments representing an (extended) piece do not overlap the line segments of the (partially filled) strip.
Placements of pieces in the strip using this representation can result in a better solution than using a grid-based method, and requires only simple calculations when compared to the no-fit polygon method or other direct representations of the pieces. Note that, while computing the NFP of two convex pieces is simple and cheap, the NFP of non-convex pieces is more complex and time-consuming, especially when the problem specification allows rotation of the pieces, and is known to have numerical problems, see \cite{WautersTony2016Jara}.

In this paper we show that a bottom-left-fill greedy placement heuristic, even when considering several rotation angles of the pieces, can be executed in a few milliseconds on a single core of a current processor, for classical benchmark data sets, when appropriate data structures are used on which the required calculations can be executed efficiently.

Of course, the simple bottom-left-fill heuristic cannot achieve high quality solutions to the nesting problem, but many approaches in the literature use metaheuristics on top of this greedy heuristic, e.g.\,\cite{23, 14, 6, 5, 20, 21, raster}. These metaheuristics can be applied regardless of the representation of the pieces and the strip.

\change{The remainder of the paper is organized as follows. Section 2 presents the precise formulation of the problem and the data structures in C++ that we use to achieve high performance. In section 3, the construction of the semi-discrete representation using a sweep-line algorithm along with the extension algorithm is explained in detail. Section 4 describes the bottom-left-fill placement heuristic and how rotations are handled. In section 5 computational results on benchmark problems from the literature are presented, providing insight in the performance of the method, i.e.\,the quality of the solution, the run-time and the scalability w.r.t.\ the number of pieces. Finally, in section 6 we draw some conclusions and direction for future research.}

\section{\change{Problem formulation and data structures}}\label{newsection2}

\change{As mentioned in section\,\ref{section1}, this paper deals with the irregular strip packing problem, a branch of `open dimension' problems, see \cite{wascher}. The aim is to place two-dimensional, possibly non-convex, polygonal pieces without overlap on a rectangular strip, while minimizing the used length of the strip. In case rotation of pieces is allowed, the placement of each piece is checked with a discrete set of rotation angles. The basic operation in most of the placement heuristics is the placement of a piece in a partially filled strip according to some criterion. 
We present a bottom-left-fill placement heuristic, where both strip and pieces are semi-discretized. 
We discretize the pieces and the strip along the horizontal $x$-axis. Each vertical line with $x$-coordinate $x_i = i \times R$, is called a resolution line and the constant distance \textit{R} between the vertical lines indicates the resolution. 
The semi-discrete representation of the pieces and the strip consists of a set of vertical line segments on the resolution lines.} 
 
\change{We assume the bottom left corner of the axis-aligned bounding box of the piece has (local) coordinates $(x,y) = (0,0)$. 
Denote the bottom and top endpoints of the \textit{j}-th line segment on the \textit{i}-th resolution line by respectively $(x_i,b_{i,j})$ and $(x_i,t_{i,j})$ with $x_i = i\times R$, $i= 0,1,2,\ldots$. The number of line segments on a resolution line varies; for a convex piece, there is only one line segment on each resolution line. 
The partially filled semi-discretized strip consists of vertical line segments, indicating space occupied by already placed pieces. We assume that the bottom left corner of the strip has (strip) coordinates (0,0). Denote the bottom and top endpoints of the \textit{l}-th line segment on the \textit{k}-th resolution line by $(x_k,b^s_{k,l})$ and $(x_k,t^s_{k,l})$ with $x_k = k\times R$, $k= 0,1,2,\ldots$, where superscript \textit{s} refers to the strip.}

\change{In the bottom-left-fill heuristic a piece is placed in the left-most and bottom-most position in the partially filled strip without overlap. For a semi-discretized piece and strip this requires to find a ‘translation vector’ $t = (x_t,y_t)$ with $x_t = m.R$ such that the \textit{y}-intervals $(b_{i,j}+y_t , t_{i,j}+y_t)$ and  $(b^s_{i+m,l} , t^s_{i+m,l})$, representing \textit{y}-coordinates of the endpoints of the line segments do not overlap $\forall i = 0,1,2,...$, $\forall j$, $\forall l$, such that $m$ is minimal and, for that $m$, $y_t$ is minimal. This requires a number of tests, involving only the \textit{y}-coordinates of the line segments of both the piece and partially filled strip, see Fig.\,\ref{fig115}. 
Ordering the line segments of both the piece and partially filled strip in each resolution line such that $b_{i,j+1} > t_{i,j}$ and $b^s_{k,l+1} > t^s_{k,l}$, limits the number of tests since overlap between $(b_{i,j}+y_t , t_{i,j}+y_t)$ and $(b^s_{i+m,q} , t^s_{i+m,q})$ with $q > l$ is not possible if $t_{i,j}+y_t < t^s_{i+m,l}$. }

\change{The required data are thus the ordered set of \textit{y}-intervals $(b_{i,j}, t_{i,j})$ for the piece and the ordered set of intervals  $(b^s_{k,l} , t^s_{k,l})$ for the partially filled strip.
To select appropriate data structures, we take into account that \textit{a)}
for each translation vector, all the line segments of a piece can be involved in the overlap tests
and \textit{b)} the run-time of a program is nowadays mainly determined by the data access pattern, rather than by the number of operations, due to the difference in time to load data from the memory hierarchy and to execute floating point operations. 
In order to optimize the cache hit ratio the \textit{y}-intervals $(b_{i,j} , t_{i,j})$, $j = 0,1,2,...$, $i = 0,1,2,...,$ representing the \textit{y}-coordinates of the line segments of the piece should be stored in consecutive memory locations. Hence a suitable data structure in C++ is an array of arrays or a vector of vectors. Considering the fact that initially the number of segments on each resolution line of the piece and strip is not known, we use vectors. Indeed, vector elements are placed in contiguous storage, as in arrays, so that they can be accessed and traversed efficiently using iterators. In addition, vectors can be initialized with a certain length, but they automatically re-scale their memory as needed. However, vectors do not reallocate memory each time an element is added. Instead, vector containers may allocate some extra storage to accommodate for possible growth in length. Accessing an arbitrary element of the vector is very efficient, as in arrays, and also inserting an element at the end of a vector is done in constant time, see \cite{vector}.
For similar reasons, the ordered set of intervals $(b^s_{k,l} , t^s_{k,l})$ representing the \textit{y}-coordinates of the line segments of the partially filled strip are also stored in a vector of vectors.}

\section{Semi-discretization of the pieces}\label{section2}

\subsection{Sweep-line algorithm} \label{sweep-line}
We implement the construction of the semi-discrete representation of a polygonal piece using a sweep-line algorithm, see \cite{10}. 
\change{The intersection of the sweep-line, i.e.\,a vertical line traversing the piece from left to right at $x_i = i\times R$, $i= 0,1,2,\ldots$, with the edges of the piece gives the \textit{y}-coordinates $(b_{i,j} , t_{i,j})$ of the endpoints of the line segments .} 
A vector \texttt{Events} is created containing the \textit{x}-coordinates of the vertices of the polygonal piece (in case several vertices have equal \textit{x}-coordinate it is added once), sorted in increasing order. The vertical sweep-line traverses the piece from left to right and the algorithm maintains a vector \texttt{ActiveEdges}, containing the information of the edges that exist for the \textit{x}-coordinates in the interval [\texttt{Events[m]} , \texttt{Events[m+1]}], sorted from bottom to top. \texttt{ActiveEdges} must be updated when the sweep-line passes through an event.

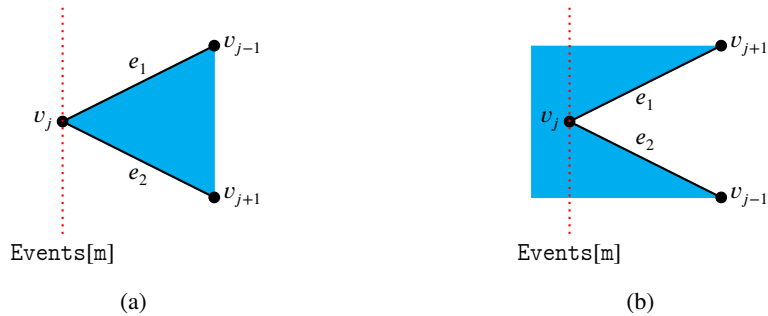
\begin{figure}[!ht]
\centering
\subfloat[\label{1a}]{%
\begin{tikzpicture}
\filldraw[cyan , thin] (0,0) -- (2,-1) -- (2,1);
\draw[black , thick] (0,0) --node[black , anchor=south] {$e_1$} (2,1);
\draw[black , thick] (0,0) --node[black , anchor=north] {$e_2$} (2,-1);
\filldraw [black] (0,0) circle (2pt) node[black , anchor=east] {$v_j$};
\draw[red , dotted , thick] (0,1.5) -- (0,-1.5) node[black , anchor=north] {$\mathtt{Events[m]}$};
\filldraw [black] (2,1) circle (2pt) node[black , anchor=west] {$v_{j-1}$};
\filldraw [black] (2,-1) circle (2pt)node[black , anchor=west] {$v_{j+1}$};
\end{tikzpicture}
}
\hfil
\subfloat[\label{1b}]{%
\begin{tikzpicture}
\filldraw[cyan , thin] (6,1) -- (4,0) -- (6,-1) -- (3.5,-1) -- (3.5,1);
\draw[black , thick] (4,0) --node[black , anchor=north] {$e_1$} (6,1);
\draw[black , thick] (4,0) --node[black , anchor=south] {$e_2$} (6,-1);
\filldraw [black] (4,0) circle (2pt) node[black , anchor=east] {$v_{j}$};
\draw[red , dotted , thick] (4,1.5) -- (4,-1.5) node[black , anchor=north] {$\mathtt{Events[m]}$};
\filldraw [black] (6,1) circle (2pt) node[black , anchor=west] {$v_{j+1}$};;
\filldraw [black] (6,-1) circle (2pt)node[black , anchor=west] {$v_{j-1}$};
\end{tikzpicture}}
\caption{\change{Updating \texttt{ActiveEdges}: case a: \texttt{Events[m]} is the \textit{x}-coordinate of the left endpoint of two edges, $e_1$ and $e_2$. Information of edges $e_1$ and $e_2$ are added to \texttt{ActiveEdges}.}}
\label{fig1}
\end{figure}

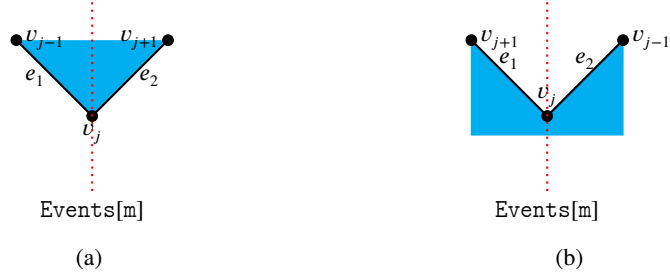
\begin{figure}[!ht]
\centering
\subfloat[\label{2a}]{%
\begin{tikzpicture}
\filldraw[cyan , thin] (0,1) -- (1,0) -- (2,1);
\draw[black , thick] (0,1) --node[black , anchor=east] {$e_1$} (1,0);
\draw[black , thick] (1,0) --node[black , anchor=west] {$e_2$} (2,1);
\filldraw [black] (1,0) circle (2pt) node[black , anchor=north] {$v_{j}$};
\draw[red , dotted , thick] (1,1.5) -- (1,-1) node[black , anchor=north] {$\mathtt{Events[m]}$};
\filldraw [black] (0,1) circle (2pt) node[black , anchor=west] {$v_{j-1}$};
\filldraw [black] (2,1) circle (2pt) node[black , anchor=east] {$v_{j+1}$};
\end{tikzpicture}
}
\hfil
\subfloat[\label{2b}]{%
\begin{tikzpicture}
\filldraw[cyan , thin] (4,1) -- (4,-1/4) -- (6,-1/4) -- (6,1) -- (5,0);
\draw[black , thick] (4,1) --node[black , anchor=south] {$e_1$} (5,0);
\draw[black , thick] (5,0) --node[black , anchor=south] {$e_2$} (6,1);
\filldraw [black] (5,0) circle (2pt) node[black , anchor=south] {$v_{j}$};
\draw[red , dotted , thick] (5,1.5) -- (5,-1) node[black , anchor=north] {$\mathtt{Events[m]}$};
\filldraw [black] (4,1) circle (2pt)node[black , anchor=west] {$v_{j+1}$};
\filldraw [black] (6,1) circle (2pt)node[black , anchor=west] {$v_{j-1}$};
\end{tikzpicture}}
\caption{\change{Updating \texttt{ActiveEdges}: case b: \texttt{Events[m]} is the \textit{x}-coordinate of the right endpoint of edge $e_1$ and the left endpoint of edge $e_2$. Information of edge $e_1$ is removed and information of edge $e_2$ is added to \texttt{ActiveEdges}.}}
\label{fig2}
\end{figure}

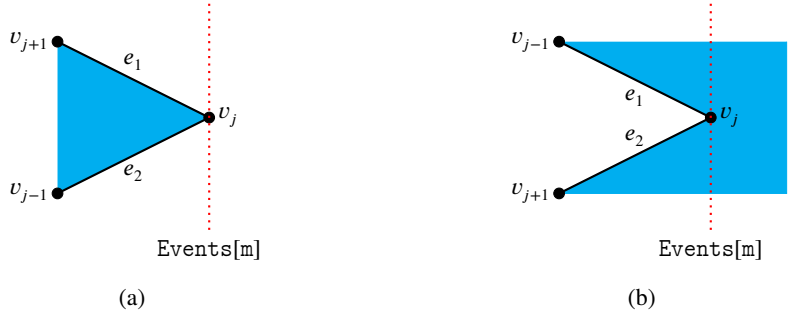
\begin{figure}[!ht]
\centering
\subfloat[\label{3a}]{%
\begin{tikzpicture}
\filldraw[cyan , thin] (0,1) -- (2,0) -- (0,-1);
\draw[black , thick] (0,1) --node[black , anchor=south] {$e_1$} (2,0);
\draw[black , thick] (0,-1) --node[black , anchor=north] {$e_2$} (2,0);
\filldraw [black] (2,0) circle (2pt) node[black , anchor=west] {$v_{j}$};
\draw[red , dotted , thick] (2,1.5) -- (2,-1.5) node[black , anchor=north] {$\mathtt{Events[m]}$};
\filldraw [black] (0,1) circle (2pt) node[black , anchor=east] {$v_{j+1}$};
\filldraw [black] (0,-1) circle (2pt)node[black , anchor=east] {$v_{j-1}$};
\end{tikzpicture}
}
\hfil
\subfloat[\label{3b}]{%
\begin{tikzpicture}
\filldraw[cyan , thin] (4,1) -- (6,0) -- (4,-1) -- (7,-1) -- (7,1);
\draw[black , thick] (4,1) --node[black , anchor=north] {$e_1$} (6,0);
\draw[black , thick] (4,-1) --node[black , anchor=south] {$e_2$} (6,0);
\filldraw [black] (6,0) circle (2pt) node[black , anchor=west] {$v_{j}$};
\draw[red , dotted , thick] (6,1.5) -- (6,-1.5) node[black , anchor=north] {$\mathtt{Events[m]}$};
\filldraw [black] (4,1) circle (2pt) node[black , anchor=east] {$v_{j-1}$};
\filldraw [black] (4,-1) circle (2pt)node[black , anchor=east] {$v_{j+1}$};
\end{tikzpicture}}
\caption{\change{Updating \texttt{ActiveEdges}: case c: \texttt{Events[m]} is the \textit{x}-coordinate of the right endpoint of two edges, $e_1$ and $e_2$. Information of edges $e_1$ and $e_2$ are removed from \texttt{ActiveEdges}.}}
\label{fig3}
\end{figure}

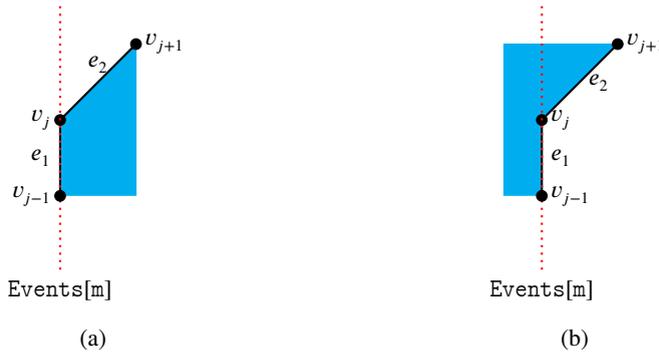
\begin{figure}[!ht]
\centering
\subfloat[\label{4a}]{%
\begin{tikzpicture}
\filldraw[cyan , thin] (0,0) -- (0,1) -- (1,2) -- (1,0);
\draw[black , thick] (0,0) --node[black , anchor=east] {$e_1$} (0,1);
\draw[black , thick] (0,1) --node[black , anchor=south] {$e_2$} (1,2);
\filldraw [black] (0,1) circle (2pt) node[black , anchor=east] {$v_{j}$};
\draw[red , dotted , thick] (0,2.5) -- (0,-1) node[black , anchor=north] {$\mathtt{Events[m]}$};
\filldraw [black] (0,0) circle (2pt)node[black , anchor=east] {$v_{j-1}$};
\filldraw [black] (1,2) circle (2pt) node[black , anchor=west] {$v_{j+1}$};
\end{tikzpicture}
}
\hfil
\subfloat[\label{4b}]{%
\begin{tikzpicture}
\filldraw[cyan , thin] (3,0) -- (3,1) -- (4,2) -- (2.5,2) -- (2.5,0);
\draw[black , thick] (3,0) --node[black , anchor=west] {$e_1$}  (3,1);
\draw[black , thick] (3,1) --node[black , anchor=west] {$e_2$}  (4,2);
\filldraw [black] (3,1) circle (2pt) node[black , anchor=west] {$v_{j}$};
\draw[red , dotted , thick] (3,2.5) -- (3,-1) node[black , anchor=north] {$\mathtt{Events[m]}$};
\filldraw [black] (3,0) circle (2pt)node[black , anchor=west] {$v_{j-1}$};
\filldraw [black] (4,2) circle (2pt)node[black , anchor=west] {$v_{j+1}$};
\end{tikzpicture}}
\caption{\change{Updating \texttt{ActiveEdges}: case d: \texttt{Events[m]} is the \textit{x}-coordinate of the left endpoint of edge $e_2$ and edge $e_1$ is vertical. Information of edge $e_2$ is added to \texttt{ActiveEdges}. }}
\label{fig4}
\end{figure}

\begin{figure}[!ht]
\centering
\subfloat[\label{5a}]{%
\begin{tikzpicture}
\filldraw[cyan , thin] (0,0) -- (1,1) -- (1,2) -- (0,2);
\draw[black , thick] (0,0) --node[black , anchor=west] {$e_1$} (1,1);
\draw[black , thick] (1,1) --node[black , anchor=west] {$e_2$} (1,2);
\filldraw [black] (1,1) circle (2pt) node[black , anchor=west] {$v_{j}$};
\draw[red , dotted , thick] (1,2.5) -- (1,-1) node[black , anchor=north] {$\mathtt{Events[m]}$};
\filldraw [black] (0,0) circle (2pt) node[black , anchor=west] {$v_{j-1}$};
\filldraw [black] (1,2) circle (2pt) node[black , anchor=west] {$v_{j+1}$};
\end{tikzpicture}
}
\hfil
\subfloat[\label{5b}]{%
\begin{tikzpicture}
\filldraw[cyan , thin] (3,0) -- (4,1) -- (4,2) -- (5,2) -- (5,0);
\draw[black , thick] (3,0) --node[black , anchor=east] {$e_1$} (4,1);
\draw[black , thick] (4,1) --node[black , anchor=east] {$e_2$} (4,2);
\filldraw [black] (4,1) circle (2pt) node[black , anchor=east] {$v_{j}$};
\draw[red , dotted , thick] (4,2.5) -- (4,-1) node[black , anchor=north] {$\mathtt{Events[m]}$};
\filldraw [black] (3,0) circle (2pt) node[black , anchor=west] {$v_{j-1}$};
\filldraw [black] (4,2) circle (2pt) node[black , anchor=west] {$v_{j+1}$};
\end{tikzpicture}}
\caption{\change{Updating \texttt{ActiveEdges}: case e: \texttt{Events[m]} is the \textit{x}-coordinate of the right endpoint of edge $e_1$ and edge $e_2$ is
vertical. Information of edge $e_1$ is removed from \texttt{ActiveEdges}.}}
\label{fig5}
\end{figure}
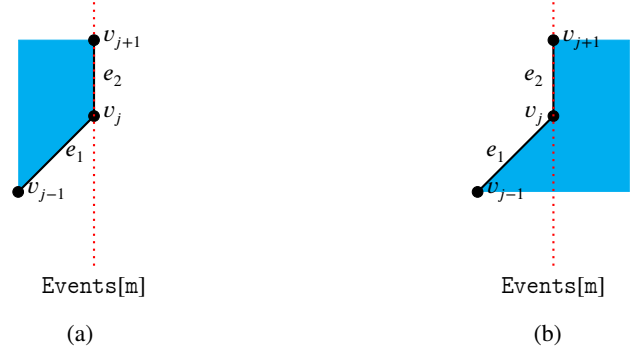

The actions required to update \texttt{ActiveEdges} when the sweep-line passes through \texttt{Events[m]} depend on the position of the vertex, associated with \texttt{Events[m]}, w.r.t.\ its adjacent edges. Five cases must be distinguished, and the corresponding actions are listed in Table\,\ref{tab:my_label} and illustrated in Figs.\,\ref{fig1}-\ref{fig5}.
If \texttt{Events[m]} is the \textit{x}-coordinate of the left endpoint of a non-vertical edge \textit{e\textsubscript{i}}, the information of edge \textit{e\textsubscript{i}} is added to \texttt{ActiveEdges}. Also, the information of the non-vertical edges which have \texttt{Events[m]} as the \textit{x}-coordinate of their right endpoint will be removed from \texttt{ActiveEdges}. In case \texttt{Events[m]} is the \textit{x}-coordinate of the endpoints of a vertical edge, the edge is not added to \texttt{ActiveEdges}.

\begin{table}[]
    \centering
    \caption{Sweep-line algorithm: different cases for the update of the \texttt{ActiveEdges} data structure}
    \begin{tabular}{|c|c|c|}
    \hline
    Case     &     Position of vertex, associated with \texttt{Events[m]},                            &   Update of \texttt{ActiveEdges} \\
             &             on its adjacent edges                                                      &         \\\hline
      a      &     \texttt{Events[m]} is the \textit{x}-coordinate of the left                        &   Add information of edges $e_1$\\    
             &     endpoint of two edges, $e_1$ and $e_2$                                             &	  and $e_2$.       \\\hline
      b      &     \texttt{Events[m]} is the \textit{x}-coordinate of the                             &   Remove information of edge $e_1$ \\
             &     right endpoint of edge $e_1$ and the left                                          &   and add information of edge $e_2$.\\
             &     endpoint of edge $e_2$.                                                            &    \\\hline
      c      &     \texttt{Events[m]} is the \textit{x}-coordinate of the right                       &   Remove information of edges  \\         
             &     endpoint of two edges, $e_1$ and $e_2$.                                            &	  $e_1$ and $e_2$.  \\\hline
      d      &     \texttt{Events[m]} is the \textit{x}-coordinate of the left                        &   Add information of edge $e_2$.   \\
             &     endpoint of edge $e_2$ and edge $e_1$                                              &               \\
             &     is vertical.                                                                       &     \\\hline
     e       &     \texttt{Events[m]} is the \textit{x}-coordinate of the right                       &   Remove information of edge $e_1$.   \\
             &     endpoint of the edge $e_1$ and edge                                                &               \\
             &    $e_2$ is vertical.                                                                  &            \\\hline
             
    \end{tabular}
    \label{tab:my_label}
\end{table}

After updating \texttt{ActiveEdges}, for every resolution line between the current event and the next event, the \textit{y}-coordinates of the intersection points of the resolution line and the edges in \texttt{ActiveEdges} are computed and saved in a vector \texttt{Ypoints}. 

Special cases occur when vertices lie on a resolution line, i.e.\,when an event occurs at $x_i = i\times R$, $i= 0,1,2,\ldots$. 
\change{In such cases, \texttt{Ypoints} is computed as follows. If vertex $v_{j}$ is the left endpoint (see Fig.\,\ref{fig1}) or the right endpoint (see Fig.\,\ref{fig3}) of both adjacent edges, the \textit{y}-coordinate of the vertex is added twice to \texttt{Ypoints} (indicating a segment of length zero when $v_{j}$ is convex). 
In case a vertical edge lies on a resolution line, the \textit{y}-coordinates of its endpoints are added to vector \texttt{Ypoints} as follows. If vertex $v_{j}$ is the upper endpoint of a vertical edge and the piece occupies the space above $v_{j}$ (see Fig.\,\ref{4b}), then the \textit{y}-value of vertex $v_{j}$ is added twice to \texttt{Ypoints}, otherwise it is added once (see Fig.\,\ref{4a}).
If vertex $v_{j}$ is the lower endpoint of a vertical edge and the piece occupies the space below $v_{j}$ (see Fig.\,\ref{5b}), then the \textit{y}-value of vertex $v_{j}$ is added twice to \texttt{Ypoints}, otherwise it is added once (see Fig.\,\ref{5a}).}

\change{Traversing the intersection points in \texttt{Ypoints} according to increasing \textit{y}-coordinate gives the \textit{y}-intervals 
($b_{i,j}, t_{i,j}$) representing the \textit{y}-coordinates of the endpoints of the line segments of the piece, where elements with even indexes in \texttt{Ypoints} ($b_{i,j}$) are the lowest points of the intervals and elements with odd indexes in \texttt{Ypoints} are the highest points ($t_{i,j}$). We assign a label to each interval, representing the position of the interval in the semi-discretized piece. This position label will be used during placement to allow that pieces touch each other, see section \ref{section3}.  
Therefore, each interval is stored as a tuple ($b_{i,j} , t_{i,j} , P$), with label \textit{P}.
Considering the position of the intervals in a piece gives three different values for \textit{P}. The position label of an interval on a resolution line between the current and the next event is \textit{M}, i.e.\,Middle.
On a resolution line coinciding with an event, a zero length interval created by a convex vertex $v_j$ has the position label $R$, i.e. Right, if both adjacent edges lie to the right of $v_j$, has the label $L$, i.e. Left, if both adjacent edges lie to the left of $v_j$. An interval corresponding to a vertical edge has the label $R$, i.e. Right, if the inside of the piece lies to the right of the edge, has the label $L$, i.e. Left, if the inside of the piece lies to the left of the edge. The tuples at $x_i = i\times R$, $i= 0,1,2,\ldots$, are saved in a vector \texttt{Piece[i]}, which is an element of the vector \texttt{Piece}. Hence at the end of the sweep-line algorithm, \texttt{Piece} contains all vectors of tuples of the piece. }

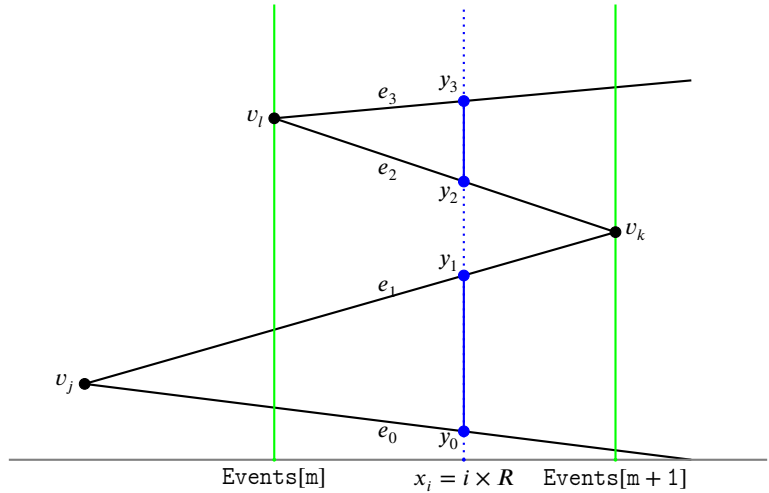
\begin{figure}[!ht]
\centering
\begin{tikzpicture}
\draw[black , thick] (-2,1) node[black , anchor=east] {$v_{j}$} -- (6,0);
\draw[black , thick] (-2,1) --  (5,3) node[black , anchor=west] {$v_{k}$};
\draw[black , thick] (5,3) -- (0.5,4.5);
\draw[black , thick] (0.5,4.5) node[black , anchor=east] {$v_{l}$} -- (6,5);
\draw[gray , thick] (-3,0) -> (7,0);
\draw[green , thick] (5,0) --(5,6);
\draw[green , thick] (0.5,0) -- (0.5,6);
\filldraw [green] (0.5,0) circle (0.5pt) node[black , anchor=north] {$\mathtt{Events[m]}$};
\filldraw [green] (5,0) circle (0.5pt) node[black , anchor=north] {$\mathtt{Events[m+1]}$};
\draw[blue , dotted , thick] (3,0) -- (3,6);
\filldraw [black] (-2,1) circle (2pt);
\filldraw [black] (0.5,4.5) circle (2pt);
\filldraw [black] (5,3) circle (2pt);
\filldraw [blue] (3,11/3) circle (2pt);
\filldraw [blue] (3,104/22) circle (2pt);
\filldraw [blue] (3,17/7) circle (2pt);
\filldraw [blue] (3,3/8) circle (2pt);
\draw[blue , thick] (3,3/8) -- (3,17/7);
\draw[blue , thick] (3,11/3) -- (3,104/22);
\node[black , thick] at (2,1/3) {$e_{0}$};
\node[black , thick] at (2,9/4) {$e_1$};
\node[black , thick] at (2,3.8) {$e_2$};
\node[black , thick] at (2,4.8) {$e_3$};
\node[black , thick] at (2.8,1/4) {$y_{0}$};
\node[black , thick] at (2.8,13/5) {$y_{1}$};
\node[black , thick] at (2.8,10.5/3) {$y_{2}$};
\node[black , thick] at (2.8,109/22) {$y_{3}$};
\filldraw [blue] (3,0) circle (0.5pt) node[black , anchor=north] {$x_i = i \times R$};
\end{tikzpicture}
\caption{Computing the intervals $(y_{0}, y_{1}, M)$ and $(y_{2}, y_{3}, M)$ of a piece at $x_i = i \times R$}
\label{fig88}
\end{figure}

\change{For example in Fig.\,\ref{fig88}, after processing \texttt{Events[m]}, \texttt{ActiveEdges} contains the information of four edges, i.e.\,$e_{0}$, $e_1$, $e_2$ and $e_3$, and \textit{y}-coordinates of the intersection points at $x_i = i\times R$, 
stored in \texttt{Ypoints}, 
are $y_{0}$,\ldots,$y_{3}$, and the tuples are $(y_{0}, y_{1} , M)$ and $(y_{2}, y_{3}, M)$. 
We also illustrate some special cases when the resolution line coincides with an event. In Fig.\,\ref{newexample}, the \textit{y}-coordinates of the intersection points at $x_i = i\times R$, coinciding with \texttt{Events[m]}, 
are $y_{0},y_{1}$, $y_{2}$ and $y_{2}$ and the tuples are $(y_{0},y_{1}, M)$ and $(y_{2}, y_{2}, R)$. In Fig.\,\ref{fig99}, the \textit{y}-coordinates of intersection points at $x_i = i\times R$, coinciding with \texttt{Events[m+1]}, are $y_0$, $y_1$, $y_1$ and $y_2$ and the tuples are $(y_0, y_1, M)$ and $(y_1, y_2, M)$.}

\begin{figure}[!ht]
\centering
\begin{tikzpicture}
\draw[black , thick] (-2,1) node[black , anchor=east] {$v_{j}$} -- (6,0);
\draw[black , thick] (-2,1) --  (5,3);
\draw[black , thick] (5,3) -- (0.5,4.5);
\draw[black , thick] (0.5,4.5) node[black , anchor=east] {$y_{2}$} -- (6,5);
\draw[gray , thick] (-3,0) -> (7,0);
\filldraw [blue] (0.5,0) circle (0.5pt) node[black , anchor=north] {$\mathtt{Events[m]}$};
\filldraw [blue] (-2,0) circle (0.5pt) node[black , anchor=north] {$\mathtt{Events[m-1]}$};
\draw[blue , dotted , thick] (-2,0) -- (-2,6);
\draw[blue , dotted , thick] (0.5,0) -- (0.5,6);
\filldraw [black] (-2,1) circle (2pt);
\filldraw [blue] (0.5,4.5) circle (2pt);
\filldraw [black] (5,3) circle (2pt)  node[black , anchor=west] {$v_{k}$};
\filldraw [blue] (0.5,24/14) circle (2pt) node[black , anchor=south] {$y_{1}$};
\filldraw [blue] (0.5,11/16) circle (2pt) node[black , anchor=north] {$y_{0}$};
\draw[blue , thick] (0.5,11/16) -- (0.5,24/14);
\node[black , thick] at (2,1/4) {$e_{0}$};
\node[black , thick] at (2,9/4) {$e_1$};
\node[black , thick] at (2,3.8) {$e_2$};
\node[black , thick] at (2,4.8) {$e_3$};
\end{tikzpicture}
\caption{\change{Computing the intervals $(y_{0},y_{1}, M)$ and $(y_{2}, y_{2}, R)$ of a piece at $x_i = i\times R$, coinciding with \texttt{Events[m]}}.
}
\label{newexample}
\end{figure}
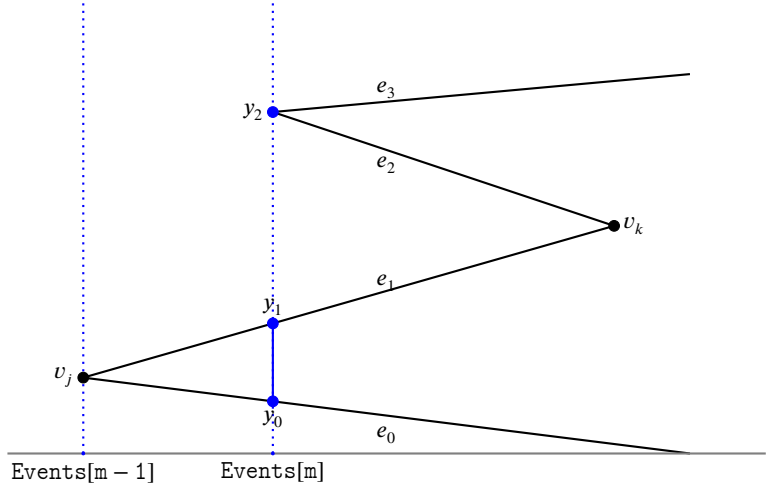\centering

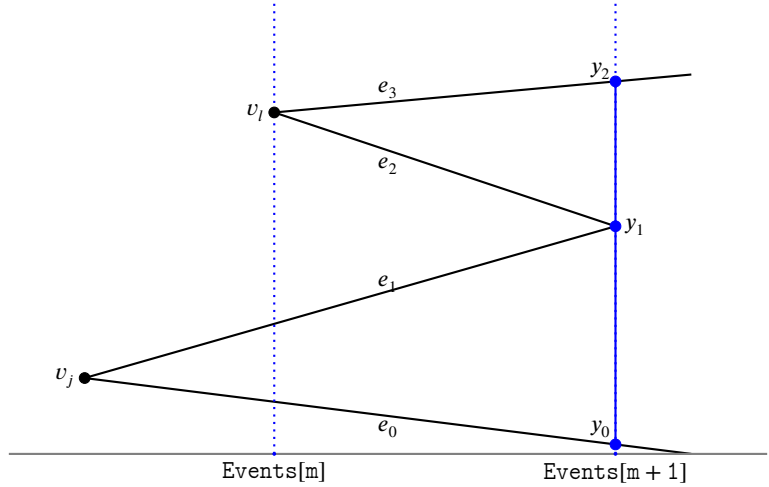
\begin{figure}[!ht]
\centering
\begin{tikzpicture}
\draw[black , thick] (-2,1) node[black , anchor=east] {$v_{j}$} -- (6,0);
\draw[black , thick] (-2,1) --  (5,3);
\draw[black , thick] (5,3) -- (0.5,4.5);
\draw[black , thick] (0.5,4.5) node[black , anchor=east] {$v_{l}$} -- (6,5);
\draw[gray , thick] (-3,0) -> (7,0);
\filldraw [blue] (0.5,0) circle (0.5pt) node[black , anchor=north] {$\mathtt{Events[m]}$};
\filldraw [blue] (5,0) circle (0.5pt) node[black , anchor=north] {$\mathtt{Events[m+1]}$};
\draw[blue , dotted , thick] (5,0) -- (5,6);
\draw[blue , dotted , thick] (0.5,0) -- (0.5,6);
\filldraw [black] (-2,1) circle (2pt);
\filldraw [black] (0.5,4.5) circle (2pt);
\filldraw [blue] (5,3) circle (2pt)  node[black , anchor=west] {$y_1$};
\filldraw [blue] (5,1/8) circle (2pt);
\filldraw [blue] (5,108/22) circle (2pt);
\draw[blue , thick] (5,1/8) -- (5,3);
\draw[blue , thick] (5,3) -- (5,108/22);
\node[black , thick] at (2,1/3) {$e_{0}$};
\node[black , thick] at (2,9/4) {$e_1$};
\node[black , thick] at (2,3.8) {$e_2$};
\node[black , thick] at (2,4.8) {$e_3$};
\node[black , thick] at (4.8,111/22) {$y_2$};
\node[black , thick] at (4.8,2/6) {$y_0$};
\end{tikzpicture}
\caption{Computing the intervals $(y_0, y_1, M)$ and $(y_1, y_2, M)$ of a piece at $x_i = i\times R$, coinciding with \texttt{Events[m+1]}.
}
\label{fig99}
\end{figure}\centering

\subsection{Extension algorithm}\label{section3.2}
\justifying The purpose of the extension algorithm is to guarantee that if there is no overlap between the line segments of two semi-discretized pieces, the corresponding original pieces do not overlap. The input of the extension algorithm is \texttt{Piece}, 
\change{with \texttt{Piece[i]} a vector of tuples $(y_1, y_2, P)$, where $(y_1, y_2)$ is an interval corresponding to a line segment of the semi-discretization of a piece on a resolution line. In the remainder of the text we will merely use the word `interval', even if we refer to the corresponding tuple or even to the corresponding line segment.}

\begin{figure}[!ht]
\centering
\subfloat[]{%
\begin{tikzpicture}
\draw[black, thick] (0,0)  -- (2,-1)  -- (3,3) -- cycle;
\draw[gray , thick] (-1,-1) -> (4,-1);
\filldraw [gray] (2.5,-1) circle (0.5pt)node[black , anchor=north] {$x_{i}$};
\filldraw [gray] (3.5,-1) circle (0.5pt)node[black , anchor=north] {$x_{i+1}$};
\draw[blue] (2.5,-1) --  (2.5,4);
\draw[blue] (3.5,-1) -- (3.5,4);
\filldraw [blue] (2.5,2.5) circle (2pt) node[black , anchor=east] {$(x_{i}, y_{A})$} node[black , anchor=west] {\textit{A}};
\filldraw [gray] (2.5,1) circle (2pt)node[black , anchor=west] {\textit{B}} node[black , anchor=east] {$(x_{i}, y_{B})$};
\filldraw [black] (3,3) circle (2pt)  node[black , anchor=south] {$v = (x_v,y_v)$};
\filldraw [blue] (3.5,3) circle (2pt) node[black , anchor=west] {$(x_{i+1}, y_{v})$};
\filldraw [blue] (2.5,3) circle (2pt) node[black , anchor=east] {$(x_{i}, y_{v})$};
\draw[gray , thick] (2.5,1) -- (2.5,2.5);
\draw[blue , ultra thick] (2.5,2.5) -- (2.5,3);
\draw[blue , dotted , thick] (2.5,3) -- (3.5,3);
\node[black , thick] at (1,1.2) {\textit{e}};
\node[black , thick] at (2.1,0) {$\Tilde{e}$};
\end{tikzpicture}
\label{samer}
}
\hfil
\subfloat[]{%
\begin{tikzpicture}
\draw[black, thick] (0,0)  -- (4,0) -- (2,3) -- cycle;
\draw[blue] (1.5,-1) -- (1.5,4);
\draw[blue] (2.5,-1) -- (2.5,4);
\filldraw [blue] (1.5,9/4) circle (2pt) node[black , anchor=east] {$(x_{i}, y_{A})$} node[black , anchor=west] {\textit{A}};
\filldraw [gray] (1.5,0) circle (2pt);
\filldraw [blue] (2.5,9/4) circle (2pt) node[black , anchor=west] {$(x_{i+1}, y_{B})$} node[black , anchor=east] {\textit{B}};
\draw[gray , thick] (0,-1) -> (4,-1);
\filldraw [gray] (1.5,-1) circle (0.5pt)node[black , anchor=north] {$x_{i}$};
\filldraw [gray] (2.5,-1) circle (0.5pt)node[black , anchor=north] {$x_{i+1}$};
\filldraw [gray] (2.5,0) circle (2pt);
\filldraw [black] (2,3) circle (2pt) node[black , anchor=south] {$v = (x_v,y_v)$};
\draw[gray , thick] (1.5,0) -- (1.5,9/4);
\draw[gray , thick] (2.5,0) -- (2.5,9/4);
\draw[blue , ultra thick] (1.5,9/4) -- (1.5,3);
\draw[blue , ultra thick] (2.5,9/4) -- (2.5,3);
\filldraw [blue] (1.5,3) circle (2pt) node[black , anchor=east] {$(x_{i}, y_{v})$};
\filldraw [blue] (2.5,3) circle (2pt)node[black , anchor=west] {$(x_{i+1}, y_{v})$};
\draw[blue , dotted , thick] (1.5,3) -- (2.5,3);
\node[black , thick] at (0.4,1) {\textit{e}};
\node[black , thick] at (3.6,1) {$\Tilde{e}$};
\end{tikzpicture}
\label{fig:subim1}
}
\hfil
\subfloat[]{%
\begin{tikzpicture}
\draw[black, thick] (3,2) -- (4.8,0) -- (5.2,4) -- cycle;
\node[black] at (5.7,-0.3) {$v_1 = (x_{v_1},y_{v_1})$};
\node[black] at (6,4.3) {$v_2 = (x_{v_2},y_{v_2})$};
\node[black] at (6.3,0.2) {$(x_{i+1}, y_{v_1})$};
\node[black] at (6.3,3.8) {$(x_{i+1}, y_{v_2})$};
\node[black] at (4.7,0.4) {\textit{A}};
\draw[blue] (5.5,-1) -- (5.5,5);
\draw[blue] (4.5,-1) -- (4.5,5);
\draw[gray , thick] (2,-1) -> (6,-1);
\filldraw [gray] (4.5,-1) circle (0.5pt)node[black , anchor=north] {$x_{i}$};
\filldraw [gray] (5.5,-1) circle (0.5pt)node[black , anchor=north] {$x_{i+1}$};
\draw[blue , ultra thick] (5.5,0) -- (5.5,4);
\filldraw [black] (4.8,0) circle (2pt);
\filldraw [black] (5.2,4) circle (2pt);
\filldraw [blue] (5.5,0) circle (2pt);
\filldraw [blue] (5.5,4) circle (2pt);
\filldraw [blue] (5.5,0) circle (2pt);
\draw[gray , thick] (4.5,0.33) -- (4.5,7/2);
\draw[blue , ultra thick] (4.5,0) -- (4.5,0.33);
\draw[blue , ultra thick] (4.5,3.35) -- (4.5,4);
\filldraw [blue] (4.5,0.33) circle (2pt) node[black , anchor=east] {$(x_{i}, y_{A})$};
\filldraw [blue] (4.5,0) circle (2pt) node[black , anchor=east] {$(x_{i}, y_{v_1})$};
\filldraw [blue] (4.5,3.35) circle (2pt) node[black , anchor=east] {$(x_{i}, y_{B})$} node[black , anchor=west] {\textit{B}};
\filldraw [blue] (4.5,4) circle (2pt)  node[black , anchor=east] {$(x_{i}, y_{v_2})$};
\draw[blue , dotted , thick] (4.5,4) -- (5.5,4);
\draw[blue , dotted , thick] (4.5,0) -- (5.5,0);
\node[black , thick] at (3.5,2.8) {\textit{e}};
\node[black , thick] at (3.4,1.2) {$e'$};
\node[black , thick] at (5.1,1) {$\Tilde{e}$};
\end{tikzpicture}
\label{fig:subim2}
}
\caption{
\change{Convex vertex $v = (x_v, y_v)$ lies between resolution lines at $x_{i}$ and $x_{i+1}$: for each edge adjacent to \textit{v} extension intervals are computed and added on both resolution lines 
(unless they overlap with already existing intervals). In \ref{samer}, both edges \textit{e} and $\Tilde{e}$ intersect the resolution line at $x_{i}$. The extension intervals due to edge $e$ are $(y_A, y_v, R)$ at $x_{i}$ and $(y_v, y_v, L)$ at $x_{i+1}$.
The extension intervals due to edge $\Tilde{e}$, i.e.\,$(y_{B}, y_{v}, R)$
at $x_{i}$ and $(y_{v}, y_{v}, L)$ at $x_{i+1}$, overlap with already existing intervals and extension intervals and will not be added again. In \ref{fig:subim1}, edge \textit{e} intersects the resolution line at $x_{i}$ and edge $\Tilde{e}$ intersects the other neighboring resolution line at $x_{i+1}$. The extension intervals due to edge $e$ are $(y_A, y_v, R)$ at $x_{i}$ and $(y_v, y_v, L)$ at $x_{i+1}$. Some of the the extension intervals due to $\Tilde{e}$, i.e.\, $(y_{v}, y_{v}, R)$ at $x_{i}$ and $(y_{v}, y_{v}, L)$ at $x_{i+1}$ overlap with already existing extension intervals. In \ref{fig:subim2}, edges \textit{e} and $e'$ intersect a resolution line at $x_{i}$ and edge $\Tilde{e}$ does not intersect any resolution line. The extension intervals due to edge \textit{e} are 
$(y_{B} , y_{v_2}, R)$ at $x_{i}$ and $(y_{v_2}, y_{v_2}, L)$ at $x_{i+1}$. The extension intervals due to edge $e'$ are 
$(y_{v_1}, y_{A}, R)$ at $x_{i}$ and $(y_{v_1}, y_{v_1}, L)$ at $x_{i+1}$.The extension intervals due to $\Tilde{e}$ are 
$(y_{v_1}, y_{v_2}, R)$, at $x_{i}$ and $(y_{v_1}, y_{v_2}, L)$ at $x_{i+1}$. However 
$(y_{v_1}, y_{A}, R)$, $(y_{A}, y_{B}, M)$ and $(y_{B} , y_{v_2}, R)$ already exist at $x_{i}$ and 
$(y_{v_2}, y_{v_2}, L)$ and $(y_{v_1}, y_{v_1}, L)$ already exist at $x_{i+1}$.}
}
\label{fig8}
\end{figure}
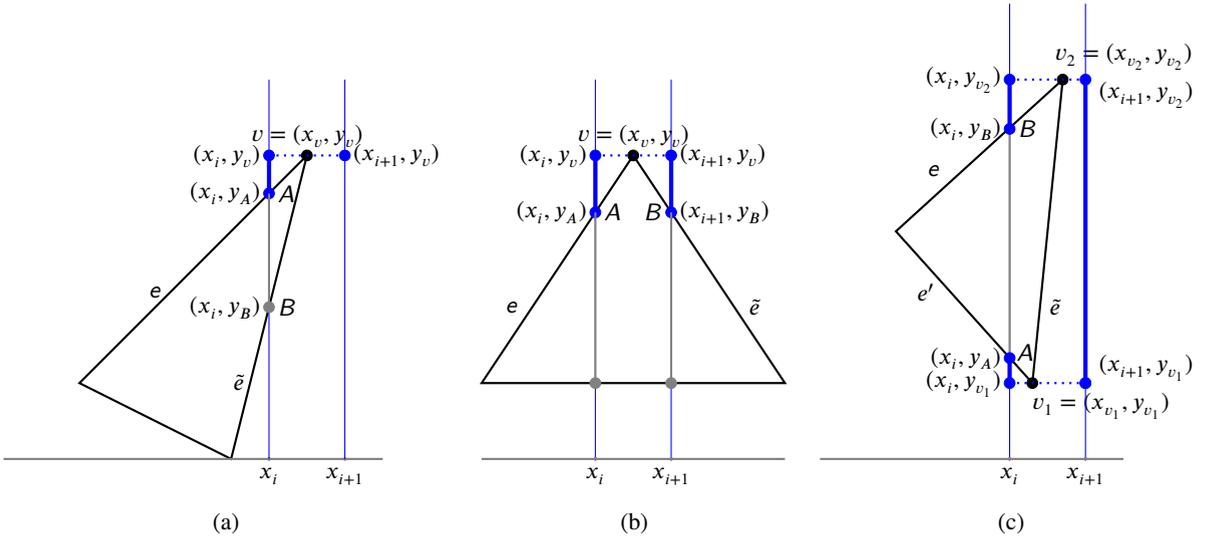

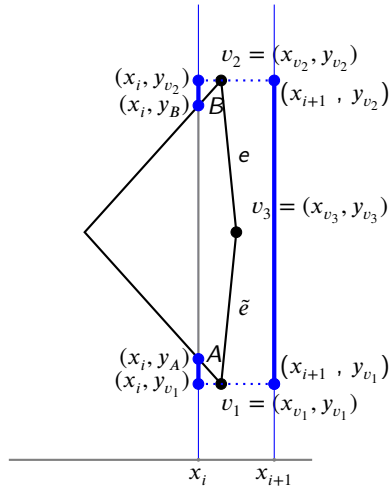
\begin{figure}[!ht]
\centering
\begin{tikzpicture}
\draw[black, thick] (3,2) -- (4.8,0) -- (5,2) -- (4.8,4) -- cycle;
\filldraw [black] (5,2) circle (2pt);
\filldraw [black] (4.8,4) circle (2pt);
\node[black] at (5.7,-0.3) {$v_1 = (x_{v_1},y_{v_1})$};
\node[black] at (5.7,4.3) {$v_2 = (x_{v_2},y_{v_2})$};
\node[black] at (6.1,2.3) {$v_3 = (x_{v_3},y_{v_3})$};
\node[black] at (6.3,0.2) {($x_{i+1}$ , $y_{v_1}$)};
\node[black] at (6.3,3.8) {($x_{i+1}$ , $y_{v_2}$)};
\node[black] at (4.7,0.4) {\textit{A}};
\draw[blue] (5.5,-1) -- (5.5,5);
\draw[blue] (4.5,-1) -- (4.5,5);
\draw[gray , thick] (2,-1) -> (6,-1);
\filldraw [gray] (4.5,-1) circle (0.5pt)node[black , anchor=north] {$x_{i}$};
\filldraw [gray] (5.5,-1) circle (0.5pt)node[black , anchor=north] {$x_{i+1}$};
\draw[blue , ultra thick] (5.5,0) -- (5.5,4);
\filldraw [black] (4.8,0) circle (2pt);
\filldraw [blue] (5.5,0) circle (2pt);
\filldraw [blue] (5.5,4) circle (2pt);
\filldraw [blue] (5.5,0) circle (2pt);
\draw[gray , thick] (4.5,0.33) -- (4.5,3.67);
\draw[blue , ultra thick] (4.5,0) -- (4.5,0.33);
\draw[blue , ultra thick] (4.5,3.67) -- (4.5,4);
\filldraw [blue] (4.5,0.33) circle (2pt) node[black , anchor=east] {$(x_{i},  y_{A})$};
\filldraw [blue] (4.5,0) circle (2pt) node[black , anchor=east] {$(x_{i}, y_{v_1})$};
\filldraw [blue] (4.5,3.67) circle (2pt) node[black , anchor=east] {$(x_{i}, y_{B})$} node[black , anchor=west] {\textit{B}};
\filldraw [blue] (4.5,4) circle (2pt)  node[black , anchor=east] {$(x_{i}, y_{v_2})$};
\draw[blue , dotted , thick] (4.5,4) -- (5.5,4);
\draw[blue , dotted , thick] (4.5,0) -- (5.5,0);
\node[black , thick] at (5.1,3) {\textit{e}};
\node[black , thick] at (5.1,1) {$\Tilde{e}$};
\end{tikzpicture}
\caption{Extension on each neighboring resolution line when none of the edges \textit{e} and $\Tilde{e}$, adjacent to a convex vertex $v_{3}$, intersect resolution lines.}
\label{fig9_dirk}
\end{figure}

Some of the piece information is lost when a vertex lies between two resolution lines, which we will call the neighboring resolution lines (of the vertex). For each convex vertex $(x_{v},y_{v})$ located between two resolution lines, extra intervals, called extension intervals, will be added to the set of intervals representing the piece in one or both of the neighboring resolution lines.
Suppose $v = (x_v,y_v)$ is a convex vertex with $x_i=i\times R<x_{v}<x_{i+1}=(i+1)\times R$. In order to insert the extension intervals in \texttt{Piece[i]} and \texttt{Piece[i+1]}, the union of the intervals which are already in these vectors and the extension intervals are computed. Hence, if the extension interval already exists in the vector, it will not be added again. For each convex vertex both adjacent edges are considered. 
For each adjacent edge $e$, we distinguish two cases. 
\begin{enumerate}
\item If edge \textit{e} intersects a neighboring resolution line, the extension will happen on both of the two neighboring resolution lines. Let $(x_{i},y_{A})$ be the intersection point of the edge and the neighboring resolution line, see Fig.\,\ref{fig8}. \change{The extension interval on the resolution line which intersects \textit{e} is equal to $(y_{A}, y_{v}, P)$, with $P = R$ if \textit{e} intersects the left resolution line and $P = L$ if \textit{e} intersects the right resolution line. On the other neighboring resolution line, which is not intersected by \textit{e}, the extension interval is equal to $(y_{v}, y_{v}, P)$ with $P = R$ if \textit{e} intersects the right resolution line and $P = L$ if \textit{e} intersects the left resolution line.} In case the other adjacent edge $\Tilde{e}$ intersects the same resolution line, as in Fig.\,\ref{samer}, a similar calculation would lead to extension intervals on both resolution lines that overlap with already existing intervals and extension intervals and hence must not be added again. However, in case $\Tilde{e}$, the other edge adjacent to vertex $(x_{v},y_{v})$, intersects the other neighboring resolution line, i.e.\,at $(x_{i+1},y_{B})$, a similar calculation results in the extension interval \change{$(y_{B}, y_{v}, P)$}, see Fig.\,\ref{fig:subim1}.

\item If the edge intersects none of the neighboring resolution lines, i.e.\,the whole edge is placed between resolution lines, the extension interval on both resolution lines is the orthogonal projection of the edge and is equal to \change{$(y_{v_1}, y_{v_2}, P)$, with $P = R$ on the left resolution line, and $P = L$ on the right resolution line}, see Figs.\,\ref{fig:subim2} and \ref{fig9_dirk}. \change{If the extension interval overlaps with an existing interval with the third element of the tuple (i.e.\,the position of the existing interval) equal to \textit{M} no update is needed.}

\end{enumerate}

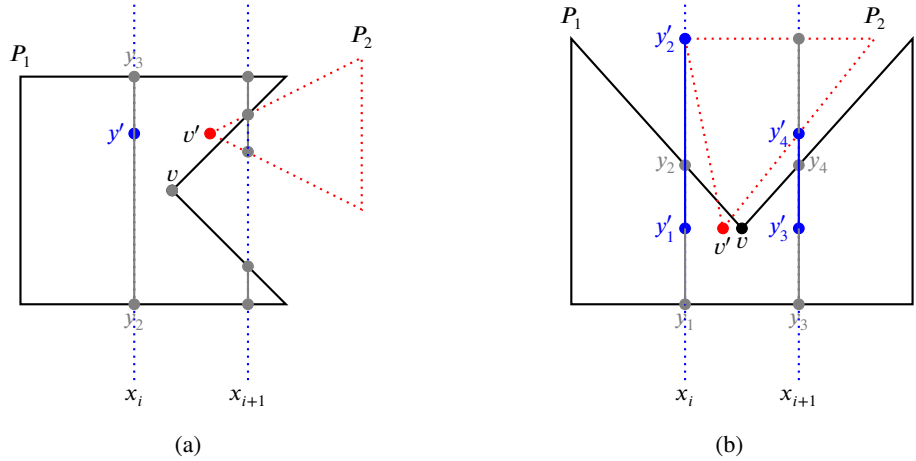
\begin{figure}[!ht]
\centering
\subfloat[\label{1144a}]{%
\begin{tikzpicture}
\draw[black, thick] (-1.5,-1)  -- (2,-1)  -- (0.5,0.5) node[black , anchor=south] {$v$} -- (2,2) -- (-1.5,2) node[black , anchor=south] {$P_1$}  -- cycle;
\draw[red, dotted, thick] (1,1.25)  -- (3,2.25) node[black , anchor=south] {$P_2$} -- (3,0.25) -- cycle;
\draw[gray , thick] (1.5,1.01) -- (1.5,1.5);
\filldraw [gray] (1.5,1.01) circle (2pt);
\filldraw [gray] (1.5,1.5) circle (2pt);
\filldraw [red] (1,1.25) circle (2pt) node[black , anchor=east] {$v'$};
\filldraw [blue] (0,1.25) circle (2pt) node[blue , anchor=east] {$y'$};;
\draw[blue , dotted, thick] (0.0,-2) node[black , anchor=north] {$x_i$} -- (0,3);
\draw[blue , dotted, thick] (1.5,-2) node[black , anchor=north] {$x_{i+1}$} -- (1.5,3);
\draw[gray , thick] (0,-1) -- (0,2);
\draw[gray , thick] (1.5,-1) -- (1.5,-0.5);
\draw[gray , thick] (1.5,1.5) -- (1.5,2);
\filldraw [black] (0.5,0.5) circle (2pt);
\filldraw [gray] (0.5,0.5) circle (2pt);
\filldraw [gray] (0,2) circle (2pt)node[gray , anchor=south] {$y_3$};;
\filldraw [gray] (1.5,1.5) circle (2pt);
\filldraw [gray] (1.5,2) circle (2pt);
\filldraw [gray] (1.5,-1) circle (2pt);
\filldraw [gray] (0,-1) circle (2pt) node[gray , anchor=north] {$y_2$};;
\filldraw [gray] (1.5,-0.5) circle (2pt);
\end{tikzpicture}
}
\hfil
\subfloat[\label{1144b}]{%
\begin{tikzpicture}
\draw[black, thick] (4,-1)  -- (8.5,-1)  -- (8.5,2.5) -- (6.25,0) node[black , anchor=north] {$v$} -- (4,2.5) node[black , anchor=south] {$P_1$} -- cycle;
\draw[red, dotted, thick] (5.5,2.5)  -- (6,0)  -- (8,2.5) node[black , anchor=south] {$P_2$} -- cycle;
\filldraw [red] (6,0) circle (2pt) node[black , anchor=north] {$v'$};
\filldraw [blue] (5.5,0) circle (2pt);
\filldraw [blue] (7,0) circle (2pt);
\filldraw [blue] (5.5,2.5) circle (2pt);
\filldraw [gray] (5.5,-1) circle (2pt) node[gray , anchor=north] {$y_1$};
\filldraw [gray] (5.5,15/18) circle (2pt) node[gray , anchor=east] {$y_2$};
\draw[blue , dotted, thick] (5.5,-2)  node[black , anchor=north] {$x_i$} -- (5.5,3);
\draw[blue , dotted, thick] (7,-2)  node[black , anchor=north] {$x_{i+1}$}-- (7,3);
\filldraw [blue] (7,5/4) circle (2pt);
\filldraw [gray] (7,5/2) circle (2pt);
\filldraw [gray] (7,-1) circle (2pt) node[gray , anchor=north] {$y_3$};
\filldraw [gray] (7,15/18) circle (2pt) node[gray , anchor=west] {$y_4$};
\draw[gray , thick] (7,-1) -- (7,15/18);
\draw[gray , thick] (5.5,-1) -- (5.5,15/18) ;
\filldraw [black] (6.25,0) circle (2pt);
\draw[gray , thick] (7,5/4) node[blue , anchor=east] {$y'_4$} -- (7,2.5);
\draw[blue , thick] (5.5,0)node[blue , anchor=east] {$y'_1$} -- (5.5,2.5) node[blue , anchor=east] {$y'_2$};
\draw[blue , thick] (7,0)node[blue , anchor=east] {$y'_3$} -- (7,5/4);
\end{tikzpicture}}
\caption{
\change{No extension is needed for a non-convex vertex $v$ of piece $P_1$. In Fig.\,\ref{1144a}, placement of the semi-discretization of another piece $P_2$, with convex vertex $v'$ between the resolution lines at $x_i$ and $x_{i+1}$, would cause overlap of the extension interval ($y', y', R$) of $v'$ with interval ($y_2, y_3, M$) belonging to $P_1$. Hence, $P_2$ would not be placed in the mentioned position.
In Fig.\,\ref{1144b}, placement of the semi-discretization of $P_2$
would cause overlap of the extension interval ($y'_1, y'_2, R$) with ($y_1, y_2, M$) and of extension interval ($y'_3, y'_4, L$) with ($y_3, y_4, M$). 
}}
\label{fig1144}
\end{figure}

\justifying
When a non-convex vertex lies between two resolution lines, no extension is needed, \change{see Fig.\,\ref{fig1144}}, since by applying the extension algorithm for all convex vertices, without extension due to non-convex vertices, overlap with another piece cannot happen.

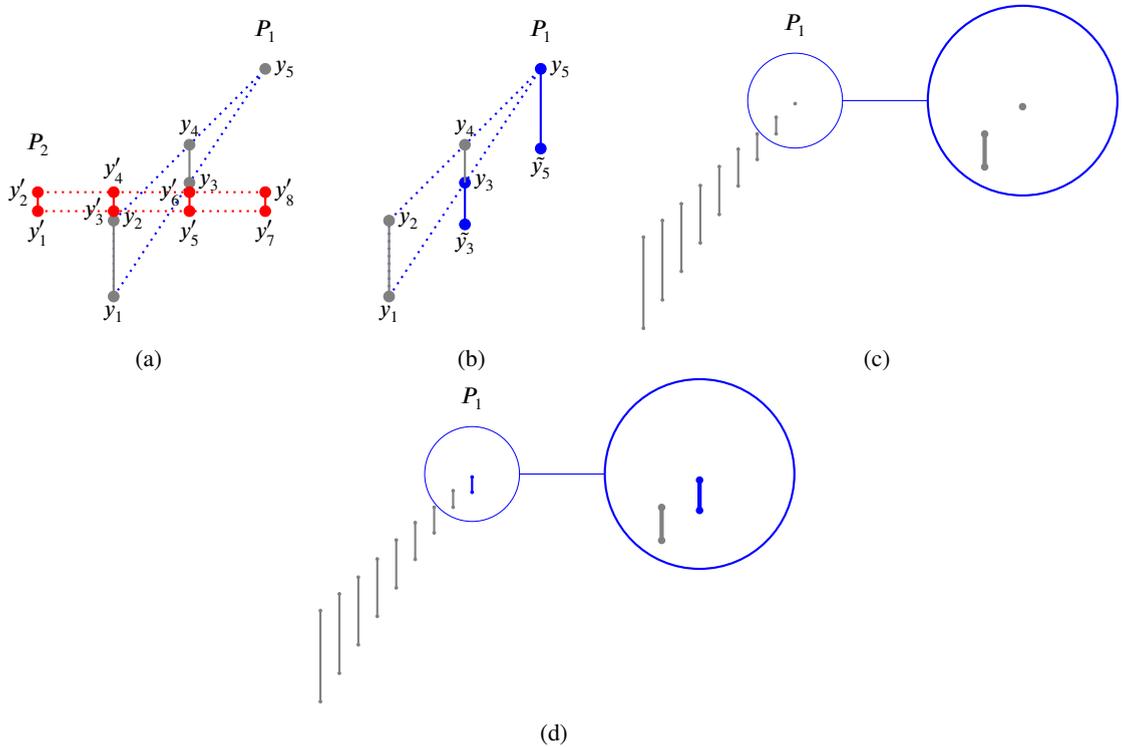
\begin{figure}[!ht]
\centering
\subfloat[\label{int1}]{%
\begin{tikzpicture}
\draw[blue, dotted, thick] (0,0)  -- (2,3) -- (0,1) -- cycle;
\node[black , thick] at (2,3.5) {$P_1$};
\node[black , thick] at (-1,2) {$P_2$};
\draw[red, dotted, thick] (-1,9/8) -- (2,9/8) -- (2,11/8) -- (-1,11/8) --cycle;
\draw[red , thick] (-1,9/8) -- (-1,11/8);
\draw[red , thick] (0,9/8) -- (0,11/8);
\draw[red , thick] (1,9/8) -- (1,11/8);
\draw[red , thick] (2,9/8) -- (2,11/8);
\filldraw [gray] (0,0) circle (2pt)node[black , anchor=north] {$y_{1}$};
\filldraw [gray] (0,1) circle (2pt)node[black , anchor=west] {$y_{2}$};
\filldraw [gray] (2,3) circle (2pt)node[black , anchor=west] {$y_{5}$};
\filldraw [gray] (1,3/2) circle (2pt)node[black , anchor=west] {$y_{3}$};
\filldraw [gray] (1,2) circle (2pt)node[black , anchor=south] {$y_{4}$};
\filldraw [red] (-1,9/8) circle (2pt)node[black , anchor=north] {$y'_{1}$};
\filldraw [red] (-1,11/8) circle (2pt)node[black , anchor=east] {$y'_{2}$};
\filldraw [red] (0,9/8) circle (2pt)node[black , anchor=east] {$y'_{3}$};
\filldraw [red] (0,11/8) circle (2pt)node[black , anchor=south] {$y'_{4}$};
\filldraw [red] (1,9/8) circle (2pt)node[black , anchor=north] {$y'_{5}$};
\filldraw [red] (1,11/8) circle (2pt)node[black , anchor=east] {$y'_{6}$};
\filldraw [red] (2,9/8) circle (2pt)node[black , anchor=north] {$y'_{7}$};
\filldraw [red] (2,11/8) circle (2pt)node[black , anchor=west] {$y'_{8}$};
\draw[gray , thick] (1,3/2) -- (1,2);
\draw[gray , thick] (0,0) -- (0,1);
\end{tikzpicture}
}
\hfil
\subfloat[\label{int2}]{%
\begin{tikzpicture}
\draw[blue, dotted, thick] (0,0)  -- (2,3) -- (0,1) -- cycle;
\node[black , thick] at (2,3.5) {$P_1$};
\filldraw [gray] (0,0) circle (2pt)node[black , anchor=north] {$y_{1}$};
\filldraw [gray] (0,1) circle (2pt)node[black , anchor=west] {$y_{2}$};
\filldraw [blue] (2,3) circle (2pt)node[black , anchor=west] {$y_{5}$};
\filldraw [blue] (1,3/2) circle (2pt)node[black , anchor=west] {$y_{3}$};
\filldraw [gray] (1,2) circle (2pt)node[black , anchor=south] {$y_{4}$};
\filldraw [blue] (2,1.95) circle (2pt)node[black , anchor=north] {$\Tilde{y_5}$};
\filldraw [blue] (1,0.95) circle (2pt)node[black , anchor=north] {$\Tilde{y_3}$};
\draw[gray , thick] (1,3/2) -- (1,2);
\draw[blue , thick] (2,3) -- (2,1.95);
\draw[blue , thick] (1,3/2) -- (1,0.95);
\draw[gray , thick] (0,0) -- (0,1);
\end{tikzpicture}}
\hfil
\subfloat[\label{int4}]{%
\begin{tikzpicture}[spy using outlines={circle, magnification=2, connect spies}]
\node[black , thick] at (2,4) {$P_1$};
\filldraw [gray] (2,2.96) circle (0.5pt);
\filldraw [gray] (1,3/2) circle (0.5pt);
\filldraw [gray] (1,2.130) circle (0.5pt);
\filldraw [gray] (0,0) circle (0.5pt);
\filldraw [gray] (0,1.2) circle (0.5pt);
\filldraw [gray] (1/2,3/4) circle (0.5pt);
\filldraw [gray] (1/2,1.64) circle (0.5pt);
\filldraw [gray] (3/2,8.9/4) circle (0.5pt);
\filldraw [gray] (3/2,20.49/8) circle (0.5pt);
\filldraw [gray] (1/4,3/8) circle (0.5pt);
\filldraw [gray] (1/4,1.42) circle (0.5pt);
\filldraw [gray] (3/4,9/8) circle (0.5pt);
\filldraw [gray] (3/4,1.88) circle (0.5pt);
\filldraw [gray] (5/4,15/8) circle (0.5pt);
\filldraw [gray] (5/4,2.36) circle (0.5pt);
\filldraw [gray] (7/4,20.5/8) circle (0.5pt);
\filldraw [gray] (7/4,2.78) circle (0.5pt);
\draw[gray , thick] (1,3/2) -- (1,2.130);
\draw[gray , thick] (0,0) -- (0,1.2);
\draw[gray , thick] (1/2,3/4) -- (1/2,1.64);
\draw[gray , thick] (3/2,8.9/4) -- (3/2,20.49/8);
\draw[gray , thick] (1/4,3/8) -- (1/4,1.42);
\draw[gray , thick] (3/4,9/8) -- (3/4,1.88);
\draw[gray , thick] (5/4,15/8) -- (5/4,2.36);
\draw[gray , thick] (7/4,20.5/8) -- (7/4,2.78);
\spy [blue, size=2.5cm] on (2,3) in node[fill=white] at (5,3);
\end{tikzpicture}}
\hfil
\subfloat[\label{int3}]{%
\begin{tikzpicture}[spy using outlines={circle, magnification=2, connect spies}]
\node[black , thick] at (2,4) {$P_1$};
\filldraw [blue] (2,2.96) circle (0.5pt);
\filldraw [blue] (2,2.76) circle (0.5pt);
\filldraw [gray] (1,3/2) circle (0.5pt);
\filldraw [gray] (1,2.130) circle (0.5pt);
\filldraw [gray] (0,0) circle (0.5pt);
\filldraw [gray] (0,1.2) circle (0.5pt);
\filldraw [gray] (1/2,3/4) circle (0.5pt);
\filldraw [gray] (1/2,1.64) circle (0.5pt);
\filldraw [gray] (3/2,8.9/4) circle (0.5pt);
\filldraw [gray] (3/2,20.49/8) circle (0.5pt);
\filldraw [gray] (1/4,3/8) circle (0.5pt);
\filldraw [gray] (1/4,1.42) circle (0.5pt);
\filldraw [gray] (3/4,9/8) circle (0.5pt);
\filldraw [gray] (3/4,1.88) circle (0.5pt);
\filldraw [gray] (5/4,15/8) circle (0.5pt);
\filldraw [gray] (5/4,2.36) circle (0.5pt);
\filldraw [gray] (7/4,20.5/8) circle (0.5pt);
\filldraw [gray] (7/4,2.78) circle (0.5pt);
\draw[gray , thick] (1,3/2) -- (1,2.130);
\draw[gray , thick] (0,0) -- (0,1.2);
\draw[gray , thick] (1/2,3/4) -- (1/2,1.64);
\draw[gray , thick] (3/2,8.9/4) -- (3/2,20.49/8);
\draw[gray , thick] (1/4,3/8) -- (1/4,1.42);
\draw[gray , thick] (3/4,9/8) -- (3/4,1.88);
\draw[gray , thick] (5/4,15/8) -- (5/4,2.36);
\draw[gray , thick] (7/4,20.5/8) -- (7/4,2.78);
\draw[blue , thick] (2,2.96) -- (2,2.76);
\spy [blue, size=2.5cm] on (2,3) in node[fill=white] at (5,3);
\end{tikzpicture}}
\caption{\change{In \ref{int1}, there are vertical gaps between the intervals of $P_1$ on the neighboring resolution lines in which the intervals of $P_2$ can be placed. In \ref{int2}, the problem is solved by extending the non-overlapping intervals to remove the vertical gaps between them. In \ref{int4}, a smaller \textit{R}-value solves the problem in interior intervals, however, interval of length zero has no overlap with the interval in the consecutive resolution line. In \ref{int3}, the problem is solved by extension of the interval of length zero.}}
\label{figinternal}
\end{figure}

\change{Besides dealing with vertices that lie between resolution lines, there is another situation where extension can be necessary. It is possible that placement of another piece causes overlap of the pieces without overlapping line segments of the semi-discretization of the pieces, see Fig.\,\ref{int1}. This can happen when there are vertical gaps between the intervals on neighboring resolution lines, if these intervals correspond to line segments lying between the same edges of the piece. In these gaps intervals of the other piece could be placed.
This situation is avoided by extending such intervals so that intervals on neighbouring resolution lines always overlap,
see Fig.\,\ref{int2}. 
The default option in our implementation is to perform this extension only for intervals of length zero ($y_v , y_v , P$) (i.e.\,a vertex lying on a resolution line), see Figs.\,\ref{int4} and \ref{int3}, if $b \geq y_v$ or $t \leq y_v$ where $b$ and $t$ denote the bottom and top endpoints of the interval on the neighboring resolution line. 
Indeed, non-overlapping intervals in neighboring resolution lines, corresponding to interior line segments, only occur for a piece with a specific geometry, i.e.\,when two consecutive edges have both a large positive or both a large negative slope, and semi-discretized with a large $R$. In such a case this extension can also be activated for the interior resolution lines or $R$ can be decreased. The latter will also improve the accuracy of the semi-discrete representation.}

\subsection{Final data structure}
\begin{figure}[!ht]
\centering
\subfloat[The original piece\label{fig:subim11}]{%
\begin{tikzpicture}
\filldraw[yellow, thick] (0,1)node[black , anchor=east] {$v_{0}$}  -- (1.5,-1) node[black , anchor=north] {$v_{1}$}  -- (2,0)node[black , anchor=west] {$v_{2}$} -- (5,0)node[black , anchor=west] {$v_{3}$} -- (2.5,1)node[black , anchor=west] {$v_{4}$} -- (5,2)node[black , anchor=west] {$v_{5}$} -- (2,2)node[black , anchor=west] {$v_{6}$} --  (2,3)node[black , anchor=south] {$v_{7}$} -- cycle;
\filldraw [black] (0,1) circle (2pt);
\draw[white , thick] (0,-2) -> (5,-2);
\filldraw [black] (1.5,-1)circle (2pt);
\filldraw [black] (2,0) circle (2pt);
\filldraw [black] (2.5,1)circle (2pt);
\filldraw [black] (5,0) circle (2pt);
\filldraw [black] (5,2) circle (2pt);
\filldraw [black] (2,2) circle (2pt);
\filldraw [black] (2,3) circle (2pt);
\end{tikzpicture}
}
\hfil
\subfloat[The semi-discretized piece with $R=1$ and extended intervals at $x=1$ and $x=2$.
\label{fig:subim22}]{%
\begin{tikzpicture}
\filldraw[yellow, thick] (5,0)  -- (6.5,-2) -- (7,-1) -- (10,-1) -- (7.5,0) -- (10,1) -- (7,1) --  (7,2) -- cycle;
\draw[gray , thick] (5,-2.5) -> (10,-2.5);
\filldraw [gray] (5,-2.5) circle (0.5pt)node[black , anchor=north] {0};
\filldraw [gray] (6,-2.5) circle (0.5pt)node[black , anchor=north] {1};
\filldraw [gray] (7,-2.5) circle (0.5pt)node[black , anchor=north] {2};
\filldraw [gray] (8,-2.5) circle (0.5pt)node[black , anchor=north] {3};
\filldraw [gray] (9,-2.5) circle (0.5pt)node[black , anchor=north] {4};
\filldraw [gray] (10,-2.5) circle (0.5pt)node[black , anchor=north] {5};
\filldraw [blue] (5,0) circle (2pt)node[black , anchor=east] {$y_{0}$};
\draw[blue , thick] (6,-2) -- (6,1);
\filldraw [blue] (6,-2) circle (2pt)node[black , anchor=north] {$y_{1}$};
\filldraw [blue] (6,1) circle (2pt)node[black , anchor=south] {$y_{3}$};
\draw[blue , thick] (7,-2) -- (7,2);
\draw[blue , thick] (8,-1) -- (8,-1/5);
\draw[blue , thick] (8,1/5) -- (8,1);
\draw[blue , thick] (9,-1) -- (9,-3/5);
\draw[blue , thick] (9,3/5) -- (9,1);
\filldraw [blue] (7,-2) circle (2pt)node[black , anchor=north] {$y_{4}$};
\filldraw [blue] (6,-4/3) circle (2pt)node[black , anchor=west] {$y_{2}$};
\filldraw [blue] (7,-1) circle (2pt)node[black , anchor=east] {$y_{5}$};
\filldraw [blue] (7,1) circle (2pt)node[black , anchor=east] {$y_{6}$};
\filldraw [blue] (7,2) circle (2pt)node[black , anchor=south] {$y_{7}$};
\filldraw [blue] (8,-1) circle (2pt)node[black , anchor=north] {$y_{8}$};
\filldraw [blue] (8,1) circle (2pt)node[black , anchor=south] {$y_{11}$};
\filldraw [blue] (9,-1) circle (2pt)node[black , anchor=north] {$y_{12}$};
\filldraw [blue] (9,1) circle (2pt)node[black , anchor=south] {$y_{15}$};
\filldraw [blue] (8,-1/5) circle (2pt)node[black , anchor=east] {$y_{9}$};
\filldraw [blue] (8,1/5) circle (2pt)node[black , anchor=east] {$y_{10}$};
\filldraw [blue] (9,-3/5) circle (2pt)node[black , anchor=south] {$y_{13}$};
\filldraw [blue] (9,3/5) circle (2pt)node[black , anchor=north] {$y_{14}$};
\filldraw [blue] (10,-1) circle (2pt)node[black , anchor=west] {$y_{16}$};
\filldraw [blue] (10,1) circle (2pt)node[black , anchor=west] {$y_{17}$};
\end{tikzpicture}
}
\caption{An example of discretization and extension of a piece.}
\label{fig114}
\end{figure}
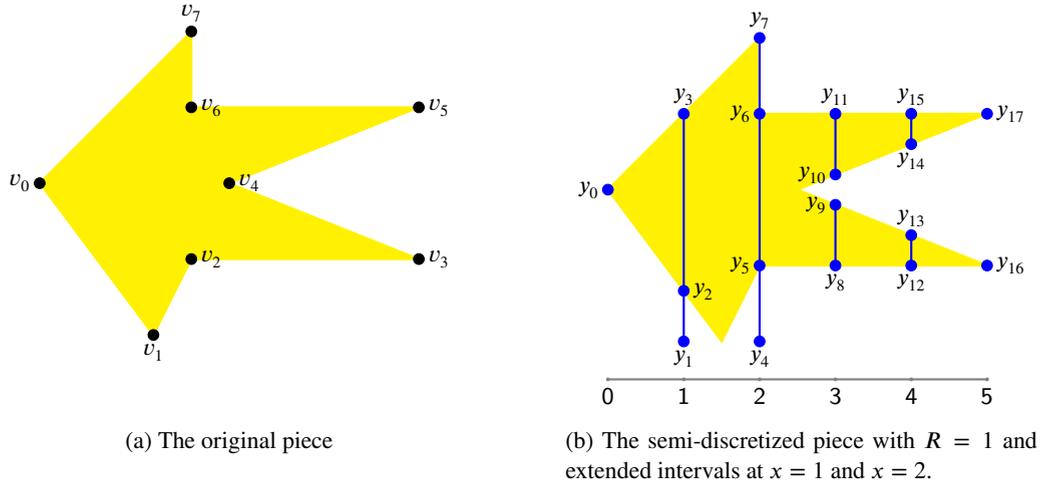

\justifying 
\change{In a last step, tuples corresponding to touching intervals and with the same third element are joined, i.e.\,when for two neighboring tuples the second element of one tuple is equal to the first element of the other tuple and they have an identical third element. The final semi-discretization of the piece, i.e.\ all tuples corresponding to intervals representing line segments of the piece, are stored in  a vector of vectors of tuples, called \texttt{Piece}. Fig.\,\ref{fig:subim22} shows a semi-discretized piece with $R=1$ and the data structure \texttt{Piece} is given by:}

\change{
$
\mathtt{
Piece =
\begin{array}{cccc} 
\Big[
\begin{bmatrix} 
(y_0 , y_0 , R)\end{bmatrix}, \begin{bmatrix} 
(y_1 , y_2 , R) , (y_2, y_3, M)
\end{bmatrix}, \begin{bmatrix} 
(y_4 , y_5 , L) , (y_5 , y_6 , M) , (y_6 , y_7 , L)
\end{bmatrix}, \\ \begin{bmatrix} 
(y_8 , y_9 , M) , (y_{10} , y_{11} , M)
\end{bmatrix}, \begin{bmatrix} 
(y_{12} , y_{13} , M) , (y_{14} , y_{15} , M)
\end{bmatrix}, \begin{bmatrix} 
(y_{16} , y_{16} , L) , (y_{17} , y_{17} , L)
\end{bmatrix}
\Big]
\end{array}
}$}

\subsection{Resolution}
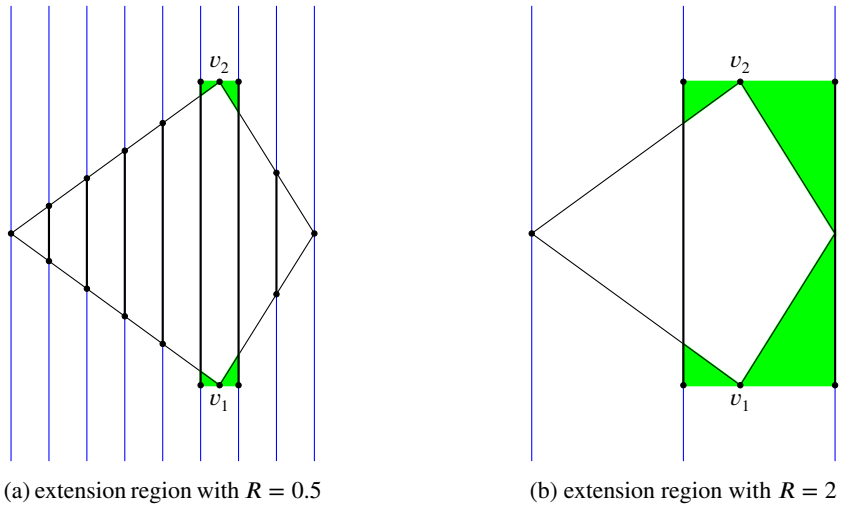
\begin{figure}[!ht]
\centering
\subfloat[extension region with $R=0.5$\label{res1}]{%
\begin{tikzpicture}
\draw[black] (0,2)  -- (2.75,0)node[black , anchor=north] {$v_{1}$}  -- (4,2)  -- (2.75,4) node[black , anchor=south] {$v_{2}$} -- cycle;
\draw[blue] (0,-1) -- (0,5);
\draw[blue] (0.5,-1) -- (0.5,5);
\draw[blue] (1,-1) -- (1,5);
\draw[blue] (1.5,-1) -- (1.5,5);
\draw[blue] (2,-1) -- (2,5);
\draw[blue] (2.5,-1) -- (2.5,5);
\draw[blue] (3,-1) -- (3,5);
\draw[blue] (3.5,-1) -- (3.5,5);
\draw[blue] (4,-1) -- (4,5);
\filldraw[green, thick] (2.75,4)  -- (2.5,42/11) -- (2.5,4) -- cycle;
\filldraw[green, thick] (2.75,0)  -- (2.5,2/11) -- (2.5,0) -- cycle;
\filldraw[green, thick] (2.75,0)  -- (3,2/5) -- (3,0) -- cycle;
\filldraw[green, thick] (2.75,4)  -- (3,18/5) -- (3,4) -- cycle;
\draw[black] (2.5,42/11) -- (2.75,4);
\draw[black] (2.5,2/11) -- (2.75,0);
\draw[black] (3,2/5) -- (2.75,0);
\draw[black] (2.75,4) -- (3,18/5);
\filldraw [black] (0,2)circle (1pt);
\filldraw [black] (2.75,0)circle (1pt);
\filldraw [black] (4,2)circle (1pt);
\filldraw [black] (2.75,4)circle (1pt);
\filldraw [black] (1/2,52/22)circle (1pt); 
\filldraw [black] (1/2,36/22)circle (1pt); 
\draw[black, thick] (1/2,52/22) -- (1/2,36/22);
\filldraw [black] (1,28/22)circle (1pt); 
\filldraw [black] (1,60/22)circle (1pt); 
\draw[black, thick] (1,28/22) -- (1,60/22);
\filldraw [black] (1.5,20/22)circle (1pt); 
\filldraw [black] (1.5,68/22)circle (1pt); 
\draw[black, thick] (1.5,20/22) -- (1.5,68/22);
\filldraw [black] (2,12/22)circle (1pt); 
\filldraw [black] (2,76/22)circle (1pt); 
\draw[black, thick] (2,12/22) -- (2,76/22);
\filldraw [black] (2.5,0)circle (1pt); 
\filldraw [black] (2.5,4)circle (1pt); 
\draw[black, thick] (2.5,0) -- (2.5,4);
\filldraw [black] (3,0)circle (1pt); 
\filldraw [black] (3,4)circle (1pt); 
\draw[black, thick] (3,0) -- (3,4);
\filldraw [black] (3.5,12/10)circle (1pt); 
\filldraw [black] (3.5,28/10)circle (1pt); 
\draw[black, thick] (3.5,12/10) -- (3.5,28/10);
\end{tikzpicture}
}
\hfil
\subfloat[extension region with $R=2$\label{res2}]{%
\begin{tikzpicture}
\draw[black] (5.5,2)  -- (8.25,0)node[black , anchor=north] {$v_{1}$}  -- (9.5,2)  -- (8.25,4) node[black , anchor=south] {$v_{2}$} -- cycle;
\draw[blue] (5.5,-1) -- (5.5,5);
\draw[blue] (7.5,-1) -- (7.5,5);
\draw[blue] (9.5,-1) -- (9.5,5);
\filldraw[green, thick] (8.25,4)  -- (7.5,38/11) -- (7.5,4) -- cycle;
\filldraw[green, thick] (8.25,0)  -- (7.5,6/11) -- (7.5,0) -- cycle;
\filldraw[green, thick] (8.25,4)  -- (9.5,2) -- (9.5,4) -- cycle;
\filldraw[green, thick] (8.25,0)  -- (9.5,2) -- (9.5,0) -- cycle;
\draw[black] (8.25,4) -- (7.5,38/11);
\draw[black] (8.25,0)  -- (7.5,6/11);
\draw[black] (8.25,4) -- (9.5,2);
\draw[black] (8.25,0) -- (9.5,2);
\filldraw [black] (5.5,2)circle (1pt);
\filldraw [black] (8.25,0)circle (1pt);
\filldraw [black] (8.25,4)circle (1pt);
\filldraw [black] (7.5,4)circle (1pt); 
\filldraw [black] (7.5,0)circle (1pt); 
\draw[black, thick](7.5,0) -- (7.5,4);
\filldraw [black] (9.5,4)circle (1pt); 
\filldraw [black] (9.5,0)circle (1pt); 
\draw[black , thick](9.5,0) -- (9.5,4);
\end{tikzpicture}}
\caption{The extension region for different \textit{R}-values. \change{It increases with increasing \textit{R}.}}
\label{fig115c}
\end{figure}

\change{The extension region which is equal to the sum of the regions of the triangles and trapezoids enclosed between the resolution lines and the edges of the piece, see Fig.\,\ref{fig115c}, is considered as the discretization error. The discretization error depends on \textit{R}; it increases with increasing \textit{R} and consequently, a larger part of the strip is not available for placement of other pieces, see Fig.\,\ref{res2}.} However, the computational cost of placing pieces on the strip decreases with \change{increasing \textit{R}} as the number of line segments representing a piece decreases and consequently the amount of computations required for checking whether a piece can be placed in a position decreases, \change{see Fig.\,\ref{res1}}. Therefore, we will use several values for \textit{R} in our tests. The base resolution \textit{R\textsubscript{b}} is set to max\{\textit{P\textsubscript{e}} , \textit{P\textsubscript{p}}/\textit{n\textsubscript{e}}\}, where \textit{P\textsubscript{e}} is the smallest \textit{x}-projection of all non-vertical edges of all pieces, \textit{P\textsubscript{p}} is the \textit{x}-projection of the smallest piece and \textit{n\textsubscript{e}} is the number of edges of the smallest piece. This leads to a good accuracy of the representation of the pieces by line segments as in most of the cases \textit{P\textsubscript{e}} is larger than \textit{P\textsubscript{p}}/\textit{n\textsubscript{e}} and this leads to at least one line segment per edge. However, for data sets with near vertical edges, $R = P_e$ would be very small, leading to an excessive computational cost. Therefore, in this case, we choose \textit{P\textsubscript{p}}/\textit{n\textsubscript{e}} as the base resolution. \change{\textit{R}-values smaller than $R_b$ are also considered}, to improve the quality of the solution by reducing the extension regions and the gaps between pieces.

\section{Placement}\label{section3}

\change{We now describe the bottom-left-fill placement of the semi-discretized pieces.
The strip is a rectangle with a fixed width and a variable length, semi-discretized along the $x$-axis with the same distance $R$ between the resolution lines as for the pieces. 
The parts of each resolution line, i.e.\,at $x_i = i\times R$, $i= 0,1,2,\ldots$, that are occupied by line segments of already placed pieces are stored in a vector of vectors of tuples, i.e.\,a similar data structure as used to represent the pieces. However, we will merely use the word `interval' instead of `tuple' (or `line segment') both for the pieces and the strip. 
The pieces are sorted, e.g. according to decreasing bounding box area, discretized and subsequently placed in the strip. If several copies of the same piece are placed consecutively, the piece is discretized only once.}

\change{To describe the placement of a semi-discretized piece, we refine the notations introduced in section\,\ref{newsection2}.
\begin{itemize}
    \item The \textit{y}-coordinates $b_{i,j}$ and $t_{i,j}$ in tuple $(b_{i,j} , t_{i,j} , P_{i,j})$ are called the local coordinates, where $b_{i,j}$ and $t_{i,j}$ denote respectively the bottom and the top endpoint of the \textit{j}-th interval on the \textit{i}-th resolution line of the piece. The label $P_{i,j}$ indicates the position of the interval in the piece, see section\ \ref{sweep-line}. The bottom left corner of the axis-aligned bounding box of the piece has (local) coordinates $(x,y) = (0,0)$.
    \item The bottom-left corner of the strip has (strip) coordinates $(x,y) = (0,0)$. The partially filled semi-discretized strip consists of tuples $(b^s_{k,l} , t^s_{k,l} , P^s_{k,l})$, indicating space occupied by already placed pieces, with $b^s_{k,l}$ and $t^s_{k,l}$ denoting respectively the bottom and the top endpoint of the \textit{l}-th interval on the \textit{k}-th resolution line, with superscript \textit{s} referring to the strip. The label $P^s_{k,l}$ indicates the position of the interval in the union of the already placed intervals.
    \item When an interval of a piece is (tentatively) placed in the strip, the coordinates of its endpoints in the strip are called the strip coordinates. The local coordinates are transformed into the strip coordinates by adding a ‘translation vector’ $t = (x_t , y_t)$ with $x_t = m.R$, $m= 0,1,2,\ldots$, such that $m$ is minimal and, for that $m$, $y_t$ is minimal.
    Hence \textit{t} indicates the position of the bottom left corner of the axis-aligned bounding box of the piece in the strip. According to the bottom-left strategy we initialize the translation vector as $t = (0,0)$.
\end{itemize}}

\begin{figure}[!ht]
\centering
\subfloat[\label{figa}]{%
\begin{tikzpicture}
\draw[blue] (0,0) -- (0,5/2);
\draw[blue] (0.5,0) -- (0.5,5/2);
\draw[blue] (1,0) -- (1,5/2)node[black , anchor=south] {\textit{Strip}};
\draw[blue] (1.5,0) -- (1.5,5/2);
\draw[blue] (2,0) -- (2,5/2);
\draw[red, thick] (0,0) -- (0,1.5/2);
\draw[red, thick] (0,2.5/2) -- (0,5/2);
\draw[red , thick] (0.5,0) -- (0.5,4.5/2);
\draw[red , thick] (1,0.5/2) -- (1,4/2);
\draw[red , thick] (1.5,1/2) -- (1.5,3.5/2);
\draw[red , thick] (4/2,1/2) -- (4/2,7/4);
\draw[blue] (5/2,0) -- (5/2,10/4);
\draw[green , thick] (0,3.5) -- (0,4);
\draw[green , thick] (0.5,3.5) -- (0.5,4)node[black , anchor=south] {\textit{Next piece}};
\draw[green , thick] (1,3.5) -- (1,4);
\node[black , thick] at (1.2,3.3) {$I_{2,0}$};
\node[black , thick] at (0.5,3.3) {$I_{1,0}$};
\node[black , thick] at (-0.1,3.3) {$I_{0,0}$};
\draw[gray , thick] (-0.5,-0.5) -> (3,-0.5);
\draw[gray , thick] (-0.5,-0.5) -> (-0.5,5.5/2);
\filldraw [gray] (-0.5,0) circle (0.5pt)node[black , anchor=east] {$0$};
\filldraw [gray] (-0.5,5/2) circle (0.5pt)node[black , anchor=east] {$y_{max}$};
\filldraw [gray] (0,-0.5) circle (0.5pt)node[black , anchor=north] {$0$};
\filldraw [gray] (0,-0.5) circle (0.5pt)node[black , anchor=north] {$0$};
\filldraw [gray] (0.5,-0.5) circle (0.5pt)node[black , anchor=north] {$R$};
\filldraw [gray] (1,-0.5) circle (0.5pt)node[black , anchor=north] {$2R$};
\filldraw [gray] (1.5,-0.5) circle (0.5pt)node[black , anchor=north] {$3R$};
\filldraw [gray] (2,-0.5) circle (0.5pt)node[black , anchor=north] {$4R$};
\filldraw [gray] (2.5,-0.5) circle (0.5pt)node[black , anchor=north] {$5R$};
\end{tikzpicture}
}
\hfil
\subfloat[\label{figb}]{%
\begin{tikzpicture}
\draw[blue] (2.5,0) -- (2.5,5/2);
\draw[blue] (3,0) -- (3,5/2);
\draw[blue] (3.5,0) -- (3.5,5/2);
\draw[blue] (4,0) -- (4,5/2);
\draw[blue] (4.5,0) -- (4.5,5/2);
\draw[blue] (5,0) -- (5,5/2);
\draw[red, thick] (2.5,0) -- (2.5,1.5/2);
\draw[red, thick] (2.5,2.5/2) -- (2.5,5/2);
\draw[red , thick] (3,0) -- (3,4.5/2);
\draw[red , thick] (3.5,0.5/2) -- (3.5,4/2);
\draw[red , thick] (4,1/2) -- (4,3.5/2);
\draw[red , thick] (4.5,1/2) -- (4.5,7/4);
\draw[gray , thick] (2,-0.5) -> (5,-0.5);
\filldraw [gray] (2.5,-0.5) circle (0.5pt)node[black , anchor=north] {$0$};
\filldraw [gray] (3,-0.5) circle (0.5pt)node[black , anchor=north] {$R$};
\filldraw [gray] (3.5,-0.5) circle (0.5pt)node[black , anchor=north] {$2R$};
\filldraw [gray] (4,-0.5) circle (0.5pt)node[black , anchor=north] {$3R$};
\filldraw [gray] (4.5,-0.5) circle (0.5pt)node[black , anchor=north] {$4R$};
\filldraw [gray] (5,-0.5) circle (0.5pt)node[black , anchor=north] {$5R$};
\draw[green , thick] (2.2,0) --node[black , anchor=east] {$I_{0,0}$} (2.2,1/2);
\draw[black] (2.4,0.4/2)--(2.6,0.6/2);
\draw[black] (2.4,0.6/2)--(2.6,0.4/2);
\end{tikzpicture}
}
\hfil
\subfloat[\label{figc}]{%
\begin{tikzpicture}
\draw[blue] (2.5,0) -- (2.5,5/2);
\draw[blue] (3,0) -- (3,5/2);
\draw[blue] (3.5,0) -- (3.5,5/2);
\draw[blue] (4,0) -- (4,5/2);
\draw[blue] (4.5,0) -- (4.5,5/2);
\draw[blue] (5,0) -- (5,5/2);
\draw[red, thick] (2.5,0) -- (2.5,1.5/2);
\draw[red, thick] (2.5,2.5/2) -- (2.5,5/2);
\draw[red , thick] (3,0) -- (3,4.5/2);
\draw[red , thick] (3.5,0.5/2) -- (3.5,4/2);
\draw[red , thick] (4,1/2) -- (4,3.5/2);
\draw[red , thick] (4.5,1/2) -- (4.5,7/4);
\draw[gray , thick] (2,-0.5) -> (5,-0.5);
\filldraw [gray] (2.5,-0.5) circle (0.5pt)node[black , anchor=north] {$0$};
\filldraw [gray] (3,-0.5) circle (0.5pt)node[black , anchor=north] {$R$};
\filldraw [gray] (3.5,-0.5) circle (0.5pt)node[black , anchor=north] {$2R$};
\filldraw [gray] (4,-0.5) circle (0.5pt)node[black , anchor=north] {$3R$};
\filldraw [gray] (4.5,-0.5) circle (0.5pt)node[black , anchor=north] {$4R$};
\filldraw [gray] (5,-0.5) circle (0.5pt)node[black , anchor=north] {$5R$};
\draw[green , thick] (2.5,3/4) --node[black , anchor=east] {$I_{0,0}$} (2.5,5/4);
\draw[green , thick] (3.4,1.5/2) -- (3.4,2.5/2);
\node[black , thick] at (3.25,1.4) {$I_{2,0}$};
\draw[black] (3.4,1.9/2)--(3.6,2.1/2);
\draw[black] (3.4,2.1/2)--(3.6,1.9/2);
\end{tikzpicture}
}
\hfil
\subfloat[\label{figd}]{%
\begin{tikzpicture}
\draw[blue] (2.5,0) -- (2.5,5/2);
\draw[blue] (3,0) -- (3,5/2);
\draw[blue] (3.5,0) -- (3.5,5/2);
\draw[blue] (4,0) -- (4,5/2);
\draw[blue] (4.5,0) -- (4.5,5/2);
\draw[blue] (5,0) -- (5,5/2);
\draw[red, thick] (2.5,0) -- (2.5,1.5/2);
\draw[red, thick] (2.5,2.5/2) -- (2.5,5/2);
\draw[red , thick] (3,0) -- (3,4.5/2);
\draw[red , thick] (3.5,0.5/2) -- (3.5,4/2);
\draw[red , thick] (4,1/2) -- (4,3.5/2);
\draw[red , thick] (4.5,1/2) -- (4.5,7/4);
\draw[gray , thick] (2,-0.5) -> (5,-0.5);
\filldraw [gray] (2.5,-0.5) circle (0.5pt)node[black , anchor=north] {$0$};
\filldraw [gray] (3,-0.5) circle (0.5pt)node[black , anchor=north] {$R$};
\filldraw [gray] (3.5,-0.5) circle (0.5pt)node[black , anchor=north] {$2R$};
\filldraw [gray] (4,-0.5) circle (0.5pt)node[black , anchor=north] {$3R$};
\filldraw [gray] (4.5,-0.5) circle (0.5pt)node[black , anchor=north] {$4R$};
\filldraw [gray] (5,-0.5) circle (0.5pt)node[black , anchor=north] {$5R$};
\draw[green , thick] (2.5,3/4) --node[black , anchor=east] {$I_{0,0}$} (2.5,5/4);
\draw[green , thick] (3.5,4/2) -- (3.5,5/2);
\node[black , thick] at (3.25,2.25) {$I_{2,0}$};
\end{tikzpicture}
}
\hfil
\subfloat[\label{fige}]{%
\begin{tikzpicture}
\draw[blue] (2.5,0) -- (2.5,5/2);
\draw[blue] (3,0) -- (3,5/2);
\draw[blue] (3.5,0) -- (3.5,5/2);
\draw[blue] (4,0) -- (4,5/2);
\draw[blue] (4.5,0) -- (4.5,5/2);
\draw[blue] (5,0) -- (5,5/2);
\draw[red, thick] (2.5,0) -- (2.5,1.5/2);
\draw[red, thick] (2.5,2.5/2) -- (2.5,5/2);
\draw[red , thick] (3,0) -- (3,4.5/2);
\draw[red , thick] (3.5,0.5/2) -- (3.5,4/2);
\draw[red , thick] (4,1/2) -- (4,3.5/2);
\draw[red , thick] (4.5,1/2) -- (4.5,7/4);
\draw[gray , thick] (2,-0.5) -> (5,-0.5);
\draw[gray , thick] (2,-0.5) -> (2,5.5/2);
\filldraw [gray] (2,5/2) circle (0.5pt)node[black , anchor=east] {$y_{max}$};
\filldraw [gray] (2.5,-0.5) circle (0.5pt)node[black , anchor=north] {$0$};
\filldraw [gray] (3,-0.5) circle (0.5pt)node[black , anchor=north] {$R$};
\filldraw [gray] (3.5,-0.5) circle (0.5pt)node[black , anchor=north] {$2R$};
\filldraw [gray] (4,-0.5) circle (0.5pt)node[black , anchor=north] {$3R$};
\filldraw [gray] (4.5,-0.5) circle (0.5pt)node[black , anchor=north] {$4R$};
\filldraw [gray] (5,-0.5) circle (0.5pt)node[black , anchor=north] {$5R$};
\draw[green , thick] (2.5,3/4) --node[black , anchor=east] {$I_{0,0}$} (2.5,5/4);
\draw[green , thick] (3.5,4/2) -- (3.5,5/2);
\node[black , thick] at (3.25,2.4) {$I_{2,0}$};
\node[black , thick] at (2.75,1.9) {$I_{1,0}$};
\draw[green , thick] (2.9,4/2)-- (2.9,5/2);
\draw[black] (2.9,4.4/2)--(3.1,4.6/2);
\draw[black] (2.9,4.6/2)--(3.1,4.4/2);
\end{tikzpicture}
}
\hfil
\subfloat[\label{figf}]{%
\begin{tikzpicture}
\draw[blue] (2.5,0) -- (2.5,5/2);
\draw[blue] (3,0) -- (3,5/2);
\draw[blue] (3.5,0) -- (3.5,5/2);
\draw[blue] (4,0) -- (4,5/2);
\draw[blue] (4.5,0) -- (4.5,5/2);
\draw[blue] (5,0) -- (5,5/2);
\draw[red, thick] (2.5,0) -- (2.5,1.5/2);
\draw[red, thick] (2.5,2.5/2) -- (2.5,5/2);
\draw[red , thick] (3,0) -- (3,4.5/2);
\draw[red , thick] (3.5,0.5/2) -- (3.5,4/2);
\draw[red , thick] (4,1/2) -- (4,3.5/2);
\draw[red , thick] (4.5,1/2) -- (4.5,7/4);
\draw[gray , thick] (2,-0.5) -> (5,-0.5);
\filldraw [gray] (2.5,-0.5) circle (0.5pt)node[black , anchor=north] {$0$};
\filldraw [gray] (3,-0.5) circle (0.5pt)node[black , anchor=north] {$R$};
\filldraw [gray] (3.5,-0.5) circle (0.5pt)node[black , anchor=north] {$2R$};
\filldraw [gray] (4,-0.5) circle (0.5pt)node[black , anchor=north] {$3R$};
\filldraw [gray] (4.5,-0.5) circle (0.5pt)node[black , anchor=north] {$4R$};
\filldraw [gray] (5,-0.5) circle (0.5pt)node[black , anchor=north] {$5R$};
\draw[green , thick] (2.9,0) -- (2.9,1/2);
\node[black , thick] at (2.75,0.7) {$I_{0,0}$};
\draw[black] (2.9,0.4/2)--(3.1,0.6/2);
\draw[black] (2.9,0.6/2)--(3.1,0.4/2);
\end{tikzpicture}
}
\hfil
\subfloat[\label{figg}]{%
\begin{tikzpicture}
\draw[blue] (2.5,0) -- (2.5,5/2);
\draw[blue] (3,0) -- (3,5/2);
\draw[blue] (3.5,0) -- (3.5,5/2);
\draw[blue] (4,0) -- (4,5/2);
\draw[blue] (4.5,0) -- (4.5,5/2);
\draw[blue] (5,0) -- (5,5/2);
\draw[red, thick] (2.5,0) -- (2.5,1.5/2);
\draw[red, thick] (2.5,2.5/2) -- (2.5,5/2);
\draw[red , thick] (3,0) -- (3,4.5/2);
\draw[red , thick] (3.5,0.5/2) -- (3.5,4/2);
\draw[red , thick] (4,1/2) -- (4,3.5/2);
\draw[red , thick] (4.5,1/2) -- (4.5,7/4);
\draw[gray , thick] (2,-0.5) -> (5,-0.5);
\filldraw [gray] (2.5,-0.5) circle (0.5pt)node[black , anchor=north] {$0$};
\filldraw [gray] (3,-0.5) circle (0.5pt)node[black , anchor=north] {$R$};
\filldraw [gray] (3.5,-0.5) circle (0.5pt)node[black , anchor=north] {$2R$};
\filldraw [gray] (4,-0.5) circle (0.5pt)node[black , anchor=north] {$3R$};
\filldraw [gray] (4.5,-0.5) circle (0.5pt)node[black , anchor=north] {$4R$};
\filldraw [gray] (5,-0.5) circle (0.5pt)node[black , anchor=north] {$5R$};
\draw[green , thick] (2.9,4/2) -- (2.9,5/2);
\node[black , thick] at (2.75,2.7) {$I_{0,0}$};
\draw[black] (2.9,4.4/2)--(3.1,4.6/2);
\draw[black] (2.9,4.6/2)--(3.1,4.4/2);
\end{tikzpicture}
}
\hfil
\subfloat[\label{figh}]{%
\begin{tikzpicture}
\draw[blue] (2.5,0) -- (2.5,5/2);
\draw[blue] (3,0) -- (3,5/2);
\draw[blue] (3.5,0) -- (3.5,5/2);
\draw[blue] (4,0) -- (4,5/2);
\draw[blue] (4.5,0) -- (4.5,5/2);
\draw[blue] (5,0) -- (5,5/2);
\draw[red, thick] (2.5,0) -- (2.5,1.5/2);
\draw[red, thick] (2.5,2.5/2) -- (2.5,5/2);
\draw[red , thick] (3,0) -- (3,4.5/2);
\draw[red , thick] (3.5,0.5/2) -- (3.5,4/2);
\draw[red , thick] (4,1/2) -- (4,3.5/2);
\draw[red , thick] (4.5,1/2) -- (4.5,7/4);
\draw[gray , thick] (2,-0.5) -> (5,-0.5);
\filldraw [gray] (2.5,-0.5) circle (0.5pt)node[black , anchor=north] {$0$};
\filldraw [gray] (3,-0.5) circle (0.5pt)node[black , anchor=north] {$R$};
\filldraw [gray] (3.5,-0.5) circle (0.5pt)node[black , anchor=north] {$2R$};
\filldraw [gray] (4,-0.5) circle (0.5pt)node[black , anchor=north] {$3R$};
\filldraw [gray] (4.5,-0.5) circle (0.5pt)node[black , anchor=north] {$4R$};
\filldraw [gray] (5,-0.5) circle (0.5pt)node[black , anchor=north] {$5R$};
\draw[green , thick] (3.4,0) -- (3.4,1/2);
\node[black , thick] at (3.25,0.7) {$I_{0,0}$};
\draw[black] (3.4,0.4/2)--(3.6,0.6/2);
\draw[black] (3.4,0.6/2)--(3.6,0.4/2);
\end{tikzpicture}
}
\hfil
\subfloat[\label{figi}]{%
\begin{tikzpicture}
\draw[blue] (2.5,0) -- (2.5,5/2);
\draw[blue] (3,0) -- (3,5/2);
\draw[blue] (3.5,0) -- (3.5,5/2);
\draw[blue] (4,0) -- (4,5/2);
\draw[blue] (4.5,0) -- (4.5,5/2);
\draw[blue] (5,0) -- (5,5/2);
\draw[red, thick] (2.5,0) -- (2.5,1.5/2);
\draw[red, thick] (2.5,2.5/2) -- (2.5,5/2);
\draw[red , thick] (3,0) -- (3,4.5/2);
\draw[red , thick] (3.5,0.5/2) -- (3.5,4/2);
\draw[red , thick] (4,1/2) -- (4,3.5/2);
\draw[red , thick] (4.5,1/2) -- (4.5,7/4);
\draw[green , thick] (3.5,2) --  (3.5,2.55);
\node[black , thick] at (3.25,2.7) {$I_{0,0}$};
\draw[gray , thick] (2,-0.5) -> (5,-0.5);
\filldraw [gray] (2.5,-0.5) circle (0.5pt)node[black , anchor=north] {$0$};
\filldraw [gray] (3,-0.5) circle (0.5pt)node[black , anchor=north] {$R$};
\filldraw [gray] (3.5,-0.5) circle (0.5pt)node[black , anchor=north] {$2R$};
\filldraw [gray] (4,-0.5) circle (0.5pt)node[black , anchor=north] {$3R$};
\filldraw [gray] (4.5,-0.5) circle (0.5pt)node[black , anchor=north] {$4R$};
\filldraw [gray] (5,-0.5) circle (0.5pt)node[black , anchor=north] {$5R$};
\end{tikzpicture}
}
\hfil
\subfloat[\label{figj}]{%
\begin{tikzpicture}
\draw[blue] (2.5,0) -- (2.5,5/2);
\draw[blue] (3,0) -- (3,5/2);
\draw[blue] (3.5,0) -- (3.5,5/2);
\draw[blue] (4,0) -- (4,5/2);
\draw[blue] (4.5,0) -- (4.5,5/2);
\draw[blue] (5,0) -- (5,5/2);
\draw[red, thick] (2.5,0) -- (2.5,1.5/2);
\draw[red, thick] (2.5,2.5/2) -- (2.5,5/2);
\draw[red , thick] (3,0) -- (3,4.5/2);
\draw[red , thick] (3.5,0.5/2) -- (3.5,4/2);
\draw[red , thick] (4,1/2) -- (4,3.5/2);
\draw[red , thick] (4.5,1/2) -- (4.5,7/4);
\draw[green , thick] (3.5,2) --  (3.5,2.55);
\node[black , thick] at (3.25,2.7) {$I_{0,0}$};
\draw[green , thick] (4.5,2) -- (4.5,2.5);
\node[black , thick] at (4.5,2.7) {$I_{2,0}$};
\draw[gray , thick] (2,-0.5) -> (5,-0.5);
\filldraw [gray] (2.5,-0.5) circle (0.5pt)node[black , anchor=north] {$0$};
\filldraw [gray] (3,-0.5) circle (0.5pt)node[black , anchor=north] {$R$};
\filldraw [gray] (3.5,-0.5) circle (0.5pt)node[black , anchor=north] {$2R$};
\filldraw [gray] (4,-0.5) circle (0.5pt)node[black , anchor=north] {$3R$};
\filldraw [gray] (4.5,-0.5) circle (0.5pt)node[black , anchor=north] {$4R$};
\filldraw [gray] (5,-0.5) circle (0.5pt)node[black , anchor=north] {$5R$};
\end{tikzpicture}
}
\hfil
\subfloat[\label{figk}]{%
\begin{tikzpicture}
\draw[blue] (2.5,0) -- (2.5,5/2);
\draw[blue] (3,0) -- (3,5/2);
\draw[blue] (3.5,0) -- (3.5,5/2);
\draw[blue] (4,0) -- (4,5/2);
\draw[blue] (4.5,0) -- (4.5,5/2);
\draw[blue] (5,0) -- (5,5/2);
\draw[red, thick] (2.5,0) -- (2.5,1.5/2);
\draw[red, thick] (2.5,2.5/2) -- (2.5,5/2);
\draw[red , thick] (3,0) -- (3,4.5/2);
\draw[red , thick] (3.5,0.5/2) -- (3.5,4/2);
\draw[red , thick] (4,1/2) -- (4,3.5/2);
\draw[red , thick] (4.5,1/2) -- (4.5,7/4);
\draw[green , thick] (3.5,2) --  (3.5,2.55);
\draw[green , thick] (4,2) --  (4,2.5);
\draw[green , thick] (4.5,2) --  (4.5,2.5);
\node[black , thick] at (3.25,2.7) {$I_{0,0}$};
\node[black , thick] at (4.5,2.7) {$I_{2,0}$};
\node[black , thick] at (3.9,2.7) {$I_{1,0}$};
\draw[gray , thick] (2,-0.5) -> (5,-0.5);
\filldraw [gray] (2.5,-0.5) circle (0.5pt)node[black , anchor=north] {$0$};
\filldraw [gray] (3,-0.5) circle (0.5pt)node[black , anchor=north] {$R$};
\filldraw [gray] (3.5,-0.5) circle (0.5pt)node[black , anchor=north] {$2R$};
\filldraw [gray] (4,-0.5) circle (0.5pt)node[black , anchor=north] {$3R$};
\filldraw [gray] (4.5,-0.5) circle (0.5pt)node[black , anchor=north] {$4R$};
\filldraw [gray] (5,-0.5) circle (0.5pt)node[black , anchor=north] {$5R$};
\end{tikzpicture}
}
\caption{\change{Placement algorithm: \ref{figa} represents the strip with already placed pieces where red indicates  filled intervals. In \ref{figb}, interval $I_{0,0}$ represented by tuple $(b_{0,0} , t_{0,0} , R)$ cannot be placed with $t = (0,0)$ so we shift it upwards. In \ref{figc}, $I_{0,0}$ is placed at $x=0$ with $t=(0,1)$, but $I_{2,0}$ represented by tuple $(b_{2,0} , t_{2,0} , L)$ cannot be placed with this translation \textit{t}. Therefore, in \ref{figd}, we shift $I_{2,0}$ upwards, \textit{t} is updated to $(0,2)$. However, in \ref{fige} $I_{1,0}$ represented by tuple $(b_{1,0} , t_{1,0} , M)$ cannot be placed with this translation \textit{t}. We update \textit{t} to shift the intervals to the right. We return to $I_{0,0}$. In \ref{figf}, \ref{figg} we cannot place $I_{0,0}$ at $x=R$. Therefore, we again update \textit{t} to shift the intervals to the right. In \ref{figi}, \ref{figj} and \ref{figk} $I_{0,0}$, $I_{2,0}$ and $I_{1,0}$ can be placed at respectively $x=2R$, $x=4R$ and $x=3R$ with $t = (2R,2)$.}}
\label{fig115}
\end{figure}
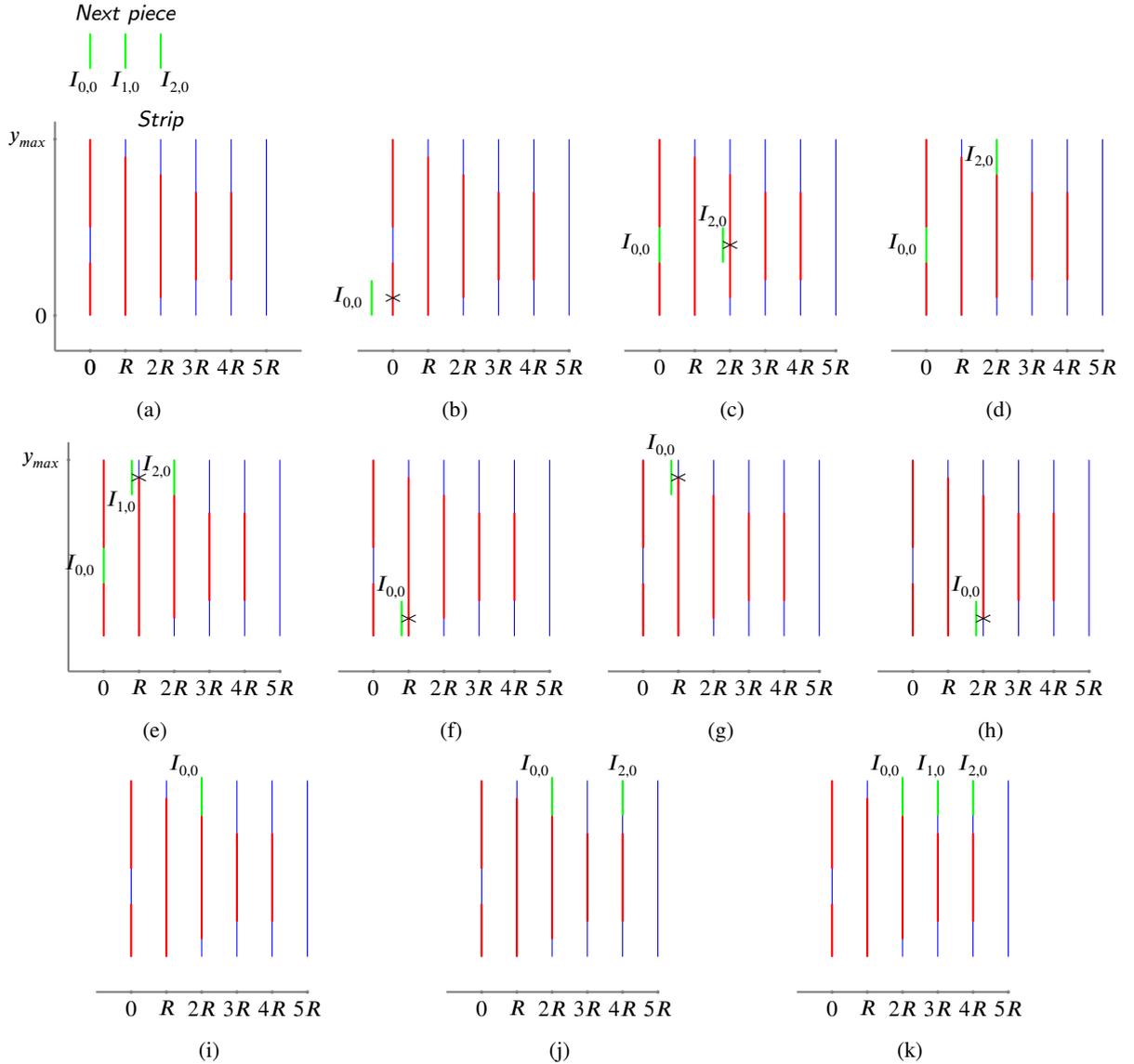

\change{Due to the semi-discretization, placing a piece in the strip without overlap corresponds to finding a translation vector \textit{t} such that the intervals of the piece do not overlap with the filled intervals in the strip (unless the labels $P_{i,j}$ and $P^s_{k,l}$ allow overlap). This requires tests with the strip coordinates of the endpoints of the intervals $(b_{i,j}+y_t , t_{i,j}+y_t)$, $i = 0, 1, 2, \ldots$, of the piece and the coordinates of the endpoints of the filled intervals in resolution lines $(b^s_{i+m,l} , t^s_{i+m,l})$ $\forall j,\forall l$. Fig.\,\ref{fig115} illustrates the placement algorithm.}  

We process the intervals of the piece in a given order, starting with the leftmost interval. For every interval, we test whether the interval can be placed in the strip with the current translation vector as follows. \change{An interval with the label $P_{i,j} = M$ can be placed if there is no overlap with an already filled interval.
An interval with label $P_{i,j} = R/L$ can be placed if there is no overlap with an already filled intervals in the strip or if there is an overlap with an already filled interval with label $P^s_{i+m,l} = L/R$.}

\change{For a given translation vector, it is not necessary to perform the tests for all intervals to decide that the current translation vector does not allow placement without overlap. Indeed, as soon as an interval cannot be placed as mentioned above, we can update the translation vector \textit{t}. Hence, we proceed as follows.}

\change{\begin{enumerate}
    \item If the interval can be placed under the above mentioned conditions, we consider the next interval.
    \item If the interval cannot be placed, we try to place it in the current resolution line, by shifting it repeatedly upwards and testing whether it overlaps with the next filled interval on the current resolution line.
    \begin{enumerate}
        \item If there is no overlap, the interval can be placed on the current resolution line. We compute the required shift in the \textit{y}-direction and add it to $y_t$ to update the translation vector and we consider the next interval.
        \item If the interval cannot be placed on the current resolution line, we break the loop over the intervals, we return to the first (leftmost) interval and we update the translation vector to $t = (x_t+R, 0)$, i.e. $x_t$ is set to the \textit{x}-value of the next resolution line and $y_t$ is set to the bottom point on that resolution line, i.e.\,$y_t = 0$.
    \end{enumerate}
\end{enumerate}}

\change{If we have performed the loop over all intervals, we must distinguish between two cases.
 \begin{enumerate}
    \item If the translation vector has not been updated during the whole loop over the intervals, then the translation vector allows to place all the intervals on the strip. The piece is then placed and the filled intervals of the strip are updated as follows.
    \begin{enumerate}
        \item If an interval is placed without overlap with the already filled intervals, the interval is added to the strip and the label of the tuple is equal to that of the piece interval.
        \item If an interval with label equal to \textit{R}/\textit{L} is placed on an already filled interval with label equal to \textit{L}/\textit{R}, the label of the overlapping interval in the strip is updated to \textit{M}, i.e.\,no more intervals can be placed on it.
    \end{enumerate}
    \item If the translation vector has been updated, the loop over the intervals must be repeated with the updated translation vector, to ensure that indeed all intervals of the piece can be placed with this translation vector.
\end{enumerate}}

The order of checking the intervals of the piece during placement can have an impact on the total number of tests and the execution time. Indeed, each test with an interval can result in an update of the translation vector. Consecutive piece intervals often have endpoints with nearly equal \textit{y}-coordinates and the same holds for the consecutive strip intervals. Checking intervals in the order of increasing \textit{x}-coordinates is therefore often not optimal, since this ordering only slowly reveals global information about the possibility to place the piece. Therefore, we check the intervals of a piece covering $L+1$ resolution lines in the order \{$0, L, L/2, L/4, 3L/4,\ldots$\} which quickly gives information about the possible placement of intervals located at the beginning, end, and middle of the piece. Therefore this can accelerate the decision about the placement of the piece in that position of the strip. 

The placement can be done more efficiently when there are several copies of the same piece in the data set. When such a piece has been placed in the strip with translation vector $t = (x_{t},y_{t})$, the search for finding an optimal position for the next copy of that piece starts with this translation vector instead of with $t = (0,0)$.

\begin{figure}[ht]
  \centering
    \includegraphics[width=.8\linewidth]{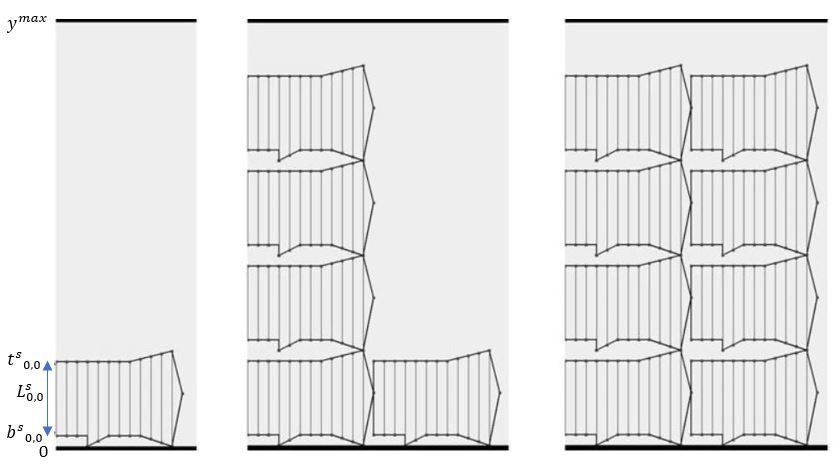}
    \caption{\change{Placement of 8 copies of a piece. Placement of each piece starts from the translation vector of the previous piece. To check placement of interval ($b_{i,j}$, $t_{i,j}$, $P_{i,j}$) with length \textit{L\textsubscript{i,j}}, interval $(y_{max} - L_{i,j}, y_{max}]$ on each resolution line of the strip is not checked.}}
    \label{firstpiece}
\end{figure}

\medskip
\change{As an example, consider the placement of eight copies of a piece from the data set `Shirts', called $P_0, \ldots, P_7$, see Fig.\,\ref{firstpiece}. Each piece covers 13 resolution lines, each with 1 interval, denoted by $I_{i,0}$, except for the 4-th resolution line where there are two intervals $I_{3,0}$ and $I_{3,1}$ (with different position labels). The placement of $P_{0}$ requires zero checks because there are no filled intervals in the strip yet. The placement of $P_{1}$ starts with translation vector $t = (0,0)$,
but for this translation the interval $I_{0,0}$, represented by tuple $(b_{0,0},t_{0,0},R)$, of $P_{1}$ overlaps with interval $I^s_{0,0}$, represented by tuple $(b^s_{0,0},t^s_{0,0},R)$ in the strip, therefore, it is shifted up over a distance $L^s_{0,0}$, see Fig.\,\ref{firstpiece}. 
Since there is no overlap, the next interval of $P_{1}$ is checked. The same amount of checks is required for intervals $I_{12,0}$, $I_{6,0}$, $I_{9,0}$, $I_{1,0}$, $I_{4,0}$, $I_{7,0}$, $I_{10,0}$, $I_{2,0}$, $I_{5,0}$, $I_{8,0}$ and $I_{11,0}$. Intervals $I_{3,0}$ and $I_{3,1}$ are checked separately and require one check with each filled intervals $I^s_{3,0}$ and $I^s_{3,1}$ of the strip. The translation vector is updated for $I_{0,0}$, $I_{3,0}$, $I_{10,0}$ and $I_{11,0}$. With the last updated translation vector, all 14 intervals of $P_{1}$ are checked again and the same amount of checks are required. In total 32 checks are required for placing $P_{1}$. The same number of interval checks and updates of the translation vector are needed to place $P_{2}$ and $P_{3}$, as the placements of $P_{2}$ and $P_{3}$ start with the translation vectors of $P_{1}$ and $P_{2}$, respectively. However, 32 and 64 extra checks are also required to search for the indexes of the filled intervals. Placement of $P_{4}$ starts by checking interval $I_{0,0}$ with the translation vector of $P_{3}$. Only one check is done in resolution line at $x = 0$ of the strip with 3 extra checks to find the index of the last filled interval. For each interval represented by tuple ($b_{i,j} , t_{i,j} , P_{i,j}$) with length $L_{i,j}$, no placement check is done in the interval $(y_{max} - L_{i,j}, y_{max}]$ of the resolution lines of the strip. Then the search starts in the next resolution line at $x = R$, i.e.\,one shift to the right. On resolution lines at $x = R$ to $x = 11R$ of the strip 48 checks are done, i.e.\,8 checks on resolution line $x =3 R$ and 4 checks on each of other resolution lines. On resolution line at $x = 12R$, the last check is done for interval $I_{0,0}$ of $P_{4}$. As the next vectors of the strip are empty, no checks are required for the other 13 intervals of $P_{4}$. In total 53 checks are required to place this piece. The placement of $P_{5}$ is similar to the placement of $P_{2}$, however, in each cycle, interval $I_{0,0}$ requires three checks at $x = 12R$, i.e.\,one check with interval $I_{12,0}$ of $P_{0}$, one check with interval $I_{12,0}$ of $P_{1}$, and one check with interval $I_{0,0}$, of $P_{4}$. In total 36 checks are required for placing $P_{5}$. The placement of $P_{6}$ is similar to the placement of $P_{5}$, however, in each cycle, interval $I_{0,0}$ requires 5 checks at $x = 12R$, 30 extra checks are also required to search for the indexes of the filled intervals. In total 70 checks are required for placing $P_{6}$. The placement of $P_{7}$ is similar to the placement of $P_{6}$, however, in each cycle, interval $I_{0,0}$ requires 7 checks at $x = 12R$, 60 extra checks are also required to search for the indexes of the filled intervals. In total 104 checks are required for placing $P_{7}$. Table \ref{counter} presents the number of checks for placement of each piece. }

\begin{table}[]
    \centering
    \caption{Number of translation vector updates and interval checks for placing eight copies of a piece.}
    \begin{tabular}{|c|c|c|c|c|}
    \hline
    Piece   &     Updates in      & Shift right&  Total number of   \\
            &  translation vector&             &      checks   \\\hline
\change{$P_{0}$} &          0         &      0      &       \change{ 0}    \\\hline             
\change{$P_{1}$} &          4         &      0      &       \change{ 32}    \\\hline    
\change{$P_{2}$} &          4         &      0      &       \change{ 64}    \\\hline    
\change{$P_{3}$} &          4         &      0      &       \change{ 96}    \\\hline    
\change{$P_{4}$} &         12         &      12     &       \change{ 53}    \\\hline    
\change{$P_{5}$} &          4         &      0      &       \change{ 36}    \\\hline    
\change{$P_{6}$} &          4         &      0      &       \change{ 70}    \\\hline    
\change{$P_{7}$} &          4         &      0      &      \change{ 104}    \\\hline    
             
    \end{tabular}
    \label{counter}
\end{table}

In case rotation of pieces is allowed, the bottom-left-fill algorithm considers all allowed rotation angles for the placement of each piece. Each piece is placed with the rotation angle that minimizes the largest strip \textit{x}-coordinate of the piece. In case several rotation angles lead to the minimum, the rotation angle that minimizes the maximum strip \textit{y}-coordinate of the piece is selected. In case several rotation angles satisfy these two criteria, the smallest rotation angle is selected.
The decision about the optimal placement of a piece with several rotation angles is made locally, hence allowing (many) rotations does not necessarily lead to a better global solution.

\section{Results}

In this section we discuss the performance of the proposed bottom-left-fill algorithm using semi-discrete representation, implemented in C++ using the Qt environment. Experiments were carried out on a single core of an Intel Core i7-7500U processor at 2.70 GHz and 16 GB of RAM. The computational experiments are done with the following ESICUP data sets: `Shirts' \cite{18}, `Han' \cite{15}, `Trousers' \cite{19}, `Jakob2' \cite{16}, `poly5b' \cite{17} and \change{`Swim' \cite{19}}. These data sets contain both non-convex and convex pieces, vary in size and in the allowed rotation angles. 
We also used a data set with 550 randomly generated pieces. The pieces are generated with 4 different diameters and 80 percent of the pieces have the smallest diameter. Most pieces are non-convex. Since the coordinates of the vertices are arbitrary floating point numbers, many intervals of the semi-discretized pieces are extended using the extension algorithm, described in \change{section\,\ref{section2}}. Table\,\ref{info} represents the information on the data sets. For the data sets with free rotation, we consider four different cases: no rotation, two rotations with $\Delta\theta=180\degree$, four rotations with $\Delta\theta=90\degree$ and eight rotations with $\Delta\theta=45\degree$. The base resolution \textit{R\textsubscript{b}} is computed for each rotation angle because rotation can change \textit{P\textsubscript{e}} (the smallest \textit{x}-projection of all non-vertical edges of all pieces). We have used several \textit{R}-values, typically up to \textit{R\textsubscript{b}}/10. For data sets with integer vertex coordinates, the \textit{R}-values are selected such that no extension occurs if the rotation angle is a multiple of $90^{\circ}$. For placement the pieces are sorted according to decreasing axis-aligned bounding box area, this allows to place small pieces in the gaps between already placed large pieces.

\begin{table}[]
    \centering
    \caption{Information on the data sets used for computational experiments shown in Tables.\,\ref{two},\ref{free},\ref{res}.}
    \begin{tabular}{|c|c|c|c|c|c|c|}
    \hline
Data set     & Number of & Number of  & Data type &  Interval of           & Allowed  & Width\\
             & pieces    & different  &    of     & \textit{x}-projection  & rotation &   of    \\ 
             &           & pieces     &coordinates&    pieces              &          &the strip\\\hline
Shirts       &   99      &      8     &    int    &      [3,12]            &   0,180  &  40   \\ \hline
Trousers     &   64      &     17     &    int    &      [6,59]            &   0,180  &  79   \\ \hline
\change{Swim}         &   \change{48}      &     \change{10}     &   \change{float}   &      \change{[359,1940]}        &   \change{0,180} &  \change{5752}\\\hline
Han          &   25      &     20     &    int    &      [3,19]            &   free   &  58  \\ \hline
Jakob2       &   25      &     25     &    int    &      [6,12]            &   free   &  70    \\ \hline
poly5b       &   75      &     75     &    int    &      [3,12]            &   free   &  40    \\ \hline
550-random   &  550      &    550     &   float   &      [1,27.2]          &   free   &  80    \\ \hline
   
    \end{tabular}
    \label{info}
\end{table}


\subsection{Results for the benchmark data sets}

Tables \ref{two} and \ref{free} show the results obtained for the five data sets: for each data set, length and time denote respectively the length of the strip and the average execution time in 100 runs. We present the time to compute the discretization and the placement time separately. 

One expects that decreasing \textit{R} leads to a smaller strip length. The results in Tables\,\ref{two} and \ref{free} show that this is indeed most often the case, but not always, as explained in \change{section \ref{section4.2}}. On the other hand, decreasing \textit{R} causes a higher computational cost. First, the computation of the discretization of the pieces for all rotation angles consists of a part for which the number of operations is inversely proportional to \textit{R} and parts, for which the cost does not depend on \textit{R} but on the number of vertices (sweep-line algorithm, extension algorithm). The timings in Tables \ref{two} and \ref{free} shows that the execution time of the discretization
is less than linear in $R^{-1}$. We assume that this is also partly due to a better cache hit ratio when \textit{R} decreases and more tuples are computed and stored in the vector data structure holding the piece (see \change{section \ref{section2}}). Second, the number of tests performed in the placement algorithm can increase more than linearly in $R^{-1}$, since the number of intervals to be placed increases linearly with $R^{-1}$, but the number of tested positions in the strip also increases with decreasing \textit{R}. Table\,\ref{two} (column \change{`check tuples'}) indicates that the actual number of tests increases slightly more than linear in $R^{-1}$. However, the timing in Tables\,\ref{two} and \ref{free} show that the actual execution times increase less than linear in $R^{-1}$ due to caching effects. 
The placement time for \textit{n} rotation angles ($n = 1, 2, 4, 8$) increases approximately linear in \textit{n}.

\begin{table}[]
\centering
\caption{\change{Performance results for the data sets with two rotation angles ($0^{\circ}$, $180^{\circ}$): length: length of the strip; time: time to compute semi-discretization for all rotation angles + time of the placement; check tuples: the number of checks for placing all pieces in the strip.}}
\change{\begin{tabular}{|l|c|c|c|c|c|c|c|}
\hline
\multicolumn{1}{|c|}{Data Set } &\multicolumn{1}{|c|}{Resolution} & \multicolumn{3}{c|}{Without Rotation} & \multicolumn{2}{c|}{With Rotation}\\
\cline{6-7}
\multicolumn{1}{|c|}{}& \multicolumn{1}{|c|}{}& \multicolumn{3}{c|}{} & \multicolumn{2}{c|}{$\Delta\theta=180\degree$} \\
\cline{3-7}
\multicolumn{1}{|c|}{}& \multicolumn{1}{|c|}{$R$}& \multicolumn{3}{c|}{$R_b = 1$}  & \multicolumn{2}{c|}{$R_b = 1$}\\\cline{3-7}
\multicolumn{1}{|c|}{} &\multicolumn{1}{|c|}{} & \multicolumn{1}{c|}{length} &\multicolumn{1}{c|}{time(ms)} &\multicolumn{1}{c|}{check tuples} & \multicolumn{1}{c|}{length} &\multicolumn{1}{c|}{time(ms)}\\
\hline
\hspace{14pt}   Shirts   &  1     & 70.0  & 0.1+0.3 &  13551   &  66.0     &  0.2+0.6 \\\cline{2-7}
\hspace{14pt} (99 pieces)&  0.5   & 69.5  & 0.2+0.5 &  25763   &  67.5     &  0.3+0.9  \\\cline{2-7}
\hspace{14pt}            &  0.2   & 68.4  & 0.3+0.9 &  72849   &  67.0     &  0.5+1.6 \\\cline{2-7}
\hspace{14pt}            &  0.1   & 68.3  & 0.5+1.9 & 156159   &  66.5     &  0.7+2.8 \\\hline
\multicolumn{1}{|c|}{}& \multicolumn{1}{|c|}{}& \multicolumn{3}{c|}{$R_b = 1 $}  & \multicolumn{2}{c|}{$R_b = 1$}\\\cline{3-7}
\multicolumn{1}{|c|}{} &\multicolumn{1}{|c|}{} & \multicolumn{1}{c|}{length} &\multicolumn{1}{c|}{time(ms)} &\multicolumn{1}{c|}{check tuples} & \multicolumn{1}{c|}{length} &\multicolumn{1}{c|}{time(ms)}\\\cline{3-7}
\hspace{14pt} Trousers  & 1      & 284.0 &  0.2+0.4  &  23127   & 284.0   &  0.3+0.8  \\\cline{2-7}
\hspace{14pt}(64 pieces)& 0.5    & 284.0 &  0.3+0.7  &  47065   & 284.0   &  0.5+1.4  \\\cline{2-7}
\hspace{14pt}           & 0.2    & 283.6 &  0.6+1.7  & 120623   & 283.6   &  1.1+2.8   \\\cline{2-7}
\hspace{14pt}           & 0.1    & 283.6 &  1.0+3.7  & 247627   & 283.6   &  2.0+6.0  \\\hline
\multicolumn{1}{|c|}{}& \multicolumn{1}{|c|}{}& \multicolumn{3}{c|}{$R_b = 36 $}  & \multicolumn{2}{c|}{$R_b = 36$}\\\cline{3-7}
\multicolumn{1}{|c|}{} &\multicolumn{1}{|c|}{} & \multicolumn{1}{c|}{length} &\multicolumn{1}{c|}{time(ms)} &\multicolumn{1}{c|}{check tuples} & \multicolumn{1}{c|}{length} &\multicolumn{1}{c|}{time(ms)}\\\cline{3-7}
\hspace{14pt} Swim      & 36.0    & 7687.4 &  0.3+0.9  &  78738   & 7255.4   &  0.5 + 1.7  \\\cline{2-7}
\hspace{14pt}(48 pieces)& 18.0    & 7597.4 &  0.5+1.3  & 119169   & 7417.4   &  0.7 + 2.4  \\\cline{2-7}
\hspace{14pt}           &  7.2    & 7651.7 &  0.9+3.2  & 294175   & 7478.6   &  1.1 + 5.2   \\\cline{2-7}
\hspace{14pt}           &  3.6    & 7565.0 &  1.4+6.1  & 610427   & 7496.6   &  1.8 +10.0  \\\hline
\end{tabular}}
\label{two}
\end{table}

\begin{table}[]
\centering
\caption{\change{Performance results for the data sets with free rotation angles: length: length of the strip; time: time to compute semi-discretization for all rotation angles + time of the placement.}}
\change{\begin{tabular}{|l|c|c|c|c|c|c|c|c|c|}
\hline
\multicolumn{1}{|c|}{Data set} & \multicolumn{1}{|c|}{Resolution } & \multicolumn{2}{c|}{Without Rotation} & \multicolumn{6}{c|}{With Rotation}\\
\cline{5-10}
\multicolumn{1}{|c|}{}& \multicolumn{1}{|c|}{$R$}& \multicolumn{2}{c|}{}  & \multicolumn{2}{c|}{$\Delta\theta=180\degree$}& \multicolumn{2}{c|}{$\Delta\theta=90\degree$}& \multicolumn{2}{c|}{$\Delta \theta=45\degree$}\\
\cline{3-10}
\multicolumn{1}{|c|}{}& \multicolumn{1}{|c|}{}& \multicolumn{2}{c|}{$R_b = 1$}  & \multicolumn{2}{c|}{$R_b = 1$}& \multicolumn{2}{c|}{$R_b = 1$}& \multicolumn{2}{c|}{$R_b = 0.5$}\\\cline{3-10}
\multicolumn{1}{|c|}{} & \multicolumn{1}{|c|}{} & \multicolumn{1}{c|}{length} &\multicolumn{1}{c|}{time(ms)}& \multicolumn{1}{c|}{length} &\multicolumn{1}{c|}{time(ms)}& \multicolumn{1}{c|}{length} &\multicolumn{1}{c|}{time(ms)} & \multicolumn{1}{c|}{length} &\multicolumn{1}{c|}{time(ms)}\\
\hline
\hspace{14pt}  Han      & 1      & 51.0& 0.1+0.1 &  49.0& 0.2+0.2 & 45.0 & 0.4+0.3 & \diagbox[width=13.5mm, height=3.4mm]{}{}  & \diagbox[width=20.5mm, height=3.4mm]{}{}\\\cline{2-10}
\hspace{14pt}(25 pieces)& 0.5    & 52.0& 0.2+0.2 &  48.5& 0.3+0.3 & 45.0 & 0.6+0.6 & 46.5               & 1.4+1.1\\\cline{2-10}
\hspace{14pt}           & 0.2    & 52.0& 0.3+0.4 &  48.2& 0.6+0.7 & 45.2 & 1.2+1.3 & 47.0               & 3.0+2.6\\\cline{2-10}
\hspace{14pt}           & 0.1    & 52.0& 0.6+0.7 &  49.9& 1.1+1.1 & 45.2 & 2.3+2.0 & 44.5               & 5.4+4.2\\\hline
\multicolumn{1}{|c|}{}& \multicolumn{1}{|c|}{}& \multicolumn{2}{c|}{$R_b = 2$}  & \multicolumn{2}{c|}{$R_b = 2$}& \multicolumn{2}{c|}{$R_b = 2$}& \multicolumn{2}{c|}{$R_b = 0.5$}\\\cline{3-10}
\multicolumn{1}{|c|}{} & \multicolumn{1}{|c|}{} & \multicolumn{1}{c|}{length} &\multicolumn{1}{c|}{time(ms)}& \multicolumn{1}{c|}{length} &\multicolumn{1}{c|}{time(ms)}& \multicolumn{1}{c|}{length} &\multicolumn{1}{c|}{time(ms)} & \multicolumn{1}{c|}{length} &\multicolumn{1}{c|}{time(ms)}\\\cline{2-10}
\hspace{14pt}Jakob2     & 2      & 30.0& 0.1+0.1 &  30.0  & 0.1+0.1 & 30.0 & 0.2+0.2 & \diagbox[width=13.5mm, height=3.4mm]{}{}  & \diagbox[width=20.5mm, height=3.4mm]{}{}\\\cline{2-10}
\hspace{14pt}(25 pieces)& 1      & 28.0& 0.1+0.1 &  28.0  & 0.2+0.2 & 31.0 & 0.3+0.3 & \diagbox[width=13.5mm, height=3.4mm]{}{}  & \diagbox[width=20.5mm, height=3.4mm]{}{}\\\cline{2-10}
\hspace{14pt}           & 0.5    & 30.5& 0.2+0.3 &  29.0  & 0.3+0.4 & 28.0 & 0.4+0.6 & 28.0             & 1.1+1.0\\\cline{2-10}
\hspace{14pt}           & 0.2    & 30.2& 0.4+0.5 &  29.2  & 0.7+0.7 & 29.2 & 1.1+1.2 & 28.4             & 2.6+2.1\\\cline{2-10}
\hspace{14pt}           & 0.1    & 30.1& 0.7+0.7 &  29.1  & 1.3+1.1 & 29.1 & 2.3+2.2 & 29.1             & 4.6+4.2\\\hline
\multicolumn{1}{|c|}{}& \multicolumn{1}{|c|}{}& \multicolumn{2}{c|}{$R_b = 1$}  & \multicolumn{2}{c|}{$R_b = 1$}& \multicolumn{2}{c|}{$R_b = 1$}& \multicolumn{2}{c|}{$R_b = 0.5$}\\\cline{3-10}
\multicolumn{1}{|c|}{} & \multicolumn{1}{|c|}{} & \multicolumn{1}{c|}{length} &\multicolumn{1}{c|}{time(ms)}& \multicolumn{1}{c|}{length} &\multicolumn{1}{c|}{time(ms)}& \multicolumn{1}{c|}{length} &\multicolumn{1}{c|}{time(ms)} & \multicolumn{1}{c|}{length} &\multicolumn{1}{c|}{time(ms)}\\\cline{2-10}
\hspace{14pt}poly5b     & 1      & 73.0& 0.3+0.6  &  70.0& 0.5+0.9  & 68.0 &1.0+ 2.7  & \diagbox[width=13.5mm, height=3.4mm]{}{}  & \diagbox[width=20.5mm, height=3.4mm]{}{}\\\cline{2-10}
\hspace{14pt}(75 pieces)& 0.5    & 69.5& 0.6+1.0  &  67.0& 0.9+1.7  & 66.5 &1.5+ 3.1  & 67.0              &  3.1+ 7.2\\\cline{2-10}
\hspace{14pt}           & 0.2    & 70.6& 0.8+2.1  &  69.4& 1.5+4.0  & 65.8 &2.8+ 7.4  & 66.0              &  6.3+15.5\\\cline{2-10}
\hspace{14pt}           & 0.1    & 72.4& 1.5+4.1  &  68.5& 2.9+7.4  & 66.0 &5.3+13.4  & 65.9              & 11.3+28.5\\\hline
\end{tabular}}
\label{free}
\end{table} 

\subsection{Effect of Resolution}\label{section4.2}
\justifying
For a detailed study of the effect of \textit{R} on the solution quality and the run-time of the algorithm, we used the `random' data set, consisting of 550 pieces, see Table\,\ref{info}, without allowing rotation of the pieces. Note that with decreasing \textit{R} the size of the extensions decrease but also the number of places where a piece can be placed increases. The results presented in Table\,\ref{res} show that the solution quality, measured by the length of the strip and the wasted space fraction \textit{W\textsubscript{f}} defined as $ W_f = 100 \times \dfrac{Area(strip) - \sum_{i=0}^{N-1} Area(piece(i))}{Area(strip)}$, is low when \textit{R} is larger than \textit{R\textsubscript{b}}, while the quality improves slowly with $R^{-1}$ for $R < R_b$. This shows that \textit{R\textsubscript{b}} is a good starting point for selecting an appropriate \textit{R}, and has diminishing returns beyond this point. The run-time of the algorithm increases less than linear in $R^{-1}$; \change{when \textit{R} decreases from 1 to 0.1 the run-time increases with a factor 3.2; when \textit{R} decreases from 0.1 to 0.01 the run-time increases with a factor 6.4.} 

\begin{figure}[ht]
\begin{subfigure}{.4\textwidth}
  \centering
  \includegraphics[width=.8\linewidth]{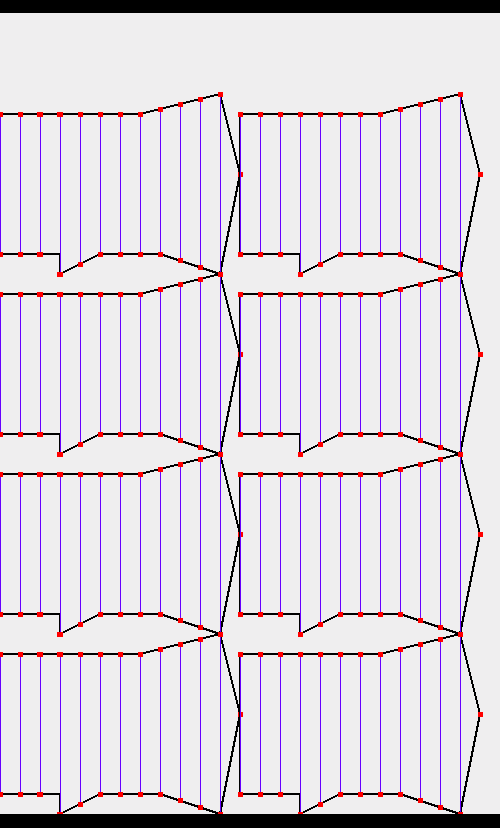}  
  \caption{R = 1, Used length of strip = 24}
  \label{fig:sub-first}
\end{subfigure}%
\hfil
\begin{subfigure}{.447\textwidth}
  \centering
  \includegraphics[width=.8\linewidth]{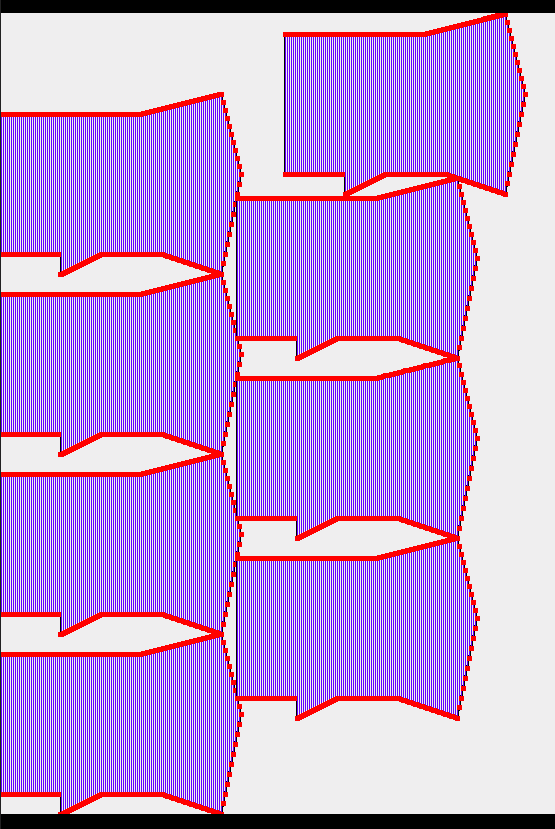}  
  \caption{Resolution = 0.1, Used length of strip = 26.2}
  \label{fig:sub-second}
\end{subfigure}
\caption{Placement of eight copies of a piece. Using a smaller \textit{R} does not always lead to a better solution.}
\label{2res}
\end{figure}

When using a simple bottom-left-fill algorithm, giving priority to `left' rather than `bottom', decreasing \textit{R} does not always lead to a smaller strip length, as shown in Fig.\,\ref{2res}, where 8 copies of a piece are placed, after discretization with a large and a small \textit{R}.

\begin{table}[]
    \centering
    \caption{\change{Placement of `random' data set. The effect of \textit{R} on the execution time of the algorithm as well as the quality of the solution: length: length of the strip; wasted space: the percentage of the wasted space in the strip; time: time to compute semi-discretization for all pieces + time of the placement.}}
    \change{\begin{tabular}{|c|c|c|c|c|}
    \hline
                    Resolution    & length     & wasted space    &    time    \\
                                  &            &   $W_f$(\%)     &    (ms)   \\\hline 
                              1   &  127.01    &    23.3         &     30    \\ \hline
                              0.5 &  119.83    &    18.8         &     42    \\ \hline
                      $R_b$ = 0.2 &  114.83    &    15.2         &     66    \\ \hline
                              0.1 &  112.63    &    13.6         &     98    \\ \hline
                              0.05&  111.93    &    13.0         &    166    \\ \hline
                              0.02&  111.23    &    12.5         &    350    \\ \hline
                              0.01&  110.74    &    12.1         &    631    \\\hline
    \end{tabular}}
    \label{res}
\end{table}

\subsection{Extension}
\justifying
The extension of the intervals of the pieces, required to avoid the overlap of the original pieces, computed by the extension algorithm (see \change{section \ref{section3.2}}) implicitly defines extension areas outside the original piece, that cannot be occupied by other (extended) pieces.
To study the effect of \textit{R} on these extension areas  we use the `poly5b' data set. We now choose \textit{R} such that many convex vertices (with integer coordinates) lie between resolution lines, causing extensions in the semi-discretized representation of the pieces. The extension area decreases with decreasing \textit{R} as indicated in \change{Section\,\ref{section2}} and illustrated in Table\,\ref{resss}. However, the extension area is small compared to the area of the gaps between the pieces, called `wasted area' in Table\,\ref{resss}, where we also show the `strip area', i.e.\,the length of the strip multiplied with the width of the strip.

\begin{table}[]
    \centering
    \caption{The effect of \textit{R} on extension for the `poly5b' data set (without rotation). Extension area vs.\ wasted area and strip area.}
    \change{\begin{tabular}{|c|c|c|c|}
    \hline
  Resolution       &  Strip area   &   Wasted area&  Extension area    \\\hline
        1.2        &   3176        &     1340     &     173                    \\ \hline
        0.6        &   2944        &     1108     &      33                     \\ \hline
        0.2        &   2824        &      988     &      10                    \\ \hline
        0.1        &   2896        &     1060     &       4                   \\ \hline
    \end{tabular}}
    \label{resss}
\end{table}

\subsection{Scalability of the algorithm}
In order to test the scalability of our algorithm, in particular when the data set contains many identical pieces, we used data sets consisting of respectively 100, 200 and 400 identical pieces, the same piece as used in Fig.\,\ref{2res} with \change{$R=1$}. In a first test we used the algorithm with the optimization mentioned in \change{section\,\ref{section3}}, namely when two identical pieces are placed consecutively, the placement of the second piece starts with the translation vector of the placement of the first piece. Doubling the size of the data set, the run-time increases only slightly. 
In a second test we removed the above mentioned optimization, such that the placement of each piece always starts at the bottom-left corner of the strip. The resulting run-time increases superlinear, but less than quadratic, in the number of pieces; \change{when the number of pieces increases from 100 to 200, the run-time increases with a factor 2.3 and when the number of pieces increases from 200 to 400, the run-time increases with a factor 2.5}.

\begin{table}[]
\centering
\caption{Scalability of the algorithm using data sets with 100, 200 and 400 copies of a piece. Execution times for placement with and without the optimization in case of identical pieces, described in \change{section \ref{section3}.}}
\change{\begin{tabular}{|l|c|c|c|c|}
\hline
\multicolumn{2}{|c|}{Placement of each piece with}& \multicolumn{2}{|c|}{Placement of each piece from}\\
\multicolumn{2}{|c|}{translation vector of previous piece}& \multicolumn{2}{|c|}{the bottom-left corner of the strip}\\\hline
\multicolumn{1}{|c|}{Data set}& \multicolumn{1}{|c|}{Time (ms)}& \multicolumn{1}{|c|}{Data set} & \multicolumn{1}{|c|}{Time (ms)} \\\hline
\hspace{14pt} 100 copies &  0.5   &   P100 &  1.4 \\\hline
\hspace{14pt} 200 copies &  0.7   &   P200 &  3.2  \\\hline
\hspace{14pt} 400 copies &  0.9   &   P400 &  8.1 \\\hline
\end{tabular}}
\label{test}
\end{table}

\subsection{Comparison with other approaches}

Only few papers use a semi-discrete representation, e.g.\,\cite{6,5,12}. Comparing our results in detail with theirs is not possible for the following reasons. The ESICUP data sets are not used in \cite{6} and \cite{5} and no detailed information on the data sets used is provided. Although \cite{12} used a semi-discrete representation, a MIP model was used for placement, which is completely different from our placement algorithm. However, even without 
computational experiments, some comparisons can be made. \cite{6} represents the pieces and the strip by axis-aligned rectangles which leads to unnecessary extensions of the pieces, compared to our extension algorithm, and the quality of the solution obtained by their bottom-left-fill algorithm (without using a metaheuristic) will be worse than the quality of the solution obtained by our algorithm. In \cite{5}, the extension rules do not cover all cases which can lead to overlap of pieces. In \cite{12}, even for problems with a small number of pieces most of the run-times are equal to the imposed time limit and the optimal solution is not found despite using a MIP method.

For the method presented in this paper, the solution quality, i.e.\,the length of the strip, presented in Tables\,\ref{two}, \ref{free}, \ref{res}, \ref{resss} and \ref{test} is limited by the simple greedy bottom-left-fill placement algorithm used. Many (meta)heuristics to improve the bottom-left-fill algorithm, such as reordering of pieces, tabu search, genetic algorithm, can easily be implemented on top of our placement algorithm. The semi-discretization of the pieces plays a role in the bottom-left-fill algorithm, described in section\,\ref{section3}, but most (meta)heuristics used in the literature are independent of the representation of the pieces and thus can use our placement algorithm as an efficient `building block'. The main advantages of the method, introduced in this paper, are the low run-time and the scalability w.r.t.\ the number of pieces. 

\change{We would like to compare the performance and mainly the execution time and the scalability  of our algorithm with other `building blocks' used in heuristics to solve nesting problems, e.g. based on the use of no-fit polygons.}

\change{In \cite{14} and \cite{144} bottom-left-fill algorithms based on respectively arc geometry and no-fit polygon are used as `building block' within a hill climbing and tabu search metaheuristic. In both papers, results are presented for the data sets mentioned above and the execution times of a single run of the bottom-left-fill algorithms, obtained on a 2 GHz Intel Pentium 4 processor with 256 MB RAM, are given, but not the strip length achieved with such a single run.}
 
In \cite{14} the positions to place a piece in the strip are limited to a set of equidistant vertical lines, as in our algorithm. The placement of each piece starts from the bottom-left corner of the strip. If the piece intersects with already placed pieces, intersections of the geometric primitives are resolved by shifting the piece upward and, if necessary to the right on the next resolution line, until a valid position is found, which is equivalent with our bottom-left-fill implementation. 
\change{In \cite{144} no-fit polygon is used to identify the intersection state of two pieces. However, the time for computing the NFP of pieces which is discussed in \cite{1444} is not considered in the timings mentioned in \cite{144}.
For several ESICUP data sets, Table\,\ref{besst} shows the run-time of a single run of the bottom-left-fill placement presented in \cite{14} and \cite{144}, the best result, obtained by using hill climbing and tabu search on top of the bottom-left-fill algorithm and the corresponding run-time reported in \cite{14}, and the best result with our algorithm, cf.\ Tables\,\ref{two} and \ref{free}, and the corresponding run-time. }

\change{When comparing the quality obtained with our algorithm and with a bottom-left-fill algorithm using continuous representation, one must make distinction between the effect of the discretization of the pieces and the discretization of the strip. When semi-discretization of a piece causes extensions, no other piece can be placed in the extension regions, but we indicated in section 5.3 that the area of the extension regions typically is small, and for pieces with integer vertex coordinates, as in most of the benchmark data sets, $R$ can be chosen such that no extension occurs. Semi-discretization of the strip limits the possible placement positions, but the results in Tables 4 and 5 show that, for sufficiently small $R$, decreasing $R$, i.e.\,increasing the number of possible placement positions does not always lead to a shorter strip length, due to the greedy nature of the bottom-left-fill heuristic.}

\change{In \cite{14} the strip length achieved with a single run of the bottom-left-fill algorithm is not mentioned, but this length should be equal to the length obtained with our bottom-left-fill algorithm, when in both \cite{14} and our method the same distance between the vertical lines in the strip and the same ordering of the pieces are used, and if no extension occurs in the semi-discretization of the pieces in our algorithm. The latter is true for the results in Table 9 for the data sets Shirts, Trousers, Jakob2 and Poly5b.}

Comparison of the execution times of both bottom-left-fill algorithms is not straightforward: \textit{R} and the ordering of the pieces are not mentioned in \cite{14} and the performance ratio of the processors used is not known, since this depends on many factors (data structures used, compiler options, cache sizes, ...). The performance ratio of the processor used for our experiments and the one used in \cite{14} \change{and \cite{144}} is not larger than 25, see the detailed analysis in Appendix. Taking this performance ratio into account, the timings for a single bottom-left-fill placement indicate that our placement algorithm using semi-discretization is two orders of magnitude faster than the bottom-left-fill placement algorithm in \cite{14}. \change{Considering the timings for computing the no-fit polygon of pieces discussed in \cite{1444}, our placement algorithm using semi-discretization is at least an order of magnitude faster than the bottom-left-fill algorithm in \cite{144}. The efficiency of our algorithm is mainly due to the limited number of tests needed on average to evaluate the placement of a piece in a certain position in the strip and the use of simple operations on data structures that allow to achieve high performance on current computers (for example, ensuring a high cache hit ratio).}

\change{We have implemented the hill climbing heuristic, used in \cite{14} and \cite{144}, on top of our bottom-left-fill algorithm (with $R = 1$). After 100 iterations we obtain, for data sets "Shirts" and "Trousers" with $R = 1$ and two allowed rotation angles, strip lengths of respectively 63.0 and 248.0 in respectively 79 ms and 105 ms.  For "Shirts" data set this result is slightly better and for "Trousers" data set this is slightly worse than the results in \cite{14} and \cite{144}. This shows that the algorithm presented in this paper can easily be used as a building block in metaheuristics to improve the solution quality.
Note that in \cite{14} several parameters of the search heuristic (probabilities for the various mutation operators, initial ordering of the pieces, etc.) are not presented and it is not clear how the execution time is determined.}

\change{We also compare our algorithm with other recently proposed `building blocks'. \cite{recent} combine exact and heuristic approaches using a dotted board model. Their method has three phases: constructive phase, improvement phase and compaction phase. The constructive phase  computes a feasible solution and can be considered as a building block. The only common data sets used in both \cite{recent} and this paper are "Jakob2" and "Trousers" for which the constructive phase results in strip lengths respectively 36.0 and 439.0, requiring respectively 34.1 seconds and 229.9 seconds. Hence our algorithm is orders of magnitude faster than the `building block' in \cite{recent} and results in a shorter strip length.}
    
\change{The recent paper \cite{onepass} propose a one-pass algorithm. The no-fit polygon of the partially packed strip and a new piece is computed. The new piece is placed at the vertex of the no-fit polygon for which a `fitting function' is maximized.
The common data sets used in both \cite{onepass} and this paper are "Jakob2", "Shirts", "Swim" and "Trousers" for which the obtained strip lengths are 22.20, 55.15, 5474.89 and 232.69 in respectively 192.53 s, 37.93 s, 530.50 s and 408.35 s. Although \cite{onepass} uses a much more powerful algorithm than bottom-left-fill, we include these results to show the best results w.r.t.\,solution quality in the literature for these data sets. }

\begin{table}[]
\centering
\caption{Comparison of the best results from Tables \ref{two} and \ref{free} with the best results obtained in \cite{14}.}
\begin{tabular}{|c|c|c|c|c|c|c|c|}
\hline
\multicolumn{1}{|c|}{ } &\multicolumn{1}{|c|}{} & \multicolumn{5}{c|}{Data sets}\\
\cline{3-7}
\multicolumn{1}{|c|}{}& \multicolumn{1}{|c|}{}& \multicolumn{1}{c|}{Shirts}& \multicolumn{1}{|c|}{Trousers}& \multicolumn{1}{|c|}{\change{Swim}}&\multicolumn{1}{c|}{Jakob2} & \multicolumn{1}{c|}{poly5b}\\
\cline{1-7}
Best results from              &    length            &   \change{ 66.0}     &    283.6       &  \change{ 7255.4}     &    \change{ 28.0}      &     \change{ 65.8}       \\ \cline{2-7}
Tables \ref{two} and \ref{free}&    \bf{run-time(ms)} &  \change{\bf{0.8} }  &   \change{ \bf{2.3}}     & \change{ \bf{2.2}}    &   \change{ \bf{0.2}}   &     \change{ \bf{10.2}}      \\ \hline \hline
\cite{14}                      &    length            &    NA       &      NA        &    NA       &      NA      &    NA      \\ \cline{2-7}
(bottom-left only)             &    \bf{run-time(ms)} &  \bf{4990}  &   \bf{7890}    & \bf{12390}  &  \bf{2130}   &  \bf{14700}        \\ \hline \hline
\cite{14}                      &    length            &    63.8     &    246.6       &  6462.40    &    25.8      &     60.5      \\ \cline{2-7}
(bottom-left + metaheuristics) &    \bf{run-time(ms)} & \bf{58360}  &  \bf{756150}   & \bf{607370} & \bf{81410}   & \bf{676610}         \\ \hline \hline
 \change{\cite{1444,144}}               &    \change{length }           &    \change{NA }      &      \change{ NA }       &    \change{ NA }      &      \change{ NA}      &   \change{  NA }     \\ \cline{2-7}
\change{(bottom-left only)}             &    \change{ \bf{run-time(ms)}} & \change{ \bf{330+770}}& \change{ \bf{730+1020}}  &\change{ \bf{6080+1240}}&\change{ \bf{5070+640}}&  \change{ \bf{141900+12620}}        \\ \hline \hline
\end{tabular}
\label{besst}
\end{table}

\section{Conclusion}
\justifying
We have shown that a semi-discretization of both the pieces and the strip can lead to a fast and scalable method to solve the 2D nesting problem. First, we presented a fast algorithm to compute the semi-discretization of possibly non-convex pieces, consisting of equidistant vertical intervals, complemented with an extension procedure, ensuring that a non-overlapping placement of the intervals guarantees that the original pieces do not overlap. The proposed extensions are minimal and complete, which is an improvement over extension techniques already proposed in the literature. Note that changing the shape of the strip from rectangular irregular, as discussed in \cite{31}, is easy in our approach, since it only requires the semi-discretization of an irregular shaped strip, without extension, which would lead to only a small amount of loss in strip space. Further, we presented an efficient bottom-left-fill placement algorithm in detail. It exploits the fact that only simple arithmetic operations are needed to detect and avoid overlap and it uses an optimised ordering of the interval overlap tests. Much attention is paid to the choice of the data structures in order to achieve high performance on current processors. 
The quality of the solution and the execution time of the algorithm greatly depend on $R$, i.e.\ the distance between the vertical intervals. However, the execution time increases less than linear in $R^{-1}$. We also show how a suitable `base resolution'  can be computed from the size and geometry of the pieces. 
Although, the quality of the solutions is limited by the simple greedy algorithm used, timings of the bottom-left-fill placement for several ESICUP data sets show that our placement algorithm using a semi-discrete representation is nearly two orders of magnitude more efficient than a similar algorithm based on arc geometry, requiring the computation of intersections of the original pieces. Moreover, since the execution times scale well with increasing number of pieces, our placement algorithm is scalable. 
Therefore, due to the low run-time and the scalability of our algorithm, the placement algorithm presented in this paper can be used as an efficient `building block' to implement various (meta)heuristics already proposed in the literature. 
\section*{Acknowledgment}

\noindent
We acknowledge the financial support of Interne Fondsen KU Leuven / Internal Funds KU Leuven, project C24/17/048.


\bibliographystyle{cas-model2-names}

\bibliography{cas-refs.bib}

\begin{thebibliography}{30}
\expandafter\ifx\csname natexlab\endcsname\relax\def\natexlab#1{#1}\fi
\providecommand{\url}[1]{\texttt{#1}}
\providecommand{\href}[2]{#2}
\providecommand{\path}[1]{#1}
\providecommand{\DOIprefix}{doi:}
\providecommand{\ArXivprefix}{arXiv:}
\providecommand{\URLprefix}{URL: }
\providecommand{\Pubmedprefix}{pmid:}
\providecommand{\doi}[1]{\href{http://dx.doi.org/#1}{\path{#1}}}
\providecommand{\Pubmed}[1]{\href{pmid:#1}{\path{#1}}}
\providecommand{\bibinfo}[2]{#2}
\ifx\xfnm\relax \def\xfnm[#1]{\unskip,\space#1}\fi
\bibitem[{Akunuru and Babu(2013)}]{5}
\bibinfo{author}{Akunuru, R.}, \bibinfo{author}{Babu, N.},
  \bibinfo{year}{2013}.
\newblock \bibinfo{title}{Semi-discrete geometric representation for nesting
  problems}.
\newblock \bibinfo{journal}{International Journal of Production Research}
  \bibinfo{volume}{51}, \bibinfo{pages}{4155–4174}.
\bibitem[{Amaro~Junior et~al.(2014)Amaro~Junior, Pinheiro, Saraiva and
  Pinheiro}]{20}
\bibinfo{author}{Amaro~Junior, B.}, \bibinfo{author}{Pinheiro, P.R.},
  \bibinfo{author}{Saraiva, R.D.}, \bibinfo{author}{Pinheiro, P.G.C.D.},
  \bibinfo{year}{2014}.
\newblock \bibinfo{title}{Dealing with nonregular shapes packing.}
\newblock \bibinfo{journal}{Mathematical Problems in Engineering}
  \bibinfo{volume}{Volume 2014, Article 548957}.
\newblock \DOIprefix\doi{https://doi.org/10.1155/2014/548957}.
\bibitem[{Baldacci et~al.(2014)Baldacci, Boschetti, Ganovelli and
  Maniezzo}]{31}
\bibinfo{author}{Baldacci, R.}, \bibinfo{author}{Boschetti, M.},
  \bibinfo{author}{Ganovelli, M.}, \bibinfo{author}{Maniezzo, V.},
  \bibinfo{year}{2014}.
\newblock \bibinfo{title}{Algorithms for nesting with defects.}
\newblock \bibinfo{journal}{Discrete Applied Mathematics.}
  \bibinfo{volume}{163}, \bibinfo{pages}{17--33}.
\bibitem[{Bennell and Oliveira(2008)}]{1}
\bibinfo{author}{Bennell, J.}, \bibinfo{author}{Oliveira, J.},
  \bibinfo{year}{2008}.
\newblock \bibinfo{title}{The geometry of nesting problem: a tutorial}.
\newblock \bibinfo{journal}{European Journal of Operational Research}
  \bibinfo{volume}{184}, \bibinfo{pages}{397–415}.
\bibitem[{de~Berg et~al.(2008)de~Berg, Cheong, van Kreveld and Overmars}]{10}
\bibinfo{author}{de~Berg, M.}, \bibinfo{author}{Cheong, O.},
  \bibinfo{author}{van Kreveld, M.}, \bibinfo{author}{Overmars, M.},
  \bibinfo{year}{2008}.
\newblock \bibinfo{title}{Computational Geometry: Algorithms and Applications}.
\newblock \bibinfo{publisher}{Springer}.
\newblock \bibinfo{note}{Chapter 2}.
\bibitem[{Burke et~al.(2006)Burke, Hellier, Kendall and Whitwell}]{14}
\bibinfo{author}{Burke, E.}, \bibinfo{author}{Hellier, R.},
  \bibinfo{author}{Kendall, G.}, \bibinfo{author}{Whitwell, G.},
  \bibinfo{year}{2006}.
\newblock \bibinfo{title}{A new bottom-left-fill heuristic algorithm for the
  two-dimensional irregular packing problem}.
\newblock \bibinfo{journal}{Operations Research} \bibinfo{volume}{54},
  \bibinfo{pages}{587–601}.
\bibitem[{Burke et~al.(2007)Burke, Hellier, Kendall and Whitwell}]{1444}
\bibinfo{author}{Burke, E.}, \bibinfo{author}{Hellier, R.},
  \bibinfo{author}{Kendall, G.}, \bibinfo{author}{Whitwell, G.},
  \bibinfo{year}{2007}.
\newblock \bibinfo{title}{Complete and robust no-fit polygon generation for the
  irregular stock cutting problem}.
\newblock \bibinfo{journal}{Operations Research} \bibinfo{volume}{179},
  \bibinfo{pages}{27–49}.
\bibitem[{Burke et~al.(2010)Burke, Hellier, Kendall and Whitwell}]{144}
\bibinfo{author}{Burke, E.}, \bibinfo{author}{Hellier, R.},
  \bibinfo{author}{Kendall, G.}, \bibinfo{author}{Whitwell, G.},
  \bibinfo{year}{2010}.
\newblock \bibinfo{title}{Irregular packing using the line and arc no-fit
  polygon}.
\newblock \bibinfo{journal}{Operations Research} \bibinfo{volume}{58},
  \bibinfo{pages}{948--970}.
\bibitem[{Cherri et~al.(2016)Cherri, Carravilla and Toledo}]{recent}
\bibinfo{author}{Cherri, L.}, \bibinfo{author}{Carravilla, M.},
  \bibinfo{author}{Toledo, F.}, \bibinfo{year}{2016}.
\newblock \bibinfo{title}{A model-based heuristic for the irregular strip
  packing problem.}
\newblock \bibinfo{journal}{Pesquisa Operacional.} \bibinfo{volume}{36},
  \bibinfo{pages}{447--468}.
\bibitem[{Cplusplus.com(2020)}]{vector}
\bibinfo{author}{Cplusplus.com}, \bibinfo{year}{2020}.
\newblock \bibinfo{title}{class template std::vector}.
\newblock
  \bibinfo{howpublished}{\url{http://www.cplusplus.com/reference/vector/vector/}}.
\newblock \bibinfo{note}{Last accessed 22 November 2020}.
\bibitem[{Dowsland et~al.(1998)Dowsland, Dowsland and Bennell}]{18}
\bibinfo{author}{Dowsland, K.}, \bibinfo{author}{Dowsland, W.},
  \bibinfo{author}{Bennell, J.}, \bibinfo{year}{1998}.
\newblock \bibinfo{title}{Jostling for position: Local improvement for
  irregular cutting patterns.}
\newblock \bibinfo{journal}{The Journal of the Operational Research Society}
  \bibinfo{volume}{49}, \bibinfo{pages}{647–658}.
\bibitem[{Gennady et~al.(2016)Gennady, Shaojuan and Abhinav}]{30}
\bibinfo{author}{Gennady, F.}, \bibinfo{author}{Shaojuan, Z.},
  \bibinfo{author}{Abhinav, S.}, \bibinfo{year}{2016}.
\newblock \bibinfo{title}{Intel {M}ath {K}ernel {L}ibrary ({I}ntel {MKL})
  benchmarks suite}.
\newblock
  \bibinfo{howpublished}{\url{https://software.intel.com/content/www/us/en/develop/articles/intel-mkl-benchmarks-suite.html}}.
\newblock \bibinfo{note}{Last accessed 22 November 2020}.
\bibitem[{Gomes(2013)}]{27}
\bibinfo{author}{Gomes, A.M.}, \bibinfo{year}{2013}.
\newblock \bibinfo{title}{Irregular packing problems: Industrial applications
  and new directions using computational geometry.}
\newblock \bibinfo{journal}{IFAC-Proceedings.} \bibinfo{volume}{46},
  \bibinfo{pages}{378--383}.
\bibitem[{Gomes and Oliveira(2002)}]{23}
\bibinfo{author}{Gomes, A.M.}, \bibinfo{author}{Oliveira, J.},
  \bibinfo{year}{2002}.
\newblock \bibinfo{title}{A 2-exchange heuristic for nesting problems.}
\newblock \bibinfo{journal}{European Journal of Operational Research}
  \bibinfo{volume}{141}, \bibinfo{pages}{359--370}.
\bibitem[{Hager and Wellein(2011)}]{Hager2011}
\bibinfo{author}{Hager, G.}, \bibinfo{author}{Wellein, G.},
  \bibinfo{year}{2011}.
\newblock \bibinfo{title}{Introduction to High Performance Computing for
  Scientists and Engineers}.
\newblock \bibinfo{publisher}{Chapman and Hall/CRC}.
\newblock \bibinfo{note}{Chapter 1-3}.
\bibitem[{Han and Na(1996)}]{15}
\bibinfo{author}{Han, G.C.}, \bibinfo{author}{Na, S.J.}, \bibinfo{year}{1996}.
\newblock \bibinfo{title}{Two-stage approach for nesting in two-dimensional
  cutting problems using neural network and simulated annealing.}
\newblock \bibinfo{journal}{Journal of Engineering Manufacture}
  \bibinfo{volume}{210}, \bibinfo{pages}{509–519}.
\bibitem[{Harper(2006)}]{28}
\bibinfo{author}{Harper, D.}, \bibinfo{year}{2006}.
\newblock \bibinfo{title}{Pentium floating-point speed test}.
\newblock
  \bibinfo{howpublished}{\url{https://www.obliquity.com/computer/speedtest.html}}.
\newblock \bibinfo{note}{Last accessed 22 November 2020}.
\bibitem[{Hopper(2000)}]{17}
\bibinfo{author}{Hopper, E.}, \bibinfo{year}{2000}.
\newblock \bibinfo{title}{Two-dimensional packing utilising evolutionary
  algorithms and other meta-heuristic methods.}
\newblock Ph.D. thesis. University of Wales, Cardiff School of Engineering.
\bibitem[{Jakobs(1996)}]{16}
\bibinfo{author}{Jakobs, S.}, \bibinfo{year}{1996}.
\newblock \bibinfo{title}{Genetic algorithms for the packing of polygons.}
\newblock \bibinfo{journal}{European Journal of Operational Research}
  \bibinfo{volume}{88}, \bibinfo{pages}{165 – 181}.
\bibitem[{Kierkosz and Luczak(2019)}]{onepass}
\bibinfo{author}{Kierkosz, I.}, \bibinfo{author}{Luczak, M.},
  \bibinfo{year}{2019}.
\newblock \bibinfo{title}{A one-pass nesting problems.}
\newblock \bibinfo{journal}{Operations Research And Decisions.}
  \bibinfo{volume}{29}, \bibinfo{pages}{37–60}.
\bibitem[{Leao et~al.(2015)Leao, Toledo, Oliveira and Carravilla}]{12}
\bibinfo{author}{Leao, A.}, \bibinfo{author}{Toledo, F.},
  \bibinfo{author}{Oliveira, J.}, \bibinfo{author}{Carravilla, M.},
  \bibinfo{year}{2015}.
\newblock \bibinfo{title}{A semi-continuous {MIP} model for the irregular strip
  packing problem}.
\newblock \bibinfo{journal}{International Journal of Production Research}
  \bibinfo{volume}{54}, \bibinfo{pages}{712–721}.
\bibitem[{Leao et~al.(2020)Leao, Toledo, Oliveira, Carravilla and
  Alvarez-Valdés}]{2}
\bibinfo{author}{Leao, A.}, \bibinfo{author}{Toledo, F.},
  \bibinfo{author}{Oliveira, J.}, \bibinfo{author}{Carravilla, M.},
  \bibinfo{author}{Alvarez-Valdés, R.}, \bibinfo{year}{2020}.
\newblock \bibinfo{title}{Irregular packing problems: A review of mathematical
  models}.
\newblock \bibinfo{journal}{European Journal of Operational Research}
  \bibinfo{volume}{282}, \bibinfo{pages}{803--822}.
\bibitem[{Ma and Liu(2007)}]{6}
\bibinfo{author}{Ma, H.}, \bibinfo{author}{Liu, C.}, \bibinfo{year}{2007}.
\newblock \bibinfo{title}{Fast nesting of {2-D} sheet parts with arbitrary
  shapes using a greedy method and semi-discrete representations}.
\newblock \bibinfo{journal}{IEEE Transactions on Automation Science And
  Engineering} \bibinfo{volume}{4}, \bibinfo{pages}{273--282}.
\bibitem[{Oliveira et~al.(2000)Oliveira, Gomes and Ferreira}]{19}
\bibinfo{author}{Oliveira, J.F.}, \bibinfo{author}{Gomes, A.M.},
  \bibinfo{author}{Ferreira, J.S.}, \bibinfo{year}{2000}.
\newblock \bibinfo{title}{A new constructive algorithm for nesting prolems.}
\newblock \bibinfo{journal}{OR Spektrum} \bibinfo{volume}{22},
  \bibinfo{pages}{263--284}.
\bibitem[{Pinheiro et~al.(2016)Pinheiro, Amaro~Junior and Saraiva}]{21}
\bibinfo{author}{Pinheiro, P.R.}, \bibinfo{author}{Amaro~Junior, B.},
  \bibinfo{author}{Saraiva, R.D.}, \bibinfo{year}{2016}.
\newblock \bibinfo{title}{A random-key genetic algorithm for solving the
  nesting problem.}
\newblock \bibinfo{journal}{International Journal of Computer Integrated
  Manufacturing} \bibinfo{volume}{29}, \bibinfo{pages}{1159–1165}.
\bibitem[{Sato et~al.(2019)Sato, Martinsa, Gomes and Tsuzuki}]{raster}
\bibinfo{author}{Sato, A.}, \bibinfo{author}{Martinsa, T.},
  \bibinfo{author}{Gomes, A.}, \bibinfo{author}{Tsuzuki, M.},
  \bibinfo{year}{2019}.
\newblock \bibinfo{title}{Raster penetration map applied to the irregular
  packing problem.}
\newblock \bibinfo{journal}{European Journal of Operational Research.}
  \bibinfo{volume}{279}, \bibinfo{pages}{657--671}.
\bibitem[{Sato et~al.(2016)Sato, Tsuzuki, Martins and Gomes}]{26}
\bibinfo{author}{Sato, A.K.}, \bibinfo{author}{Tsuzuki, M.S.G.},
  \bibinfo{author}{Martins, T.C.}, \bibinfo{author}{Gomes, A.M.},
  \bibinfo{year}{2016}.
\newblock \bibinfo{title}{Study of the grid size impact on a raster based strip
  packing problem solution.}
\newblock \bibinfo{journal}{IFAC-PapersOnLine.} \bibinfo{volume}{49},
  \bibinfo{pages}{143--148}.
\bibitem[{Toledo et~al.(2013)Toledo, Carravilla, Ribeiro, Oliveira and
  Gomes}]{dotted-board}
\bibinfo{author}{Toledo, F.}, \bibinfo{author}{Carravilla, M.},
  \bibinfo{author}{Ribeiro, C.}, \bibinfo{author}{Oliveira, J.},
  \bibinfo{author}{Gomes, A.M.}, \bibinfo{year}{2013}.
\newblock \bibinfo{title}{The dotted-board model: A new {MIP} model for nesting
  irregular shapes.}
\newblock \bibinfo{journal}{International Journal of Production Economics.}
  \bibinfo{volume}{145}, \bibinfo{pages}{478--487}.
\bibitem[{Wascher et~al.(2007)Wascher, Haußner and Schumann}]{wascher}
\bibinfo{author}{Wascher, G.}, \bibinfo{author}{Haußner, H.},
  \bibinfo{author}{Schumann, H.}, \bibinfo{year}{2007}.
\newblock \bibinfo{title}{An improved typology of cutting and packing
  problems}.
\newblock \bibinfo{journal}{European Journal of Operational Research}
  \bibinfo{volume}{183}, \bibinfo{pages}{1109–1130}.
\bibitem[{Wauters et~al.(2016)Wauters, Uyttersprot and
  Esprit}]{WautersTony2016Jara}
\bibinfo{author}{Wauters, T.}, \bibinfo{author}{Uyttersprot, S.},
  \bibinfo{author}{Esprit, E.}, \bibinfo{year}{2016}.
\newblock \bibinfo{title}{{JNFP}: a robust and open-source {J}ava based nofit
  polygon generator library}.
\newblock \bibinfo{note}{13th {ESICUP} meeting
  \url{https://paginas.fe.up.pt/~esicup/extern/esicup-13thMeeting/uploads/Conference/13th_ESICUP_Meeting_booklet_2016.pdf}}.

\end{thebibliography}
\appendix
\section{Appendix}\label{appendix} 

 In order to compare the performance of one core of the Intel Core i7-7500U processor, used for our experiments, with the Pentium 4 processor, used in \cite {14}, we rely on actual performance benchmark results for the former and published results for the latter processor. 
 A complete specification of the Intel Core i7-7500U processor can be found at  https://ark.intel.com/content/www/us/en/ark/compare.html?productIds=95451: 2 cores, 4 threads, 2.7 GHz base frequency, 3.5 GHz max.\ turbo frequency, 4 MB Intel Smart Cache, 16 Gbyte memory, 34.1 GB/s max.\ memory bandwidth.
We do not have the complete specification (cache size, memory bandwidth, ...) of the Pentium 4 (2 GHz) processor used in \cite {14}).
 
 The number of floating-point operations achieved on the Linpack benchmark can be considered as the peak performance that a processor can achieve on an actual application code. 
 In \cite{30} performance of a Pentium 4 (2 GHz) processor on the Linpack benchmark is presented for matrix sizes up to 1804 KB. Assuming that the matrix is stored in double precison (64 bit words) the latter corresponds to a matrix of dimension 475.
 The highest floating point rate (0.66 Gflops = $0.66 \cdot 10^9$ floating points operations per second) is achieved for matrix dimension  $\approx 155$. For matrices of dimension 100 the floating point rate is $\approx 0.55$ Gflops, while for matrices larger than 280 the floating point rate is $\approx 0.25$ Gflops.
 
 On the Intel Core i7-7500U processor with 2 cores (4 threads) we have used the optimized Linpack benchmark code downloaded from https://software.intel.com/content/www/us/en/develop/articles/intel-mkl-benchmarks-suite.html. For matrices of dimension 100 we measured a floating point rate of $\approx 10$ GFlops. For larger matrix dimensions the floating point rate increases, e.g.\ for matrix dimension 10000 we measured a floating point rate of $\approx 88$ GFlops.
 Hence, for small matrix sizes, the floating point performance of the Intel Core i7-7500U processor with 2 cores is approximately a factor 20 higher than of a 2 GHz Pentium 4 processor.
 However, since we performed the computational experiments on one core of this processor we may assume that the floating point performance on one core is approximately a factor 10 higher than the performance of a 2 GHz Pentium 4 processor.
 
  Performance on the Linpack benchmark is not representative for the performance on the placement procedures, since in the latter case only few operations are performed per data element fetched from memory. Therefore we also compare the performance of both processors on a C code, designed to run under Linux, performing simple operations on elements of two arrays, downloaded from \cite{28}, where also the performance obtained on a 2.8 GHz Pentium 4 processor is given. 
  On the latter processor, performing 100 times the addition of two arrays of length $10^7$ requires 6.73 seconds, which corresponds to 148 Mflops. Based on this we estimate the floating point rate of a 2 GHz Pentium processor for this operation at 106 Mflops.
  
  On a single core of the Intel Core i7-7500U processor the same operation requires 1.14 seconds, which corresponds to 876 Mflops (code compiled using gcc with optimisation level `O3'). Comparison of the performance of also other operations on elements of two arrays on both processors, indicates that the floating point performance ratio of a single core of the Intel Core i7-7500U processor and a 2 GHz Pentium 4 processor is approximately a factor 8.
 
 Since these performance comparisons may give an incomplete picture of the performance difference between both processors for the implementation of the placement algorithms, we have used in section 4 the factor 25 for the performance ratio, which is probably an over-estimation.

\begin{figure}[!ht]
    \centering
    \includegraphics[scale=.5]{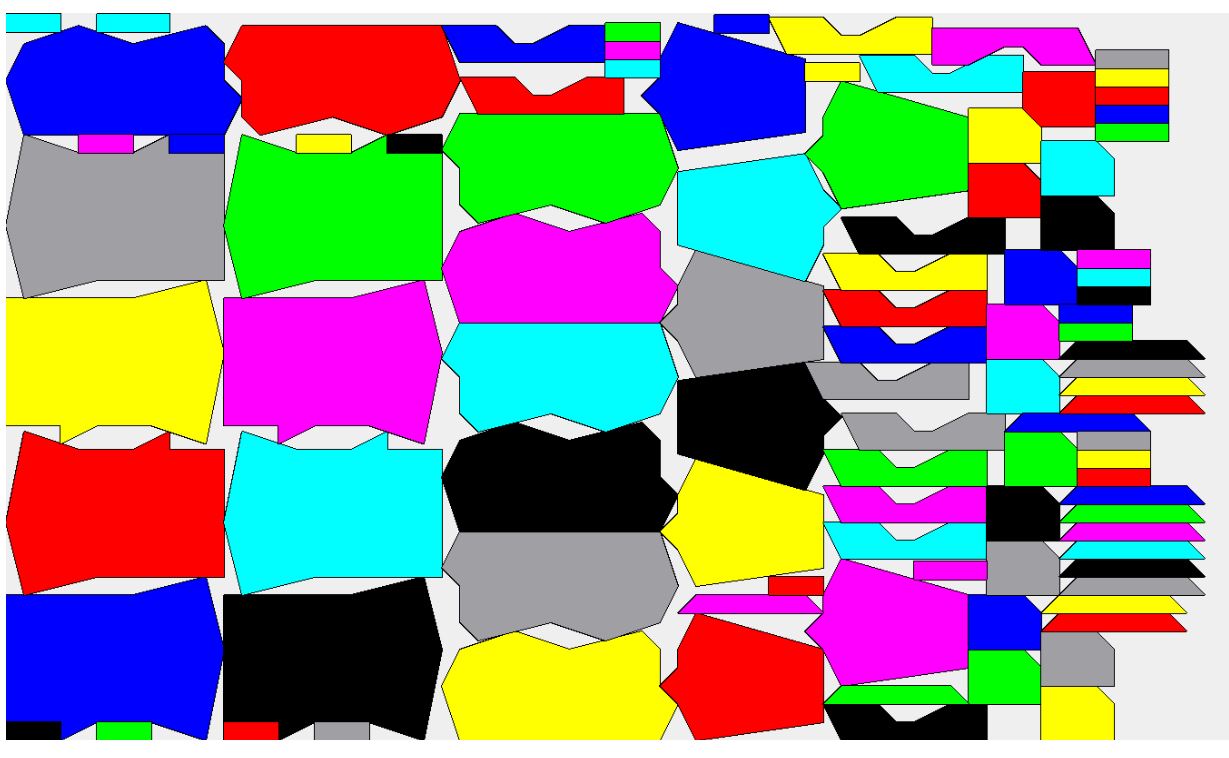}
    \caption{\change{Placement result for dataset `Shirts' semi-discretised with $R= 1$ and 2 allowed rotations ($\Delta\theta=180\degree$). Resulting strip length = 66.}}
    \label{fig16}
\end{figure}

\begin{figure}[!ht]
    \centering
    \includegraphics[scale=.4]{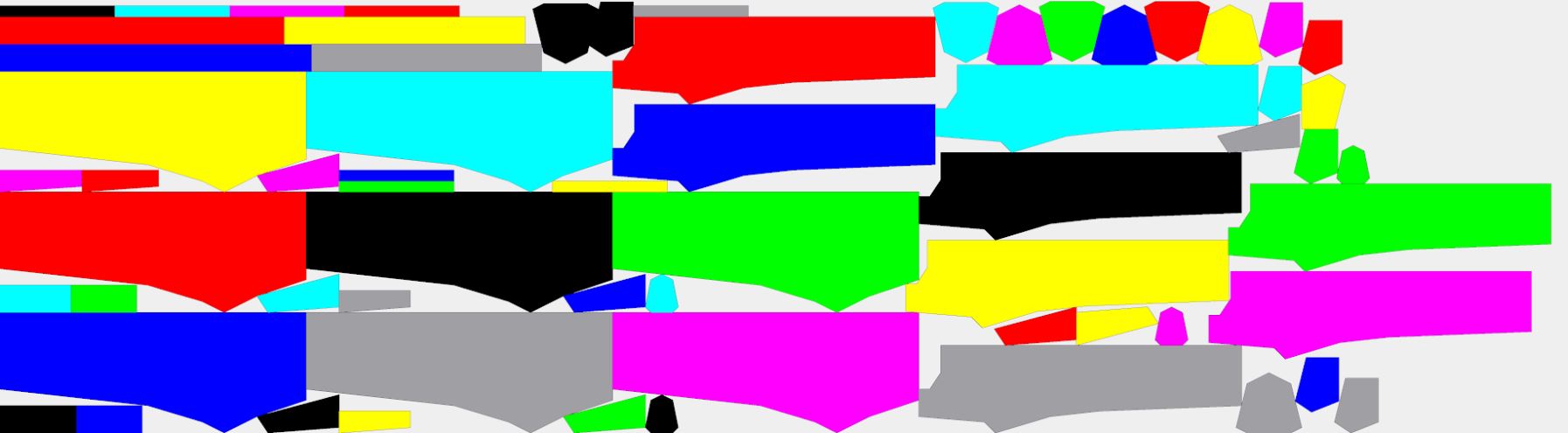}
    \caption{\change{Placement result for dataset `Trousers' semi-discretised with $R= 0.2$ and 2 allowed rotations ($\Delta\theta=180\degree$). Resulting strip length = 283.6.}}
    \label{fig16c}
\end{figure}

\begin{figure}[!ht]
    \centering
    \includegraphics[scale=.5]{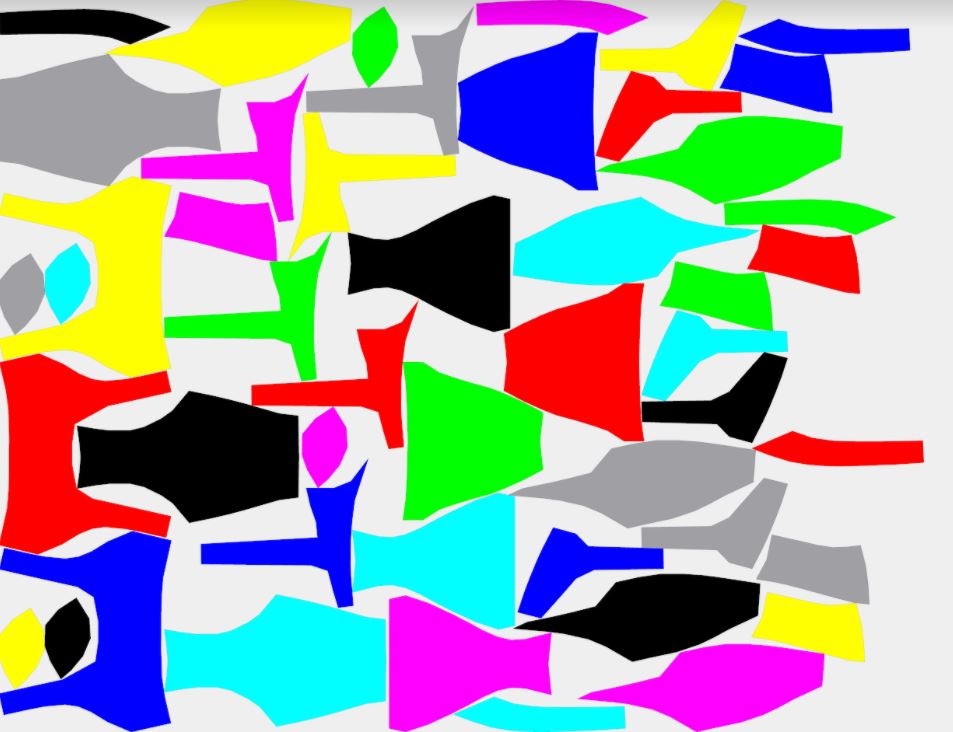}
    \caption{\change{Placement result for dataset `Swim' semi-discretised with $R= 36$ and 2 allowed rotations ($\Delta\theta=180\degree$). Resulting strip length = 7255.4.}}
    \label{fig16p}
\end{figure}

\begin{figure}[!ht]
    \centering
    \includegraphics[scale=.7]{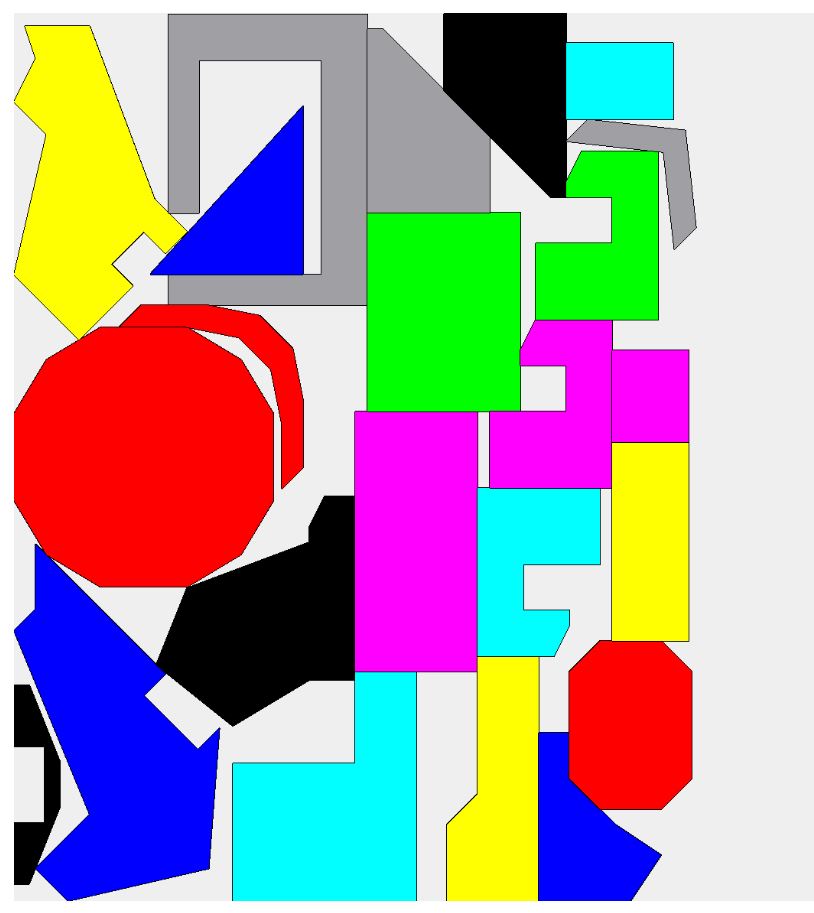}
    \caption{\change{Placement result for dataset `Han' semi-discretised with $R= 0.1$ and 8 allowed rotations ($\Delta\theta=45\degree$). Resulting strip length = 44.5.}}
    \label{fig16d}
\end{figure}

\begin{figure}[!ht]
    \centering
    \includegraphics[scale=.7]{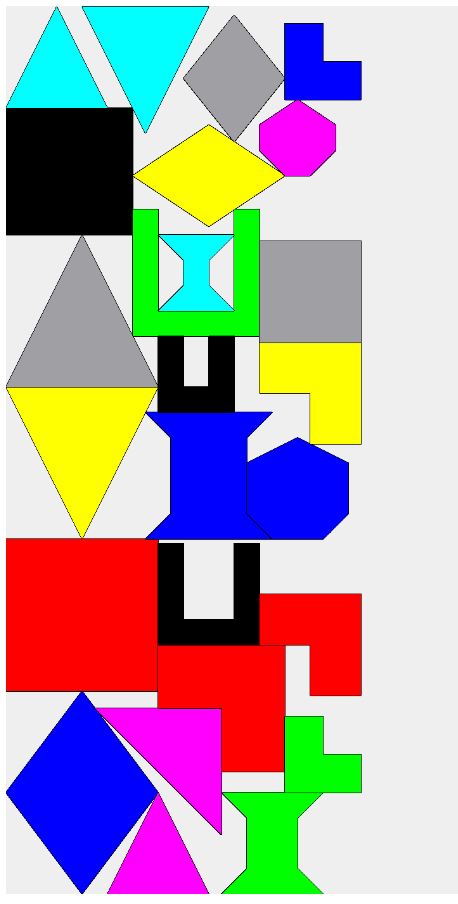}
    \caption{\change{Placement result for dataset `Jakob2' semi-discretised with $R= 1$ and no rotation allowed. Resulting strip length = 28.0.}}
    \label{fig16e}
\end{figure}

\begin{figure}[!ht]
    \centering
    \includegraphics[scale=.5]{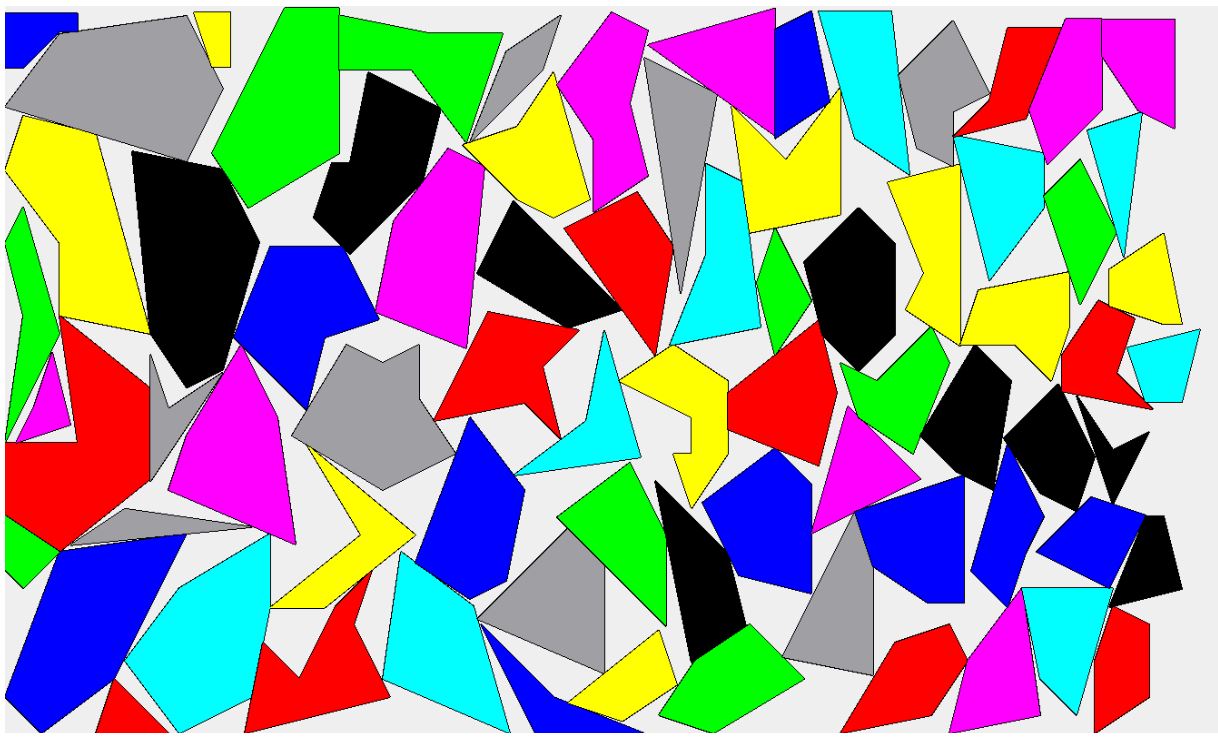}
    \caption{\change{Placement result for dataset `poly5b' semi-discretised with $R= 0.2$ and 4 allowed rotations ($\Delta\theta=90\degree$). Resulting strip length = 65.8.}}
    \label{fig16f}
\end{figure}

\begin{figure}[!ht]
    \centering
    \includegraphics[scale=.7]{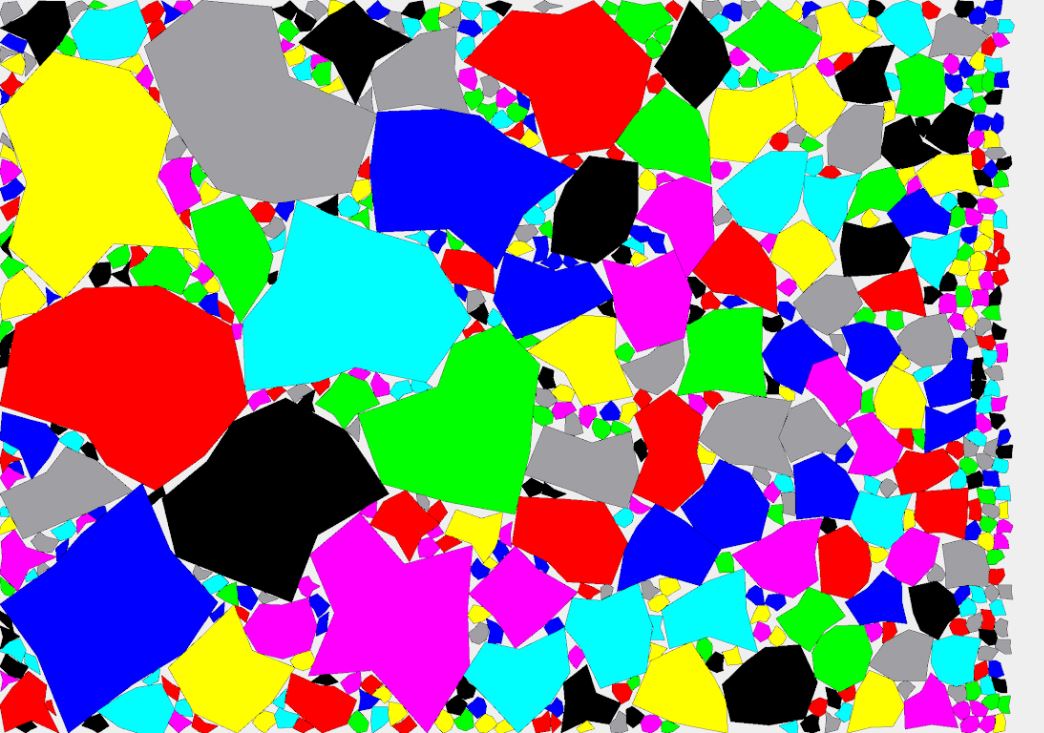}
    \caption{\change{Placement result for random data set with 550 pieces, semi-discretised with $R = 0.01$ without rotations allowed. Resulting strip length = 110.74.}}
    \label{fig16g}
\end{figure}

\end{document}